\documentclass[12pt,a4paper,twoside,openright]{report}
\input{StyFiles/packages.sty}

\usepackage[subpreambles=false]{standalone}
\usepackage{textgreek}

\begin{document} 

\definecolor{color1}{rgb}{1,0.733333333333333,0.470588235294118}
\definecolor{color0}{rgb}{0.0901960784313725,0.745098039215686,0.811764705882353}
\definecolor{color2}{rgb}{0.580392156862745,0.403921568627451,0.741176470588235}

\definecolor{color1}{RGB}{27,158,119}
\definecolor{color0}{RGB}{217,95,2}
\definecolor{color2}{RGB}{117,112,179}


\pagenumbering{roman}			

\begin{titlepage}
\newgeometry{top=3cm, bottom=3cm,
			left=2.25 cm, right=2.25cm}	
			
\AddToShipoutPicture*{\backgroundpic{-4}{56.7}{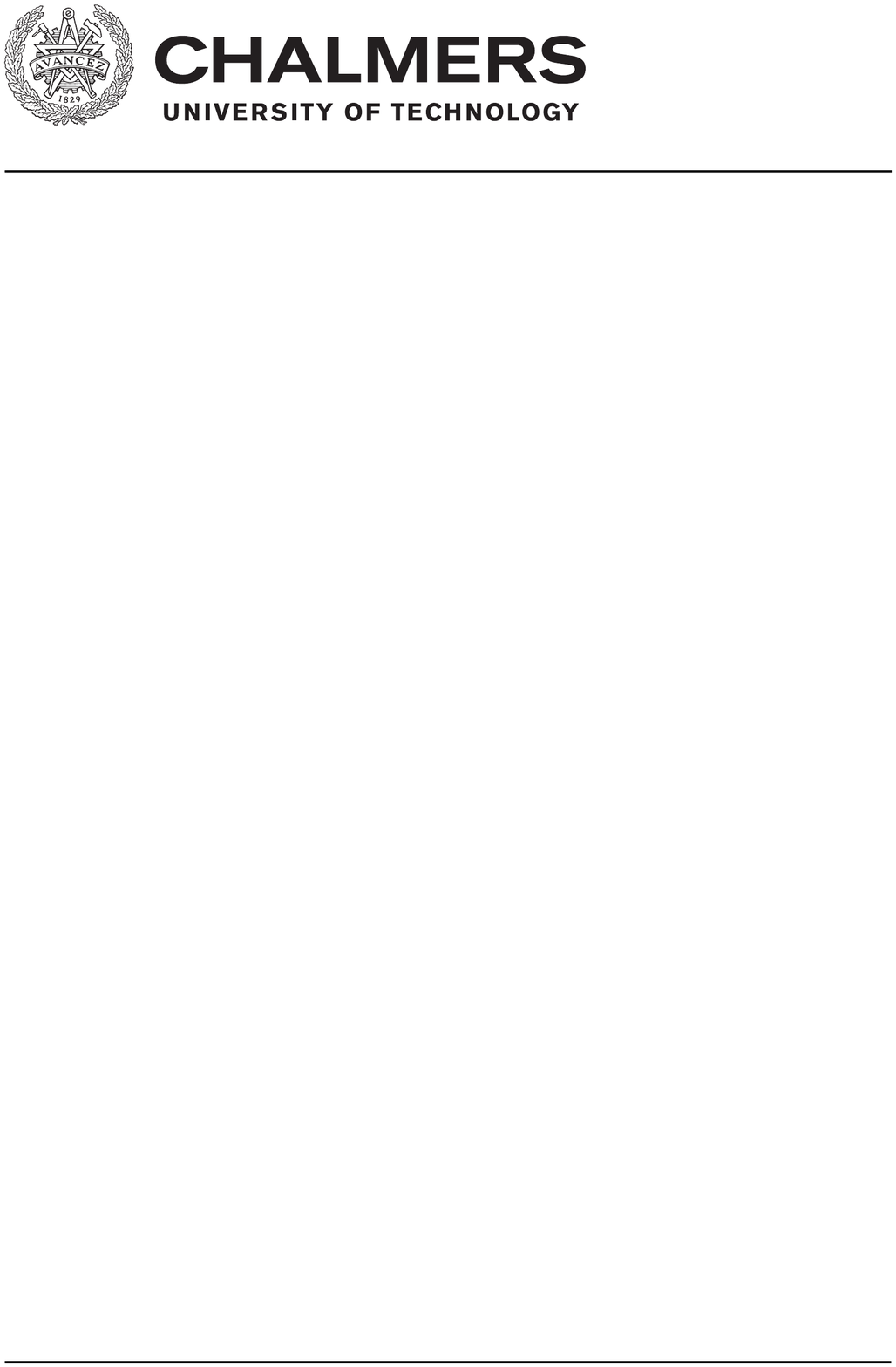}}
\addtolength{\voffset}{2cm}

\begin{figure}[H]
\centering
\vspace{1cm}	
\setlength\figureheight{0.25\textheight}
\setlength\figurewidth{\textwidth}
\input{Chapters/Front/critPlot.tex}\\\vspace{1cm}

\setlength\figurewidth{1.2\textwidth}
\setlength\figureheight{0.3\textheight}
\hspace{-4.3cm}
\input{Chapters/Front/front2d1.tex}
\end{figure}

\mbox{}
\vfill
\renewcommand{\familydefault}{\sfdefault} \normalfont 
\noindent
\textbf{{\Huge 	The role of fast ions in stabilising the ion temperature gradient mode}} 	\\[0.5cm]
\\[0.5cm]
Master's thesis in Physics and Astronomy \setlength{\parskip}{1cm}

{\Large AYLWIN IANTCHENKO} \setlength{\parskip}{2.9cm}

Department of Physics \\
\textsc{Chalmers University of Technology} \\
Gothenburg, Sweden 2017

\renewcommand{\familydefault}{\rmdefault} \normalfont 
\end{titlepage}

\newpage
\restoregeometry
\thispagestyle{empty}
\mbox{}

\newpage
\thispagestyle{empty}
\begin{center}
	\textbf{\Large The role of fast ions in stabilising the ion temperature gradient mode} \\[1cm]
	{\large AYLWIN IANTCHENKO}
	
	\vfill	
	\begin{figure}[H]
	\centering
	\includegraphics[width=0.2\pdfpagewidth]{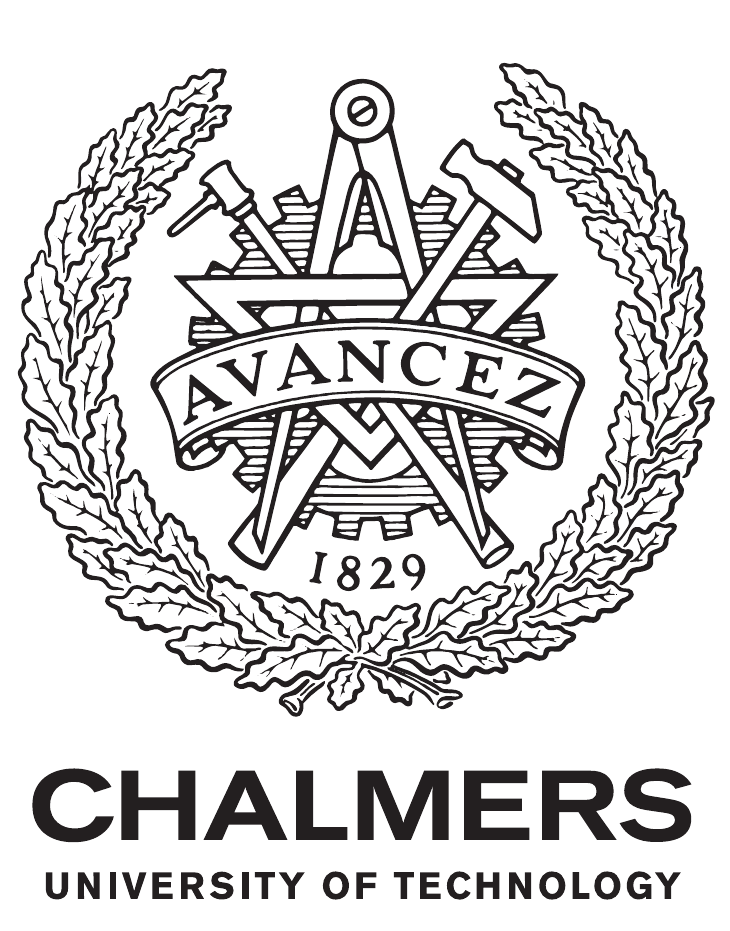} \\	
	\end{figure}	\vspace{5mm}	
	
	Department of Physics \\
	\emph{Division of Subatomic and Plasma physics}\\
	Plasma Theory\\
	\textsc{Chalmers University of Technology} \\
	Gothenburg, Sweden 2017 \\
\end{center}

\newpage
\thispagestyle{plain}
\vspace*{4.5cm}
The role of fast ions in stabilising the ion temperature gradient mode\\
AYLWIN IANTCHENKO \setlength{\parskip}{1cm}

\copyright ~ AYLWIN IANTCHENKO, 2017. \setlength{\parskip}{1cm}

Supervisor: George Wilkie, Department of Physics\\
Examiner: T{\"u}nde F{\"u}löp, Department of Physics \setlength{\parskip}{1cm}

Department of Physics\\
Division of Subatomic and Plasma physics\\
Plasma Theory\\
Chalmers University of Technology\\
SE-412 96 Gothenburg\\
Telephone +46 31 772 1000 \setlength{\parskip}{0.5cm}

\vfill
Cover: Illustration of growth rate of the ion temperature gradient instability (top) and 2d plot (bottom) of growth rates versus the fast ion density gradient ($x$-axis) and total plasma $\beta$ ($y$-axis).\setlength{\parskip}{0.5cm}

Typeset in \LaTeX \\
Gothenburg, Sweden 2017

\newpage

The role of fast ions in stabilising the ion temperature gradient mode\\
AYLWIN IANTCHENKO\\
Department of Physics\\
Chalmers University of Technology \setlength{\parskip}{0.5cm}

\thispagestyle{plain}			
\setlength{\parskip}{0pt plus 1.0pt}
\section*{Abstract}

Thermonuclear fusion is a potential candidate for providing a clean source of energy and satisfying the high electricity demands of the future. The fuel in a typical reactor is heated to a very high temperature forming a gas of charged particles known as a plasma. The fusion reactions in the tokamak have to reach a self-sustaining regime to minimise the input power required to drive the reactor. Reaching this regime demands a sufficiently low transport of energy, which remains one of the biggest challenges in plasma physics today. Turbulence driven by small scale instabilities causes large heat and particle transport and is a major limiting factor of current fusion devices. Above a critical value, the ion temperature gradient leads to the growth of a microinstability -- the ion temperature gradient mode -- that often dominates the ion energy transport. 

It has recently been discovered that energetic ions generated by auxiliary heating may reduce the growth of this instability. By applying the gyrokinetic formalism and performing linear simulations using the local continuum gyrokinetic code GS2, we explore the linear physics of this stabilising effect. In order to isolate important effects due to the presence of fast ions, we make use of the flexibility of GS2 to change the plasma and magnetic geometry parameters independently. We assess the possibility to neglect magnetic geometry changes to simplify the analysis, by investigating its contribution to the stabilising effect. For the cases studied we find that the Shafranov shift and safety factor profile might have to be taken into account. For fixed fast ion density and temperature a destabilising influence of their density gradient is found, while the high fast ion temperature gradient is stabilising, both as predicted by theoretical models. A large part of the observed stabilisation comes from the fast ion contribution to the plasma $\beta$ which is the ratio of the total thermal to magnetic pressure. In addition, the effect of $\beta$ is enhanced because of the large density and temperature gradients of the fast ions. We investigate the role of hot ion mass and charge in order to evaluate the stabilisation of different types of hot ions. The charge enhances the destabilising effect of the hot ion density gradient, while increasing the mass improves stability in general. Also the possibility of adjusting the electron and ion profiles to account for the presence of fast ions without including them as a kinetic species, is considered. We find the changes because of fast ions accessible by modifying electron and ion profiles of comparable importance as the fast ion gradients. Finally, quasi-linear theory is invoked for linking linear results to saturated values of the nonlinear heat fluxes. 

\vfill
Keywords: fusion, turbulence, fast ions, ion temperature gradient mode, quasi-linear theory

\newpage				
\thispagestyle{empty}
\mbox{}

\newpage

\setlength{\parindent}{1cm}
\thispagestyle{plain}			
\section*{Acknowledgements}
There are many people who I would like to thank, and who have contributed to this work directly and indirectly. First of all I would like to express my deepest gratitude to my supervisor George for his patience in teaching me the fundamentals of gyrokinetics, turbulence and fast ions. Throughout this work he has provided me with clear and illuminating solutions to the various problems I have had, proposed insightful interpretations, suggested interesting physical problems to tackle and directly influenced in making this thesis self-contained and interesting to read. He has definitely inspired me to continue to explore this fascinating field of turbulence and plasma physics. 

I am also grateful for the time Edmund spent in answering the myriads of questions I have had on various topics encountered during this work. A significant part of this thesis, particularly the puzzling role of geometry, has relied on the tools and data provided by Edmund not to mention the comprehensive discussions on this subject. 

To István for constantly being available to provide incredibly pedagogical and insightful answers to my every question, be it in plasma physics, physics in general or practically anything else. 

Special thanks to George, István and Stefan for proofreading this thesis and for their excellent comments which has improved this work significantly.

I also would like to thank Ian for his help, Ronald for providing the data that has allowed me to verify and compare my results and Jonathan for providing essential input files for the geometry chapter.

To T{\"u}nde, not only for inviting me to her team and giving me the opportunity to work on a topic I am very fascinated about, but also for the extraordinary care and support she has shown me and co-workers alike. To her and the rest of the plasma theory group at Chalmers I want to thank for creating the perfect environment for research, one that is characterised by collaboration, humour and where you are motivated to perform at your best.

Finally, my warmest gratitude to my family for their concern, support and encouragement. I am especially in debt to my wonderful sister Alexandra for her selfless care and patience. 

\vspace{1.5cm}
\hfill
Aylwin Iantchenko, Gothenburg, June 2017

\newpage				
\thispagestyle{empty}
\mbox{}

\newpage
\tableofcontents


\cleardoublepage
\setcounter{page}{1}
\pagenumbering{arabic}			
\setlength{\parskip}{0pt plus 1pt}


\setlength{\parindent}{1cm}
\setlength{\parskip}{0pt} 
\chapter{Introduction}\label{sec:Introduction}
The ever increasing demand of energy and evidence of the negative environmental impacts of current means of energy production calls for a sustainable energy resource to be found in the near future. A possible solution is to use fusion.

In fusion reactions two nuclei fuse to generate a large amount of energy. The easiest to initiate is the deuterium - tritium reaction generating a helium nucleus (alpha particle) and a neutron~\cite{ref:Freidberg2007}. The excess energy can be used to produce steam which drives a turbine to yield electric power. To fuse, the two light particles have to overcome the potential (Coulomb) barrier generated by electrostatic interactions between the particles. This means that the fuel in a fusion reactor has to be heated to a very high temperature in the order of $\sim 10^8 \unit{K}$. At this temperature a gas of charged particles is formed, known as a \textit{plasma}. 

To use fusion as a source of energy we have to produce more power than what is required to drive the fusion reactor, that is, to heat the plasma. Ideally we want a self-sustaining plasma, a plasma that heats itself using only the energy generated in the fusion reactions. The critical point when the plasma becomes self-sustaining is known as \textit{ignition}. The condition for reaching ignition is given by the Lawson criteria which for a given main ion density $n_i$, temperature $T_i$ and confinement time $\tau_E$ sets a minimum value of the fusion triple product $n_iT_i\tau_E$. Here and throughout this work we use subscript $i$ as a label for main (deuterium) ions. For fixed ion temperature, if the ion density is high the confinement time $\tau_E$ can be relatively low. This regime is implemented in inertial confinement fusion~\cite{ref:Keefe}. Magnetic confinement fusion instead uses low density and try to maximise the confinement time.

The confinement time  $\tau_E$ is a measure of the time the energy stays in the fusion plasma. It is equal to the ratio between the stored thermal energy and the power loss. Making $\tau_E$ sufficiently high remains one of the biggest challenges in plasma physics today. Confinement with walls is not possible since there is to date no material which can withstand the high temperatures needed in a fusion reactor. Instead, since the particles inside are charged, specially designed magnetic field configurations can be used. 

\begin{figure}[H]
    \centering
    \includegraphics[width=0.7\textwidth]{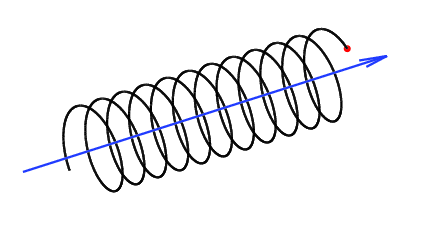}
    \caption{The motion of a charged particle in a spatially constant magnetic field (Source~\cite{ref:Bsc}).}
    \label{fig:intro:mot1}    
\end{figure}

In the presence of a magnetic field $\vecB$ and an electric field $\vecE$ a particle of charge $Ze$ and velocity $\vecv$ is influenced by the Lorentz force
\eqre{
\label{eq:intro:F1}
\Fb = Ze\left(\vecE + \frac{1}{c}\vecv\times \vecB\right),
}
where $c$ is the speed of light.
Solving the equations of motion for $\vecE = \mathbf{0}$ and constant $\vecB$ leads to a circular motion perpendicular to the field as shown in Fig.~\ref{fig:intro:mot1}. The particle traces out a circle of (thermal) Larmor radius $\rho = v_{t}/\Omega$ and rotates with the gyration frequency $\Omega = ZeB/cm$ known as the Larmor frequency. Here $m$ is the mass and for a temperature $T$ the thermal speed is given by $v_{t} = \sqrt{2T/m}$. All physical quantities in formulas in this work are written in Gaussian units. Actual values of temperature are given in keV and the mass in units of the proton mass. 

In mirror machines the parallel motion may be constrained by the straight magnetic field, but the loss cone makes those devices strongly unstable~\cite{ref:Freidberg2007}. In other configurations straight magnetic fields will not be enough to constrain the parallel motion of the particles, but the fields have to be bent to form a torus. This magnetic topology may be achieved in variety of ways. In a stellarator~\cite{ref:Hawryluk} the field line geometry is generated predominantly by external magnets that usually have complicated non-planar structure to produce an optimised magnetic field. The most recently built stellarator is the Wendelstein 7-X in Greifswald, Germany~\cite{ref:Wen}. The current thesis focuses more on the tokamak concept where a twist of the toroidal magnetic field is achieved by driving a toroidal current in the plasma. The toroidal current generates the poloidal field which together with the toroidal magnetic field creates the desired torus topology. The International Thermonuclear Experimental Reactor (ITER), currently under construction in Cadarache, France, will be the largest tokamak experiment yet. The aim of the experiment is to demonstrate the physical feasibility of fusion energy production, by generating an output power ten times the input power~\cite{ref:Bateman}. The plasma parameters considered in this thesis are from The Joint European Torus (JET)~\cite{ref:Jet1992} tokamak in Culham, UK which is currently the largest tokamak in the world. 
  
\begin{figure}
    \centering
    \includegraphics[width=0.4\textwidth]{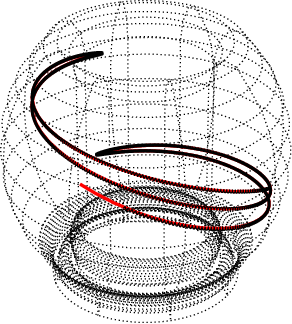}
    \caption{The banana shaped trajectory of a trapped particle moving in a spatially varying magnetic field inside the ITER reactor (Source~\cite{ref:Bsc}).}
    \label{fig:intro:mot2}    
\end{figure}

Even though these magnetic field configurations are designed to confine energy, it may still escape confinement due to various transport mechanisms.  In classical transport theory energy and particles gradually move across field lines as a consequence of Coulomb collisions between species. The process is further complicated by various drifts arising from Eq.~\eqref{eq:intro:F1} when both spatially varying electric and magnetic fields are considered. Additionally, some particles may become trapped in which case their trajectories takes a banana shaped form. An example is depicted in Fig.~\ref{fig:intro:mot2}. Including these various effects and collisions is the basis of the well studied neoclassical transport theory~\cite{ref:Helander}. Although in the edge neoclassical transport can play a more important role~\cite{ref:Catto11,ref:Catto13}, in the core of reasonable sized tokamak reactors neoclassical theory predicts a tolerable level of heat transport~\cite{ref:Galambos}. The much higher level of transport observed experimentally, is instead caused by a completely different mechanism, namely \textit{turbulence}~\cite{ref:Wootton}. 

\section{Turbulence}
Turbulence is a phenomena we encounter every day: when we study currents in water, the smoke from a chimney or stir milk into our coffee. We see it as the unpredictable coherent structures which appear after disturbing a fluid or a gas. Even in simple fluids, predicting the formation of these turbulent structures is a very difficult task and turbulence in general remains one of the unsolved problems of classical physics today. In fusion the picture is even more complicated since we need to study turbulence in plasmas, involving chaotic fluctuations in density, temperature, electric and magnetic fields, all interacting with each other. 

Microturbulence consists of small scale fluctuations with wavelengths in the order of the ion Larmor radius $\rho_i$.  Ion or electron density and temperature gradients may cause exponential growth in amplitude of perturbations in density, temperature and fields. Nonlinear interactions between the fields saturate the growth of these instabilities generating turbulence and saturates the heat flux  $\vecQ_s = \int \epsilon_s f_s \vecv\text{d}^3\vecv$ where $\epsilon_s = \frac{1}{2}m_sv^2$ is the energy of the particle species $s$ with mass $m_s$ moving at a speed $v$. The distribution function $f_s$ describes the probability of finding a particle at spatial coordinate $\vecx$ and with a velocity $\vecv$. The time evolution of this distribution function is described by the Fokker-Planck equation which will be presented in Section~\ref{sec:studTurb:gk}. Higher energy flux leads to power loss and therefore a shorter confinement time $\tau_E$ making ignition more difficult to reach. The loss is given by the surface integral of the $\int_{\partial \Omega_p} \vecQ \cdot \dif{\vecS}$, integrated over the plasma boundary $\partial \Omega_p$. To use fusion as a main source of energy now or in the future, reduction of this heat flux and turbulence is required.  Recently it has been discovered that energetic ions generated by auxiliary heating or produced by the fusion reactions themselves may have such an impact on the turbulence~\cite{ref:Citrin}.

\section{Fast ions}
The ignition criteria on the fusion triple product $n_iT_i\tau_E$ essentially breaks down the condition for fusion reactors into two separate parts. The first, is reaching a sufficiently high thermal pressure $p_i = n_iT_i$. In a tokamak the density is relatively small and the high pressure is mainly reached by increasing the temperature of the ion population. This requires the plasma to be heated. Some of the heat comes from the internal currents, but a large fraction is provided by means of auxiliary heating such as neutral beam injection (NBI) and ion cyclotron resonance heating (ICRH). These types of heating introduces small populations of particles that are much more energetic than the thermal populations. These particles are known as \textit{fast ions} (here ``fast'' refers to the high temperature of this ion population). Even with their low density the high temperature of the fast ion population leads to a significant contribution to the total thermal pressure. Collisions transfer the energy of the fast ions to the thermal species and leads therefore to an increase in the first part of the fusion triple product $p_i\tau_E$. This is their primary, and well known effect of adding fast ions in the plasma. There is also a second, less studied benefit, which is the increase in the second parameter, the confinement time $\tau_E$.

\begin{figure}[H]
\setlength\figureheight{0.3\textheight}
\setlength\figurewidth{\textwidth}
    \centering
    \input{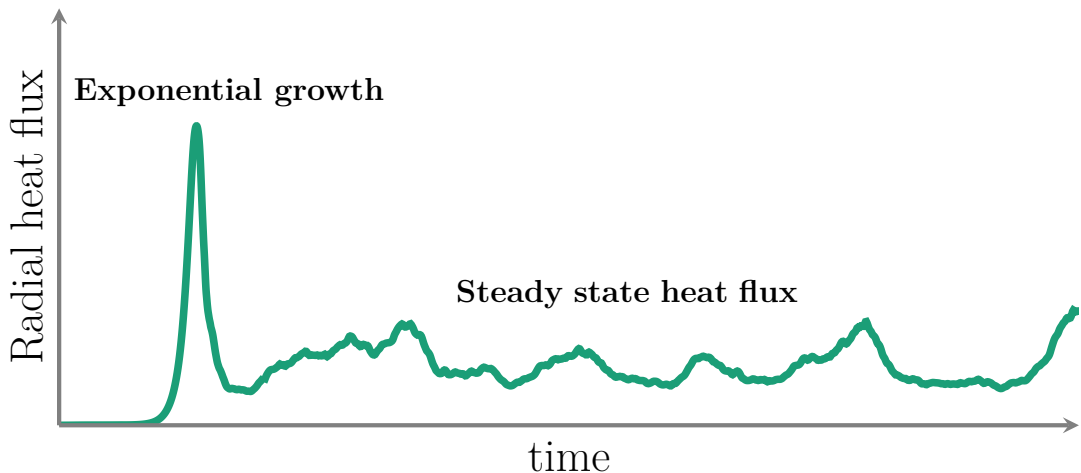} 
    \caption{An illustration of the radial heat generated by turbulent fluctuations in a plasma.  Linear physics causes exponential growth of the heat flux which is saturated at later times by nonlinear mechanisms.}
\label{fig:intro:fluxes}
\end{figure}

An increase in the confinement time is equivalent to a reduced heat flux. In Fig.~\ref{fig:intro:fluxes} we illustrate the nonlinear heat flux generated from turbulent fluctuations in a plasma. In the beginning the heat flux grows exponentially but as the time is increased, it saturates around some steady state value. Citrin et al.~\cite{ref:Citrin} compared the steady state heat flux generated in a plasma with and without fast ions, and found a strong stabilising effect when the fast ions had been included. The effect was particularly strong for the 73224 JET discharge~\cite{ref:Bravenec}. A reduction in the heat flux is equivalent with a decrease in the ion energy transport and therefore an increase in the confinement $\tau_E$. The effect is of particular interest for the fusion product in a deuterium-tritium plasma, because of the fast fusion-born alpha particles. If these particles have a suppressing effect on the turbulence a self-reinforcing loop might arise. The alpha particles heat the plasma, triggering more fusion reactions which generates more fast ions which in turn reduce the turbulent transport leading to longer confinement time and an even hotter plasma. The benefit of alpha particles increases the more fusion reactions that takes place. In ITER they may therefore play an important role in increasing the confinement time and reaching ignition. 

Information about the species and the magnetic geometry which describe the fusion plasma in the simulations presented in this work, is taken from the 73224 JET discharge~\cite{ref:Bravenec}. The choice of this discharge is motivated by the strong reduction in the heat flux reported by Citrin et al.~\cite{ref:Citrin}.

While the reduction in the heat flux demonstrated by Citrin et al. is interesting by itself, we wish to go one step further and explain \textit{why} fast ions have this stabilising effect. The most direct approach would be to evaluate the heat flux for different fast ion characteristics, varying the density, temperature, charge etc. But there is a caveat with this strategy, being the computational complexity in calculating the nonlinearly saturated heat fluxes. This requires nonlinear physics to be included leading to expensive and complex simulations that are difficult to interpret. Another option to consider, is to exclude the nonlinear physics, and focus on the linear mechanisms instead. In this case, the heat fluxes will not be saturated and reach the steady state observed at later times in Fig.~\ref{fig:intro:fluxes} but will grow exponentially with some growth rate $\gamma$, corresponding to the initial peaks in the figure. These phases of exponentially increasing fluctuation amplitudes correspond to a growth of \textit{microinstabilities} which are the aformentioned unstable modes with wavelengths in the order of the ion Larmor radius $\rho_i$, driven by ion or electron temperature and density gradients~\cite{ref:Waltz}. Since the equation describing the linear physics of these instabilities is not explicitly dependent on time, the time evolution of a fluctuating quantity $\phi$ may be written in normal mode expansion~\cite{ref:Freidberg2007}
\eqre{
\label{eq:StudTurb:NormMod}
\phi(t) = \phi_0e^{-i\tilde\omega t}. 
}
If the frequency of oscillation $\tilde \omega = \omega + i\gamma$ is purely real ($\gamma = 0$) the wave is simply oscillating around some equilibrium. If it also contains an imaginary part ($\gamma \ne 0$) it will grow (or decay if $\gamma < 0$) in time, with a growth rate $\gamma$. On the other hand, if different modes interact with each other, the dynamics of $\phi$ has to be described by a nonlinear PDE and its time evolution cannot be constructed by independent frequencies $\omega$, as is done in Eq.~\eqref{eq:StudTurb:NormMod}.

\begin{figure}[H]
 \setlength\figureheight{0.25\textheight}
 \setlength\figurewidth{0.9\textwidth}
       \centering
    \input{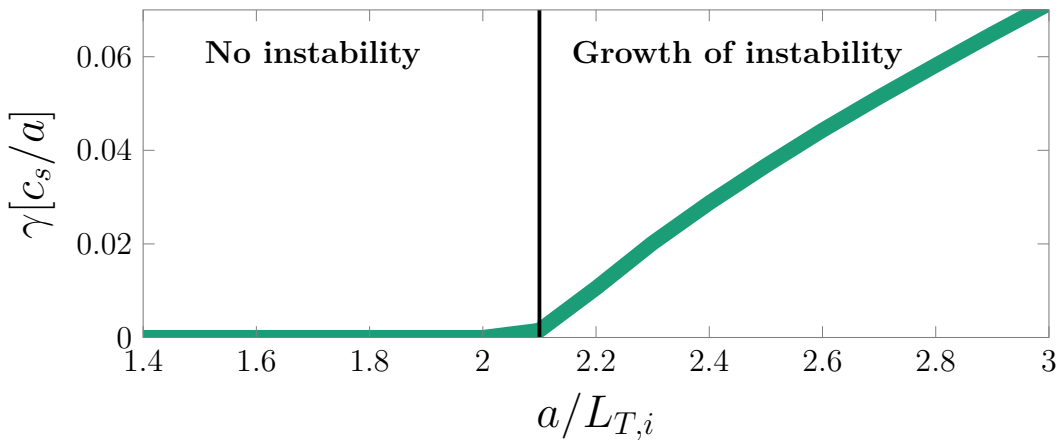}
    \caption{Growth rates for increasing main ion temperature gradient. Simulations have been performed using the gyrokinetic continuum code \texttt{GS2} presented in Chapter~\ref{CH:studTurb}. Growth rates are presented in units of the ion sound speed $c_s = \sqrt{T_e/m_i}$ over the tokamak minor radius $a$ defined in the next Chapter. Here $T_e$ is the electron temperature, and $m_i$ is the main ion mass. Also $L_{T,i} = -\dif{\ln T_i}/\dif{r}$ is the main ion temperature gradient length scale. No instability exists before a critical value of the ion temperature gradient, after which a roughly linear scaling is seen.}
\label{fig:intro:Ref_tg}
  \end{figure}

A particular microinstability is believed to dominate the ion heat transport in the core of a conventional tokamak. It is the ion temperature gradient (ITG) mode~\cite{ref:Estrada}, which above some critical point, is driven by the ion temperature gradient. An example of the scaling of growth rates with the ion temperature gradient is shown in Fig~\ref{fig:intro:Ref_tg}. In the figure we see that the growth rates are zero until some critical value of the ion temperature gradient is reached. Above this value the growth rate scales roughly linearly with the ion temperature gradient. 
The increase of ITG growth rates leads to the unwanted phenomenon known as profile stiffness. This means that the temperature profile inside the fusion reactor cannot be made significantly steeper by simply heating the main ions. More heating leads to larger temperature gradients and therefore higher growth rates which in turn will lead to larger heat fluxes, transporting the injected energy that heats the main ions, out of the reactor. The size of a fusion reactor implies certain temperature gradients required to sustain a high temperature in the core while minimising the temperature load on the surrounding walls and for achieving efficient energy production.  Profile stiffness is therefore a major limitation in how small fusion reactors can be made.

Including only the linear physics, the simulations are made significantly cheaper to run, and we can perform the desired scan in the fast ion parameters to evaluate their role on the ITG growth rates. Still, in the end, the ultimate goal is to explain the reduction in the heat flux, not the growth rates. We need therefore to link the exponential growth of the fluxes, to their final, saturated values. This can be achieved with \textit{quasi-linear} theory. 

Using the \textit{mixing-length} argument for estimating the time until turbulent structures decorrelate and \textit{mix} with the environment, the quasi-linear theory predicts the nonlinear saturated heat fluxes to scale roughly as $Q \propto \gamma/k_{\perp}^2$~\cite{ref:Casati}. Here $\gamma$ is the growth rate of the unstable mode and $k_{\perp}$ the perpendicular component (with respect to the magnetic field) of the wave vector. Quasi-linear theory is used as an important component in interpretive simulations of tokamak plasmas, to include the transport because of turbulence, without having to run complex nonlinear simulations.  An example is the code QuaLiKiz~\cite{ref:Bourdelle2007}  which is using quasi-linear theory to predict results from experiments.

Instead of running nonlinear simulations, we may evaluate the role of fast ions on the growth of ITG, and then use the quasi-linear model to predict their role on the saturated heat flux, in an attempt to explain the result shown by Citrin et al. Recently, many studies have been done in the attempt of explaining similar figures. These have both been of linear kind, where reduction in growth rates is investigated, and nonlinear where the saturated heat and particle fluxes instead are considered. See for example the work done by Citrin et al.~\cite{ref:Citrin}, Liljeström et al.~\cite{ref:Liljestrom1990} and Garcia et al.~\cite{ref:Garcia}. 

To properly understand the role of fast ions in stabilising the turbulence the various global effects and changes in the plasma and magnetic geometry parameters, have to be captured and understood. One of these parameters is $\beta = 8\pi nT/B^2$ which is the ratio of the thermal to magnetic pressure. Here $B$ is the magnitude of the magnetic field and we have summed over all species in the plasma to obtain the total thermal pressure $p = \sum_j n_jT_j$, where $n_j$ and $T_j$ are the species density and temperature respectively. Fast ions contribute to $\beta$ by increasing the total thermal pressure $p$.

Bravenec et al.~\cite{ref:Bravenec} have observed that increasing $\beta$ is very effective in reducing growth rates. It is observed that the stabilising effect of fast ions is strongest at high $\beta$ close to the point when another instability start to dominate over ITG. This instability is the kinetic ballooning mode instability (KBM)~\cite{ref:Citrin2} (note that beta is unlikely to be sufficiently high to destabilize KBMs in tokamak cores, except in plasmas with extended low magnetic shear regions, such as hybrid discharges~\cite{ref:Moradi14}). Citrin et al.~\cite{ref:Citrin,ref:Citrin3} claimed that this stabilising influence is enhanced by high density and temperature gradients of the fast ions.

In a tokamak, the magnetic field ``confines'' the particles with magnetic fields which constrains their thermal motion. A change in the pressure will lead to a change in the magnetic field configuration required to balance the pressure force of the particles. Therefore the fast ion contribution to the thermal pressure will affect the  magnetic configuration. In turn the changes in this configuration are known to affect the stiffness of the temperature profiles~\cite{ref:Citrin}. Bravenec et al.~\cite{ref:Bravenec} also showed that NBI injected fast ions may generate a parallel velocity gradient (PVG) which has a strong destabilising effect.

In this work we present a careful linear analysis on the role of fast ions on the ITG turbulence. In our treatment we model the fast ions with a Maxwellian distribution at a high temperature $T_f$. We neglect plasma rotation and perform our analysis in a small region of the plasma such that variation of equilibrium quantities can be expanded linearly in a radially local description.

The remainder of this thesis is organised as follows. First in Chapter~\ref{CH:FIin} we introduce the concepts of fast ions and discuss the main sources before exploring the global effects they have on the plasma. Their high contribution to the pressure leads to a change in the magnetic geometry, which we investigate using a numerical tool to calculate the magnetic geometry for a given pressure profile. In Chapter~\ref{CH:studTurb} the concepts of turbulence and microinstabilities are presented. To study turbulence and microinstabilities in plasmas, we employ the \textit{gyrokinetic} formulation and solve the gyrokinetic equation for the fluctuating turbulent distribution function. The theory takes advantage of the properties of turbulent fluctuations, to separate it from both very rapidly varying quantities but also slowly varying equilibrium parameters. We use the gyrokinetic formalism and outline the main steps in deriving the gyrokinetic equation. In this work we solve the equation numerically using the gyrokinetic continuum code \texttt{GS2} for which we show the relevant input definitions and normalisations.  We then focus on the most important microinstability for perpendicular ion heat transport: the ITG mode. For a two species plasma we derive a dispersion relation for describing main characteristics of this mode. The effect of fast ions is then also included. Although very approximate, we show that the derived expression agrees well with corresponding results from \texttt{GS2}.
In Chapter~\ref{CH:analysis} we present a careful analysis on the effect of fast ions on the ITG turbulence. We begin with a general overview and then explain the observed behaviour by isolating the effect of temperature, density, their gradients, mass and charge  of the fast ions. We end this chapter by invoking the quasi-linear model and examine if our linear results are sufficient to explain the strong nonlinear reduction in the heat flux, for the modelled 73224 JET discharge.  Finally in Chapter~\ref{CH:discussion} we discuss and summarise our findings.

Simulations were performed on the supercomputer Marconi, at CINECA in Italy.
\clearpage
\setlength{\parskip}{0pt} 

\chapter{Global effects}\label{CH:FIin}
Before we attempt to explain the role of fast ions in stabilising ITG turbulence, we have to take a step back and define what we mean with a ``fast ion''. What are the characteristics of these particles? This was done to some extent in the previous chapter, but here we do this more thoroughly. Once we have outlined the properties of this energetic species we have to understand, what kind of changes we have to include to properly account for their presence in the fusion plasma. Their actual main purpose is heating the plasma but we will consider other effects. Answering these questions and clarifying the most important global effects of fast ions, is the purpose of this chapter.  

We begin with describing the sources of energetic ions focusing on neutral beam injection (NBI) and ion cyclotron resonance heating (ICRH). We carefully examine the most important changes that occur in the plasma because of fast ions. The pressure is important in terms of the magnetic geometry which will be described in this chapter. We define the concept of magnetic geometry in Section~\ref{sec:Geometry:mgeom} by introducing an equation describing the magnetohydrodynamic equilibrium, the Grad-Shafranov (G-S) equation and discussing its implications. 

Given a change in the pressure and current we use a numerical tool: \texttt{CHEASE} to solve the G-S equation numerically. This will generate all necessary information about the magnetic geometry. For quantitative comparison of geometry for a change in pressure we present the Miller model for performing a local parametrisation of the equilibrium. The model will reappear again in Chapter~\ref{CH:analysis} where gyrokinetic simulations of the turbulence will be performed. 
By implementing the described tools and parametrisation, in Section~\ref{ref:geom:Eff} we investigate the changes in the magnetic geometry because of fast ions. 

\begin{figure}[H]
    \centering
    \centering\resizebox{4.0in}{!}{\input{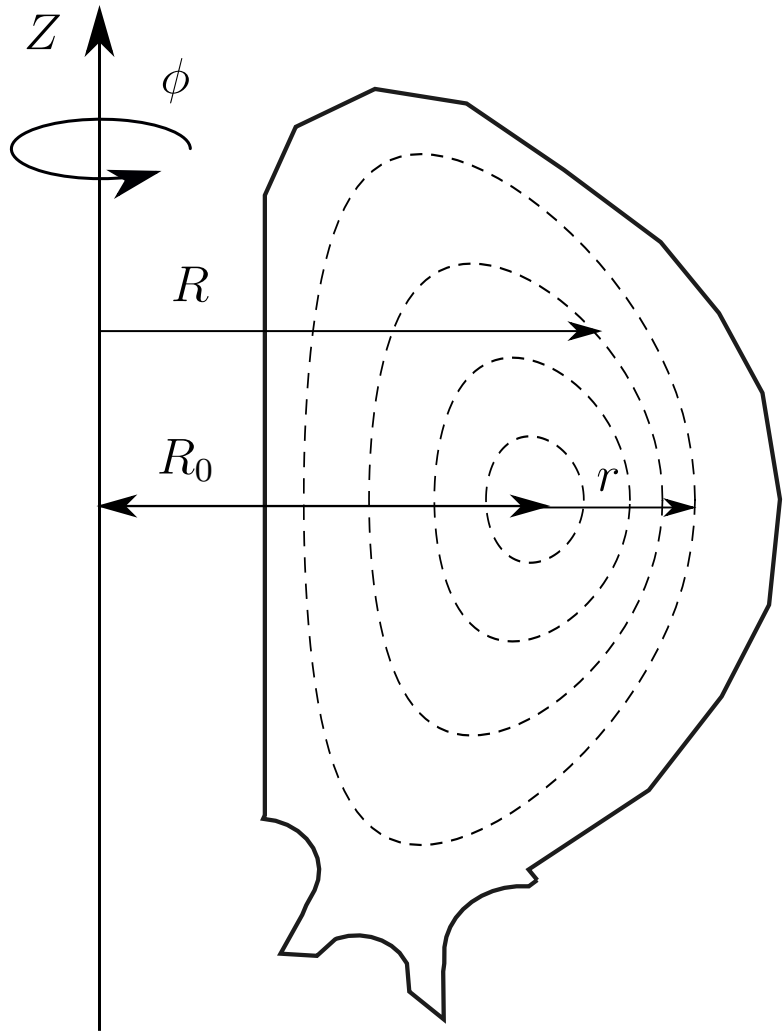}}
    \caption{Definition of coordinates used in this chapter for describing the tokamak geometry. The radial coordinate from the vertical $Z$ axis of symmetry is the major radius $R$. Also shown are some example flux surfaces to be defined in Section~\ref{sec:Geometry:mgeom}. These flux surfaces converge towards the \textit{magnetic axis} located at $R = R_0$. The minor radius $r$ is used to label the flux surfaces and $\phi$ is the toroidal (azimuthal) angle with positive direction into the page.}
\label{fig:geom:tokCoord}
\end{figure}

Before we begin, it is convenient to introduce a suitable coordinate system for describing the tokamak geometry. In Fig.~\ref{fig:geom:tokCoord} we present the poloidal cross-section of a tokamak device, along with an illustration of flux surfaces (i.e. surfaces on which the magnetic field lines lie). These will be defined in Section~\ref{sec:Geometry:mgeom}.

The major radius $R$ measures radial distance from the axis of symmetry, $Z$ is the vertical coordinate and $\phi$ is the toroidal angle describing a rotation around the $Z$-axis, with positive direction into the page. The flux surfaces converge towards the \textit{magnetic axis} located at $R = R_0$. The minor radius $r$ is used to label these surfaces. It is defined as the half-width of a flux surface at the height of the magnetic axis, and is consequently, constant on a flux surface.  We label the minor radius at the last closed flux surface (LCFS) with $a$.
\newpage
\section{Sources of fast ions}
For fusion reactions to take place the fusing particles needs to have enough energy to overcome the Coulomb barrier. To reach these high energies the plasma has to be heated. In a tokamak, some of the heating power is provided by the induced current used to create the poloidal magnetic field. This is known as resistive (Ohmic) heating and is dependent on the number of collisions occurring in the plasma. But the plasma is a peculiar medium where the collision frequency reduces rather rapidly with the temperature. The power from resistive heating can therefore only raise the plasma temperature to $T\approx \SI{3.3}{keV}$. Ignition requires $T\approx \SI{15}{keV}$ and therefore additional auxiliary heating has to be provided~\cite{ref:Freidberg2007}. This type of heating is the source of fast ions. We will focus mainly on two types of auxiliary heating: neutral beam injection (NBI) and ion cyclotron resonance heating (ICRH) which both create fast ions of different characteristics. 

\subsection{Neutral beam injection}
NBI is one of the popular choices of heating used in conventional tokamak devices. In this scheme a beam of highly energetic atoms are injected in the plasma. Since the plasma will deflect the motion of any charged particle because of the electric and magnetic fields, the injected particles have to be neutral. After travelling some distance collisions will ionise the neutrals in the beam. When the highly energetic beam becomes ionised we will have a small fraction of fast particles constrained by the fields in the plasma. After a short duration the fast ion population will slow down by means of collisions and form a thermalised population at or close to the bulk ion temperature. Transfer of the fast ion energy to the bulk species (i.e. thermal population) raises the temperature of the thermal species. In the presence of a constant source of NBI ions the new ion population will have a tail in velocity space. It is this tail which we are interested in and which we will refer to as \textit{fast ions}. When created by NBI these fast ions tend to develop large density gradients rather than large temperature gradients.

\subsection{Ion cyclotron resonance heating}
Instead of injecting fast particles in the plasma another choice is to have an antenna inject radio frequency waves. If the frequency of these waves $\omega$ is a multiple of the ion cyclotron frequency $\omega = k\Omega_i, k = 1, 2,  \dotsc$, the wave may give up its energy to heat these plasma ions. For $k = 1$ this is known as the first harmonic, $k = 2$ second harmonic etc. Again a fast ion population is introduced which collide with colder particles and gives up its energy to raise the temperature. What is different for ICRH is that instead of injecting a density of new particles ICRH takes an existing population and pushes it towards the higher velocity regime.  These type of fast ions will instead have a high temperature gradient. 

 Distinction between NBI and ICRH generated fast ions will be important in the following chapters where the fast ion effects on the turbulence will be studied. We will explicitly make a distinction between ICRH heated fast ions from NBI by considering either particles with a high temperature or density gradients. 

\subsection{Alpha particles}\label{sec:geom:alpha}
As we heat the plasma more and more using Ohmic and auxiliary heating, the number of fusion reactions increases. Each of these reactions creates an alpha particle at \SI{3.5}{MeV}. Since these are charged particles they will stay confined in a large\footnote{The plasma volume has to be large since the alpha particles, because of their high energy, will have very wide banana orbits. In a small plasma they would simply escape the plasma directly.} plasma and give up their excess energy as they thermalise with the surrounding population. Once sufficient temperatures are reached the heating power from the alpha particles will start to dominate over possible losses and the fusion plasma will become self-sustained. This is known as \textit{ignition}~\cite{ref:Freidberg2007}.

Part of this thesis is aimed at understanding whether the effects on the ITG turbulence obtained from NBI and ICRH heated fast ions works favourably for alpha particles as well, that is, that they also stabilise ITG turbulence. A reduction in ITG turbulence would imply a decrease in transport the more fusion reactions that occur. Consequently ignition could then be reached more easily than without considering the favourable effects of the fast ion populations. This is another motivation for why fast ion effects are important to study. 

Similarities in how the alpha particles are generated with NBI suggests that these would also have a high density gradient. Their features, roughly speaking, should therefore be similar to the NBI heated population, apart from possible velocity anisotropies for NBI ions.

\section{The global effects of fast ions: an overview}\label{sec:geom:equil}
\begin{figure}[H]
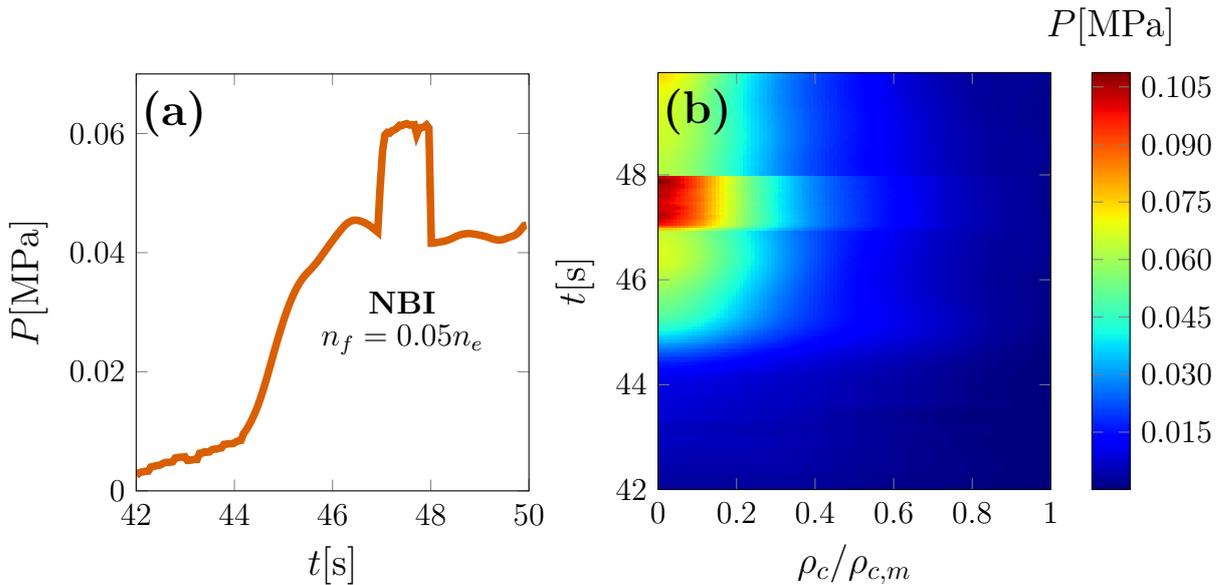

 \setlength\figureheight{0.3\textheight}
 \setlength\figurewidth{0.45\textwidth}
    \begin{subfigure}[b]{0.47\textwidth}
    \input{../../figures/stuff/ptot.tex}
    \phantomcaption
        \label{fig:Geometry:PevolT_a}
    \end{subfigure}
    \begin{subfigure}[b]{0.47\textwidth}
    \input{../../figures/stuff/ptot_2d.tex}
        \phantomcaption
    \label{fig:Geometry:PevolT_b}
    \end{subfigure}    
    \caption{Time evolution of the pressure at one radial coordinate $\rho_c/\rho_{c,m} = 0.3$ (left) and the corresponding full radial distribution of the pressure in the plasma (right). Fast deuterium is injected in the 47-48 \unit{s} time-slot. The radial coordinate $\rho_c/\rho_{c,m}$ is defined with $\rho_c = \sqrt{\Psi_t \pi B_0}$ where $B_0$ is the vacuum magnetic field at some reference major radius $R_c$ and $\Psi_t$ is the toroidal magnetic flux. Then $\rho_{c,m}$ is the value of $\rho_c$ at the last closed flux surface (to be defined in Section~\ref{sec:Geometry:mgeom}). We should note that the time-evolution is also including all the initialisations. The discharge itself lasts for only about six seconds (44-\SI{50}{s}). The data was generated using the \texttt{CRONOS} suite~\cite{ref:CRONOS}.}
\label{fig:Geometry:PevolT}
\end{figure}

Imagine we have initiated a tokamak discharge and managed to create a plasma consisting of only deuterium ions and electrons with densities $n_D, n_e$. For a certain duration the NBI device is turned on, injecting fast ions, and then is turned off again. The fast ion density is $n_f \approx 0.05n_e$, where $n_e$ is the electron density. If the total pressure in the plasma is measured, we would obtain something similar to Fig.~\ref{fig:Geometry:PevolT} where the time evolution of the total pressure (Fig.~\ref{fig:Geometry:PevolT_a}), and its radial distribution (Fig.~\ref{fig:Geometry:PevolT_b}), is shown. Even with their small density $n_f = 0.05n_e$, their high temperature leads to a significant increase in the pressure. As their purpose is to heat the part of the plasma where fusion reactions occur, they are concentrated close to the centre. 

The plasmas we consider here are quasineutral which means that the total charge averaged over a macroscopic volume of the plasma is very close to zero. For our case this means that the total charge contribution from all the electrons $-ne$, main ions $Z_Den_D$ and the fast ions $Z_{f}en_f$ has to add up to zero. In addition to their direct effect -- the increase in the total pressure -- they also change the densities of the bulk species. Which of the two, electrons or deuterium ions, should be used to satisfy quasineutrality is unclear. In NBI new electrons are introduced as the fast ions are created from the ionisation of the neutral beam particles. For ICRH injected fast ions a certain amount from the other colder ion population is taken while the electron densities should stay more or less the same. In reality both the ion and the electron profiles will change when heating is turned on. In our theoretical study we can consider two extreme situations, when either the ion or the electron density is kept constant. This should hopefully bracket the experimental situation. As seen in Fig.~\ref{fig:Geometry:PevolT} in this case the electron population has been increased to account for the excess positive charge introduced by the fast ions.  The shape of the $n_e$ profile has also been modified in particular close to the centre ($\rho_c/\rho_{c,m} = 0$) where $n_f$ is the largest.

\begin{figure}
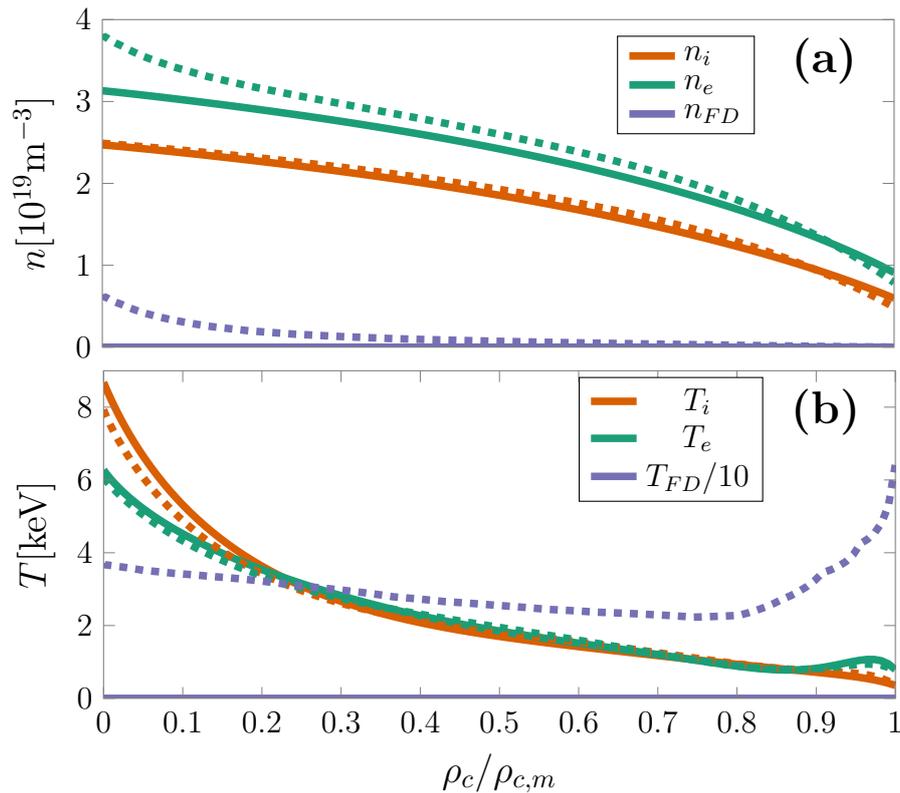

 \setlength\figureheight{0.25\textheight}
 \setlength\figurewidth{0.8\textwidth}
     \begin{subfigure}[b]{\textwidth}
    \centering
    \input{../../figures/GS2IN/dens_sp.tex}
    \phantomcaption    
            \label{fig:Geometry:sp_NT_a}
    \end{subfigure}\\
    \vspace{-0.5cm}
    \hspace{0.1cm}
    \begin{subfigure}[b]{\textwidth}
            \centering
    \input{../../figures/GS2IN/temp_sp.tex}
    \phantomcaption
        \label{fig:Geometry:sp_NT_b}
    \end{subfigure}
    \caption{Density (top) and temperature (bottom) profiles for a \texttt{CRONOS} model of a plasma with only NBI heated fast deuterium ions. The temperature of the fast particle has been rescaled by a factor of 10 for convenience. The radial coordinate is same as in Fig.~\ref{fig:Geometry:PevolT}. Solid lines represent the case without any fast ions, while dashed have small fraction of fast deuterium ions, modelling the presence of NBI heating.}
\label{fig:Geometry:sp_NT}
\end{figure}

In Fig~\ref{fig:Geometry:sp_NT} we distinctly see the high temperature of the fast ion population (rescaled by a factor of ten). Although they have a small density they contribute to a significant part of the total pressure as shown in Fig.~\ref{fig:Geometry:PevolT}. The profiles shown in both Fig.~\ref{fig:Geometry:PevolT} and in Fig.~\ref{fig:Geometry:sp_NT} have been generated with the \texttt{CRONOS}~\cite{ref:CRONOS} suite of numerical codes for interpretive simulations of tokamak discharges\footnote{Thanks to E. Highcock for generating the \texttt{CRONOS} data and J. Citrin for the input files.}.

To this point we have the following effects of fast ions: (1) redistribution of the pressure and (2) radial distribution of either deuterium or electrons density profiles because of quasineutrality which in turn also have an effect on the total pressure and its shape.

\section{Magnetic geometry}\label{sec:Geometry:mgeom}
The effects explained in the previous section are only a few of the numerous changes introduced by the fast ions. We know that for a fusion plasma to be feasible the charged particles should be confined. This is realised with a specially shaped magnetic field which exerts a force to restrict the charged particle motion and sustain a stable macroscopic equilibrium. The more energetic the particle population is, the higher velocity and therefore temperature. A higher temperature means a higher pressure. If the pressure profile of the particles change, the magnetic field also has to be modified in terms of magnitude and geometry to confine the motion. To characterise the various possibilities of stable geometries one may solve the equilibrium equations from the magnetohydrodynamic (MHD) model
\eqre{
\label{eq:Geometry:MHD1}
\vecJ \times \vecB = \nabla p,
}
\eqre{
\label{eq:Geometry:MHD2}
\nabla \times \vecB = \frac{4\pi}{c}\vecJ,
}
\eqre{
\label{eq:Geometry:MHD3}
\nabla \cdot \vecB = 0,
}
derived  under assumption of an equilibrium with parameters constant in time and without equilibrium flow~\cite{ref:Freidberg2007}. 

The pressure profile sets restrictions on the possible magnetic geometries which are stable. The relationship between magnetic field and the pressure can be noticed by taking the scalar product of $\vecB$ with Eq.~\eqref{eq:Geometry:MHD1} to obtain
\eq{
\vecB \cdot \nabla p = 0.
}
By definition of $\nabla p$ this implies that magnetic fields lines will lie on (flux) surfaces that are contours of constant pressure. We label these flux surfaces with $\Psi = RA_{\phi}$ where $A_{\phi}$ is the magnetic vector potential in the azimuthal (toroidal) direction related to the magnetic field via $\vecB = \nabla \times \vecA$~\cite{ref:Freidberg2007} and $R$ is the major radius previously defined in Fig.~\ref{fig:geom:tokCoord}. The type of magnetic geometry is often illustrated by the cross section of these flux surfaces, describing one set of magnetic field lines.

If we evaluate the poloidal flux of the magnetic field going through a horizontal plane $D$ that goes through the magnetic axis, radially limited by $R = R_1$ and $R = R_2$ with $R_2 > R_1$, it can be observed that
\eqre{
\label{eq:geom:pFlux}
\int_{D} \vecB \cdot \text{d}\vecS = 2\pi\int_{R_1}^{R_2} R B_Z \text{d}R = 2\pi \int_{R_1}^{R_2} \frac{1}{R}\dpd{}{R}(RA_{\phi})R\text{d}R = \\ = 2\pi\left(\Psi(R_2) - \Psi(R_1)\right),
}
where in the second equality we have used that the vertical component of the magnetic field is $\left(\vecB\right)_Z = \left(\nabla \times \vecA\right)_Z = R^{-1}\partial A_{\phi} /\partial R$. Here $\phi$ is the azimuthal coordinate from Fig.~\ref{fig:geom:tokCoord}. Since the tokamak is a toroidally symmetric device, the derivatives with respect to $\phi$ vanish. From Eq.~\eqref{eq:geom:pFlux} it is evident that $\Psi$ is proportional to the poloidal flux, hence the name: \textit{flux surface}.

\begin{figure}[H]
\setlength\figureheight{0.4\textheight}
\setlength\figurewidth{0.5\textwidth}
    \centering
    \input{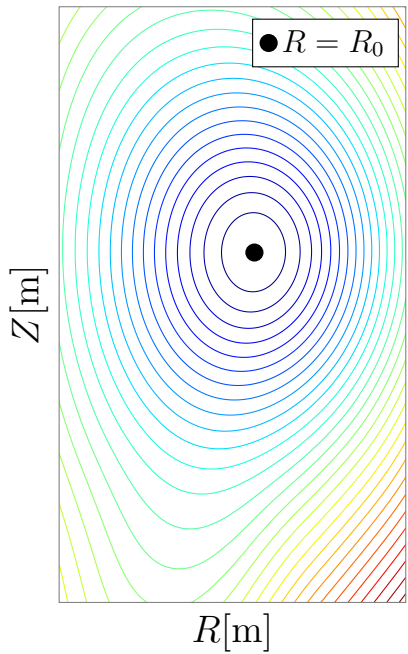}   
    \caption{Example of flux surfaces in the poloidal plane. The figure represents flux surfaces generated by the MHD equilibrium solver \texttt{CHEASE} presented in Section~\ref{sec:geom:param}. The closer we get to the centre, the more damped are higher order corrections to the flux surface shape. In the end they become completely elliptical (elongation penetrates down to the magnetic axis) as seen in the figure.  The flux surfaces converge towards one point. This is known as the magnetic axis located at the radial coordinate $R = R_0$.}
\label{fig:PreSim:Geometry:FluxEx}
\end{figure}

The shape of these flux surfaces characterises the \textit{magnetic geometry} and represents the balance between the $\nabla p$ and the $\vecJ \times \vecB$ force in Eq.~\eqref{eq:Geometry:MHD1}. Example shapes in the poloidal plane are shown in Fig.~\ref{fig:PreSim:Geometry:FluxEx}.
\noindent
The flux surfaces converge towards the \textit{magnetic axis}, where the poloidal field is zero. The boundary is given by the last closed flux surface and is usually denoted by $\Psi_0$. The equation describing the magnetic geometry and solves for $\Psi(R, Z)$ is known as the Grad-Shafranov equation. In toroidally symmetric systems such as the tokamak it is given by~\cite{ref:Grad1958}
\eqre{
\label{eq:PreSim:Geometry:GS}
\Delta^* \Psi = -\frac{4\pi}{c}r^2p'-ff'.
}
On the right hand side is the pressure gradient $p' = \text{d}p/\text{d}\Psi$. The pressure is a function of the poloidal flux alone with $p \equiv p(\Psi)$. The same holds for the other quantity $f \equiv f(\Psi)$ with $f = RB_{\phi}$ where $B_{\phi}$ is the magnetic field in the toroidal direction. On the left hand side we have the operator $\Delta^* = \partial^2/\partial Z^2 - \left(1/R\right)\partial/\partial R + \partial^2/\partial R^2$. As such, the differential Eq.~\eqref{eq:PreSim:Geometry:GS} has an interesting form: $\Psi$ is both a free variable on the RHS and is acted on by the differential operator $\Delta^*$ on the LHS. Since the fast ions change the pressure gradient, they will also change $\Psi$ as stated in the G-S equation. But since $f$ and $f'$ are functions of $\Psi$, the fast ions will in turn also affect these quantities. This shows the nonlinearity of the equation which makes it nontrivial to solve in the complicated magnetic geometry of a tokamak.  Solutions of the G-S equation corresponding to real experimental cases requires implementation of numerical tools. 

In this work we have made use of the the Cubic Hermite Element Axisymmetric Static Equilibrium (\texttt{CHEASE})~\cite{ref:CHEASE} code for solving the G-S Eq.~\eqref{eq:PreSim:Geometry:GS} numerically. The underlying principles of \texttt{CHEASE} are similar to standard finite element procedures. First the G-S equation is rewritten in a variational form. The poloidal flux $\Psi$ is then expanded in a convenient choice of finite elements defined on a rectangular grid. The result is a discretised nonlinear system of equations which is solved iteratively. For solving the G-S equation \texttt{CHEASE} requires information of $p'$, $ff'$ and boundary conditions in form of the LCFS $\Psi_0$. These can be specified in a variety of different functional forms or through a set of data points as is done here. These sets of data points have been obtained from running the tokamak discharge modelling tool \texttt{CRONOS}~\cite{ref:CRONOS}. In addition to this data \texttt{CHEASE} reads a namelist of variables where other information, for instance specifications of the solver accuracy, are given. The non-default values used in this work are presented in the Appendix~\ref{app:geom:1}. For a full description of the functionality of the code we refer to the complete user manual~\cite{ref:CHEASE}. 

With \texttt{CHEASE} we may solve the Grad-Shafranov equation and obtain all the necessary  information to evaluate the effect of fast ions on the magnetic geometry. Because of the often complex magnetic structure of the geometry this information might be to detailed to be able to draw exact conclusions. Instead we seek a way of summarising the information about the magnetic geometry in a few selected parameters. This can be done by \textit{parameterising} the equilibrium.

\section{Parametrisation of the magnetic equilibrium}\label{sec:geom:param}
All models for parametrising local properties of the MHD equilibrium should represent a solution to the G-S Eq.~\eqref{eq:PreSim:Geometry:GS}. A local parametrisation is generally performed for some choice of flux surface $\Psi$ and associated MHD equilibrium quantities are evaluated in a neighbourhood of that flux surface.  One of the simplest models where all these parameters are specified is the shifted circle model~\cite{ref:Miller98}. Here all the flux surface shapes are assumed to be of circular shape but incorporate also possible radial shift of the flux surface centres $R_0(\Psi)$ known as the Shafranov shift. The shifted circle model can be accurate in non-elongated plasmas for flux surfaces close to the centre of the tokamak, where the aspect ratio  $A = R_0/r$ is large.  But in many experimentally relevant scenarios $A$ is often not small and the shapes of the flux surfaces are far from circular. A more accurate model, even away from the magnetic axis, is the Miller parametrisation often used in stability studies, which also is the choice here.
\subsection{Miller parametrisation}\label{sec:Geometry:Miller}
Miller~\cite{ref:Miller98} generalised the shifted circle model to be valid for finite $A$ and incorporate non-circular flux surface features. The model is based on the localised studies by the Mercier-Luc formalism~\cite{ref:Miller98}, and is characterised by a set of nine parameters to describe a realistically shaped equilibrium: $A, \delta, \kappa, s_{\delta}, s_{\kappa}, \partial_r \Delta, q, \hat s, \alpha$. Five of these determines the shape of the flux surfaces: the aspect ratio $A$, triangularity $\delta$, elongation $\kappa$ and derivatives of these defined as $s_{\delta} = a\partial\delta/\partial r$ and $s_{\kappa} = a\partial \kappa/\partial r$ respectively. 

The Shafranov shift $\Delta = R_0(\Psi)-R_0(\Psi=0)$ describes the shift of flux surface centres as a function of the minor radius. In the Miller model the radial derivative of this quantity is specified. The safety factor $q = \partial \Psi_t/\partial\Psi$ essentially counts the number of toroidal turns it takes for the field lines to perform a full poloidal turn. Here $\Psi_t$ is the toroidal flux.  The parameter $q$ is important for MHD stability equilibrium~\cite{ref:Freidberg2007} where $q > 1$ is desirable. Finally the model is complete with the magnetic shear $\hat s = r \partial \ln q/\partial r$ and $\alpha$ which is proportional to the pressure gradient.

\begin{figure}[H]
\setlength\figureheight{0.35\textheight}
\setlength\figurewidth{0.5\textwidth}
    \centering
    \input{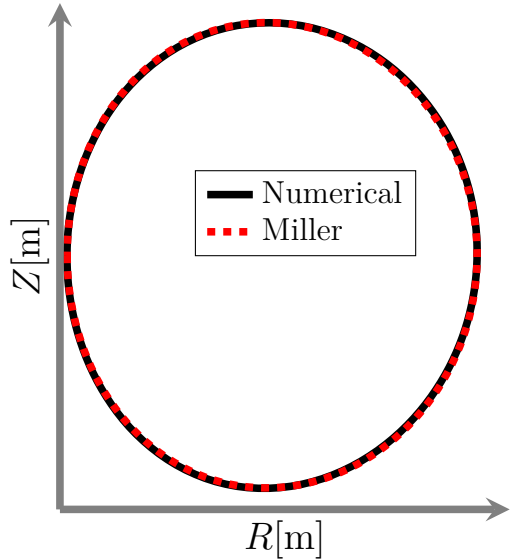}
    \caption{Comparison between the flux surface shape obtained from a Miller parametrisation and an actual flux surface obtained from the MHD equilibrium solver \texttt{CHEASE} as explained in the text.}
\label{fig:PreSim:Geometry:MillerFlux}
\end{figure}

A demonstration of the Miller parameters is shown in Figure~\ref{fig:PreSim:Geometry:MillerFlux}, where the flux surface obtained from a Miller parametrisation together with the numerical, non-parametrised flux surface is shown. The equilibrium has been generated with \texttt{CHEASE}. We see in the figure that the Miller parametrisation of the flux surface shape agrees well with the corresponding numerical solution. Fig.~\ref{fig:PreSim:Geometry:FluxEx} indicates that close to the edge of the plasma the flux surface shapes may have complicated structure for instance, up-down asymmetry, sharp variations close to an X-point etc. As we move closer to the centre higher order shaping effects are damped and the shapes may be well described by a triangularity (how ``triangular'' a flux surface is) and elongation (how elongated along the vertical $Z$-axis a flux surface is). Even closer to the centre also triangularity becomes negligible and the flux surface shapes become elliptical. Therefore the accuracy in using the Miller parameters for describing experimental flux surface shapes increases towards the core of the plasma, where higher order corrections to the circular flux surface shapes are damped. 

In this Chapter we use the Miller parameters for conveniently quantifying the fast ion effects on the geometry. Later in Chapter~\ref{CH:analysis} we will use the model in gyrokinetic simulations to identify individual effects of fast ion parameters. To facilitate future interpretation of the result it is important to understand the implication of the model and how it relates to using full numerical equilibrium. 

For a change of the total pressure gradient from the G-S Eq.~\eqref{eq:PreSim:Geometry:GS} we see that the magnetic geometry then also has to be different. In order to be consistent we therefore have to recalculate the equilibrium whenever the pressure gradient is changed. This method is complicated because it has an effect on other local parameters as well. We label this case as a global solution to the G-S equation since in this case the magnetic geometry reflects the global changes in the pressure profile. The boundary condition for this case is then the last closed flux surface $\Psi_0$. We should note that the global solution is representative for what is actually done in experiments where global rather than local changes in the profiles are performed. 

Another possibility is to use the Miller model to parametrise a local flux surface $\Psi$ which is included in the full global solution. We should point out that the Miller parametrisation will not exactly correspond to $\Psi$ which corresponds to \textit{the} local solution of the G-S equation for a given global pressure profile. Instead Miller yields \textit{a} local solution $\Psi_M$ since it assumes the flux surface shape to be fully captured with a triangularity $\delta$ and elongation $\kappa$ which is an approximation of the actual shape. To find optimal values of the triangularity, elongation and flux surface centre essentially a fit of the global solution is performed. In a conventional tokamak plasma typically $\Psi_M$ will be an accurate description of the actual $\Psi$ close to the centre of the plasma. 

\section{The fast ion effect on the magnetic geometry}\label{ref:geom:Eff}
\begin{figure}[H]
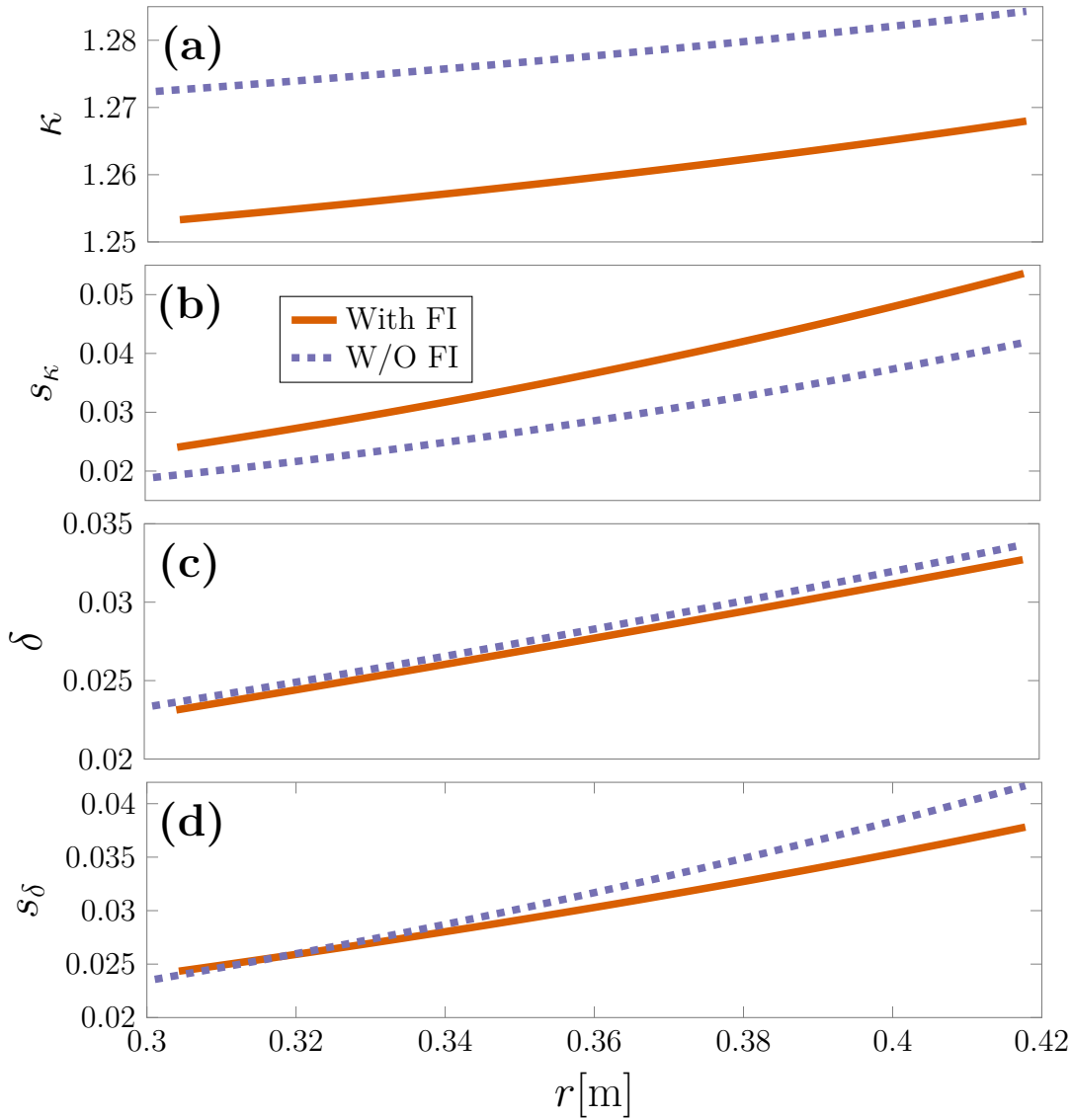

 \setlength\figureheight{0.2\textheight}
 \setlength\figurewidth{0.9\textwidth}
    \centering
        \begin{subfigure}[b]{\textwidth}
        \centering    
    \input{../../figures/geometry/FD_No_FD/kappa.tex}
    \phantomcaption
    \label{fig:geom:kappa}
    \end{subfigure}\\
    \hspace{-0.3cm}
        \begin{subfigure}[b]{\textwidth}
            \centering
     \input{../../figures/geometry/FD_No_FD/skappa.tex}
         \phantomcaption
    \label{fig:geom:skappa}         
         \end{subfigure}\\
 \hspace{-0.6cm}
        \begin{subfigure}[b]{\textwidth}
            \centering
    \input{../../figures/geometry/FD_No_FD/delta.tex}
    \phantomcaption    
    \label{fig:geom:delta}    
        \end{subfigure}
    \vspace{-0.1cm}
    \hspace{-0.005cm}    
        \begin{subfigure}[b]{\textwidth}
            \centering
    \input{../../figures/geometry/FD_No_FD/sdelta.tex}
    \phantomcaption    
    \label{fig:geom:sdelta}    
        \end{subfigure}
\caption{The calculated elongation (a), triangularity (c) and their derivatives (b and d) respectively. Results are presented for the case with $n_f = 0.05n_e$ (solid orange) and without (dashed purple) fast ions.}
\label{fig:geom:Miller1}
\end{figure}

We have now introduced all the tools for studying the effects of fast ions on the geometry. The analysis is presented in terms of the pressure profiles generated by \texttt{CRONOS} previously shown in Fig.~\ref{fig:Geometry:PevolT}. We separate these in two pieces. The first is just before fast ions are injected at roughly $\SI{46}{s}$ which we will label as the ``without fast ion'' case. Next we take a time average over the part of the profiles when NBI has been turned on. This corresponds to the pressure peak between 46-$\SI{48}{s}$ in Fig.~\ref{fig:Geometry:PevolT} which we labels as the case ``with fast ions''. For each of these two cases we utilise \texttt{CHEASE} and generate two sets of consistent global MHD equilibria from which we compute the nine Miller parameters. The Miller parameters are computed using a script that is part of the \texttt{GENE} code~\cite{ref:GENE}. In this work only the part of \texttt{GENE} used to calculate the Miller parameters is used, and not the entire code itself. The Miller parameters are computed for minor radii $0.3 \le r \le \SI{0.42}{m}$, local to the flux surface which is used later in our gyrokinetic simulations.  
We begin with the parameters describing the shape of the flux surfaces. The fast ions in our case leave the shape of the flux surfaces essentially unchanged as shown in Figs.~\ref{fig:geom:kappa} and \ref{fig:geom:delta}. A slightly larger difference is obtained for the derivatives of $\kappa$ (Fig.~\ref{fig:geom:skappa}) and $\delta$ (Fig.~\ref{fig:geom:sdelta}). 

\begin{figure}[h]
 \setlength\figureheight{0.17\textheight}
 \setlength\figurewidth{0.9\textwidth}
    \centering
    \input{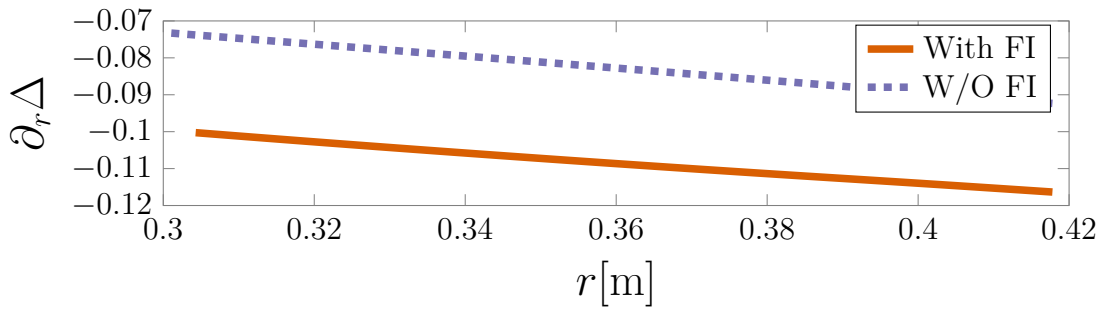}        
    \caption{The change in the radial derivative of the Shafranov shift because of fast ions. Results are presented for the case with (solid orange) and without (dashed purple) fast ions.}
\label{fig:geom:growth3}
  \end{figure}

The radial derivative of the Shafranov shift, i.e the change in the shift of the flux surface centres is shown in Fig.~\ref{fig:geom:growth3}. The radial derivative of the shift reduces for larger values of the minor radius $r$. Adding fast ions reduces this parameter by roughly 30\%.

\begin{figure}[h]
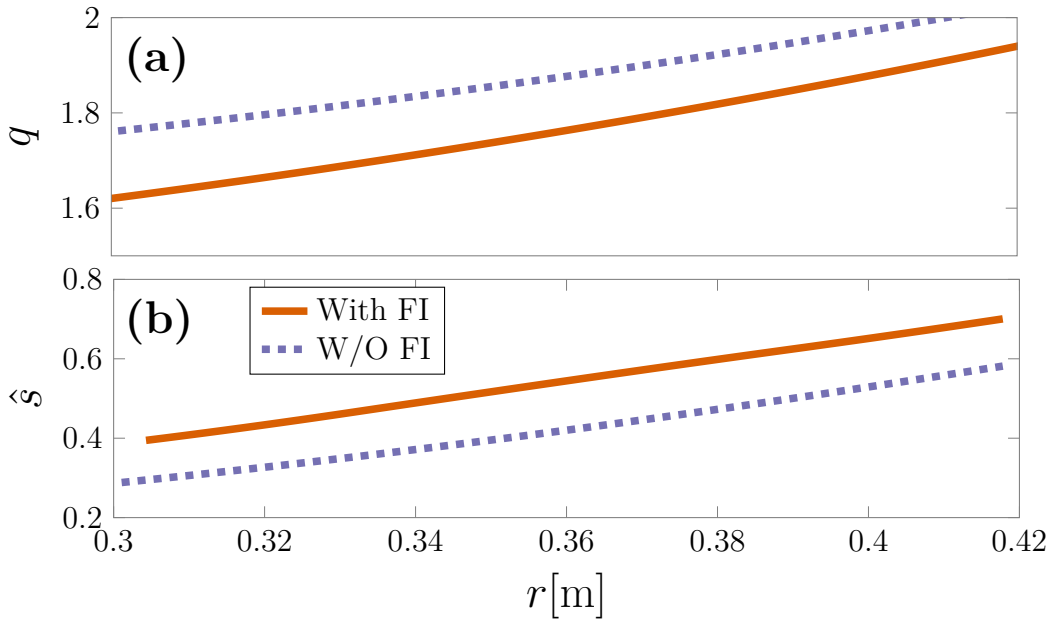

 \setlength\figureheight{0.2\textheight}
 \setlength\figurewidth{0.9\textwidth}
         \centering
        \begin{subfigure}[b]{\textwidth}
            \centering
        \input{../../figures/geometry/FD_No_FD/qprofile.tex}
    \phantomcaption    
    \label{fig:geom:qprof}         
        \end{subfigure}\\
        \hspace*{0.17cm} 
        \begin{subfigure}[b]{\textwidth}
            \centering
    \input{../../figures/geometry/FD_No_FD/shear.tex}  
        \phantomcaption    
    \label{fig:geom:shear} 
 \end{subfigure}    
\caption{The change in the safety factor (top) and the magnetic shear (bottom). Results are presented for the case with (solid orange) and without (dashed purple) fast ions.}
\label{fig:geom:Miller3}
\end{figure}

One of the most important quantities for both MHD stability and for suppressing turbulence is the safety factor profile. This includes both $q$ itself and its derivative, the magnetic shear. The safety factor profile has a large influence on the confinement time of a fusion device~\cite{ref:Levinton1995}. The influence of fast ions on the safety factor and the magnetic shear are shown in Figs.~\ref{fig:geom:qprof} and~\ref{fig:geom:shear}.  For our case we note a small reduction in the safety factor $q$ and a slight increase in the magnetic shear.

Given the changes in the various geometrical parameters presented here we now wish to answer the question: do we have to include the change in the magnetic geometry in gyrokinetic simulations to properly account for the presence of fast ions? Even for the slightest change, the corresponding effect on the ITG growth rates could be significant.
The influence of $\kappa$ and $\delta$ has been studied by Waltz et al.~\cite{ref:Waltz1999} where they discovered a decrease in the growth rate with increasing $\kappa$. Growth rates first increase with triangularity until a peak is reached and after which $\gamma$ start to reduce. In their studies they included the effect of sheared flows. These are flows of equilibrium scale and with steep gradients across the flux surfaces~\cite{ref:Edmund2012}. The shaping influences the value of the flow shear rate which if comparable to the maximum $\gamma$ may fully suppress the turbulence~\cite{ref:Waltz1999}. 

To complete our study of the magnetic geometry we choose to report the effect on ITG growth rates. The results of the gyrokinetic simulations for the Miller parameters can be found in Appendix~\ref{app:geom:2}.

For the given change in the shaping parameters, we do not find the growth rate to change much. The fact that we do not see an effect when modifying the shaping parameters might seem alarming, since previous studies have found the shaping to be of greater importance. As pointed out by Waltz et al.~\cite{ref:Waltz1999}, the result very much depends on the other quantities describing the equilibrium. For example elongation $\kappa$ has a stronger effect at low values of the safety factor $q$, and a much smaller effect for higher $q$. What seems to be the key difference is that here we do not include flow shear in the simulations. If the shaping parameters change the growth rate mainly because of the flow shear, our results are not surprising. 

In their study Jhowry et al.~\cite{ref:Jhowry2004} found a varying dependence on the shift $\Delta$. Decreasing the shift can both have a negligible effect, may slightly increase or decrease the growth rate depending on values of elongation along with ion temperature and density gradients. For the roughly 20\% decrease in the derivative of the Shafranov shift because of fast ions, we find a roughly 15\% increase in the growth rates. For larger fast ion densities the effect of Shafranov shift may therefore have to be included to properly account for fast ions. 

The magnetic shear $\hat s$ is responsible for twisting the magnetic field lines from one flux surface to the next. This leads to a large variation in the gradients of quantities and may have a large influence on the stability. Low values of the magnetic shear is known to stabilise the turbulence~\cite{ref:Nordman2007,ref:Nordman07_2}. For the roughly 20\% increase in the magnetic shear, we find a slight destabilising effect indicated by a $\sim 14\%$ increase in the growth rates.

Later on in this work, we will perform further gyrokinetic simulations comparing plasmas with and without fast ions, to investigate their role on ITG growth rates. A proper comparison between a case with and without fast ions might require the magnetic geometry to be appropriately adjusted. We will use a much larger fast ion density ($n_f = 0.13n_e$) compared to the value used here ($n_f = 0.05n_e$). Since simulations in our case show a negligible effect on the growth rates even for larger variations of the shaping parameters $\delta, \kappa, s_{\delta}, s_{\kappa}$, these can safely be kept constant even for the larger fast ion density.  This leaves the safety factor $q$, the magnetic shear and the radial derivative of the Shafranov shift $\partial_r\Delta$ which although here had a relatively small effect, might be important for $n_f = 0.13n_e$. To properly account for these parameters we need to recalculate these parameters of the Miller model, whenever we change the fast ion density and/or temperature profiles. Including the change in geometry whenever we modify any of the fast ion characteristics complicates the analysis significantly since isolated parameter scans will be very difficult to perform. To simplify future analysis we also choose to keep these Miller parameters constant, in addition to the shapes of the flux surfaces. Given the results in this chapter we should bear in mind that our findings may not be consistent with cases where full numerical geometry has been used. In addition, the large reduction in the heat flux reported by Citrin et al.~\cite{ref:Citrin} did not include the change in the magnetic geometry. Clearly there are other, perhaps more important effects of fast ions to study, which might be discovered even for a constant magnetic geometry. In the following chapters we will therefore make the additional (and not entirely justified) simplification and keep the magnetic geometry fixed.

\clearpage

\chapter{Studying turbulence and the ITG mode}\label{CH:studTurb}

Transport of particle densities $n_s$ is often characterised by a diffusion coefficient $D\sim \text{step size}^2\times\text{frequency}$. This measures how fast particles are transported out from the plasma. Similarly the thermal diffusion coefficient  $\chi$ is used to measure the transport of heat.

In a plasma various mechanisms lead to a finite value of these diffusion coefficients and limits the confinement of particles and heat. Some contribution is due to Coulomb collisions. As particles collide, the position of their guiding centres is shifted and they are gradually transported outside the plasma. A more realistic value of the diffusion coefficients is retrieved if one in addition to collisions also takes the various drifts into account, as is done in neoclassical transport theory. It turns out the value of the perpendicular transport of heat measured experimentally is still far away even from the neoclassical prediction. The missing part is referred to as \textit{anomalous diffusion} and is a consequence of rapid fluctuations known as turbulence. Fast ions may reduce this type of transport which implies a significant increase in the confinement time $\tau_E$.

We know from Chapter~\ref{CH:FIin} that fast ions are introduced via ICRH and NBI heating, and have seen their various global effects on the plasma such as a change in the magnetic geometry. Without actually having introduced the concepts of turbulence, growth of microinstabilities and how they can be simulated, we presented the effect of geometry on growth rates of the ion temperature gradient mode. In this chapter we clarify these various concepts and introduce the tools for studying turbulence and calculating growth rates.

We will start with the general characteristics of turbulence which may be described by the Fokker-Planck equation. The equation captures a very wide range of effects, many of which we are not interested in studying here as for instance electromagnetic wave propagation in the plasma. Under certain assumptions of the turbulence characteristics we may separate the Fokker-Planck equation into a slow equilibrium part and a rapidly fluctuating part. Since the fluctuations are much slower than the gyration of plasma particles, this time scale separation can be used to simplify the problem. We expand the Fokker-Planck equation and arrive at the gyrokinetic equation for studying the turbulence.  This equation is solved numerically with the gyrokinetic continuum code \texttt{GS2}.  An introduction to the code including normalisations and calculation of input parameters relevant for this work is given in Section~\ref{sec:studTurb:GS2}.

The turbulence described here is driven by microinstabilities\footnote{Subcritical turbulence~\cite{ref:Edmund2012} may exist even without instabilities, but is not treated here.}. There are many instabilities but we will focus on the ion temperature gradient mode (ITG), driven by ion temperature gradients. Above some critical threshold the ion temperature gradient generates an instability. This ITG instability grows exponentially in amplitude with some growth rate $\gamma$ and generates exponential growth of the heat flux. The exponential growth is captured by linear physics but nonlinear interactions are required to obtain the saturated values of the heat flux leading to finite diffusion coefficients. We simplify the computational requirements by only including the linear physics, allowing us to perform parameter scans and evaluate the fast ion effects on the growth rates. The linear growth rates of the mode are related to the saturated values of the heat flux and values of the diffusion coefficient with quasi-linear theory~\cite{ref:Casati}. One quasi-linear model predicts  $D \sim \gamma/k_{\perp}^2$ where $\gamma$ is the growth rate of the unstable mode and $k_{\perp}$ its perpendicular wavenumber. Reduced growth rates $\gamma$ are therefore directly related to reduced diffusion coefficients and reduced transport.

In Section~\ref{sec:studTurb:ITG} the physical principles behind the growth of the ITG instability is presented. Given the gyrokinetic equation we derive a \textit{dispersion relation} relating mode frequencies $\tilde \omega = \omega + i\gamma$ to wavenumbers $\veck$. A positive imaginary part indicates an instability which grows with the exponential growth rate $\gamma$. First we present an expression of the ITG growth rate for a plasma consisting of only electrons and singly charged deuterium ions. We consider then also a single species of fast ions and investigate how the expression for the critical ion temperature gradient leading to positive growth rates is modified. Finally, the theoretical predictions are compared with actual gyrokinetic simulations of the ITG growth rates, and we investigate the role of fast ion temperature and density gradients. For simplicity only fluctuating electric fields are treated without including possible fluctuations in the magnetic fields. For the gyrokinetic simulations this assumption is relaxed in the next chapter.

\section{The gyrokinetic formulation of turbulence}\label{sec:studTurb:gk}

Turbulence in essence is described by rapid fluctuations on different scales in the density, temperature, electric field, magnetic field and associated quantities. It may be triggered by microinstabilities which disturbs the plasma and leads to the formation of eddies at various scales.  The eddies create local gradients which lead to rapid anomalous diffusion of particles, energy and an evolution of the distribution function $f_s$ for the species $s$. The equation which describes this evolution is the kinetic or Fokker-Planck equation~\cite{ref:Abel}
\eqre{
\dpd{f_s}{t} + \vecv \cdot \nabla f_s + \frac{Z_se}{m_s}\left(\vecE + \frac{1}{c}
\vecv \times {\vecB} \right) \dpd{f_s}{\vecv} = C[f_s],
\label{eq:studTurb:FP}
}
where the functional on the RHS is the collision operator that captures the effect of collisions between particles. 

\begin{figure}[H]
    \centering
    \includegraphics[width=\textwidth]{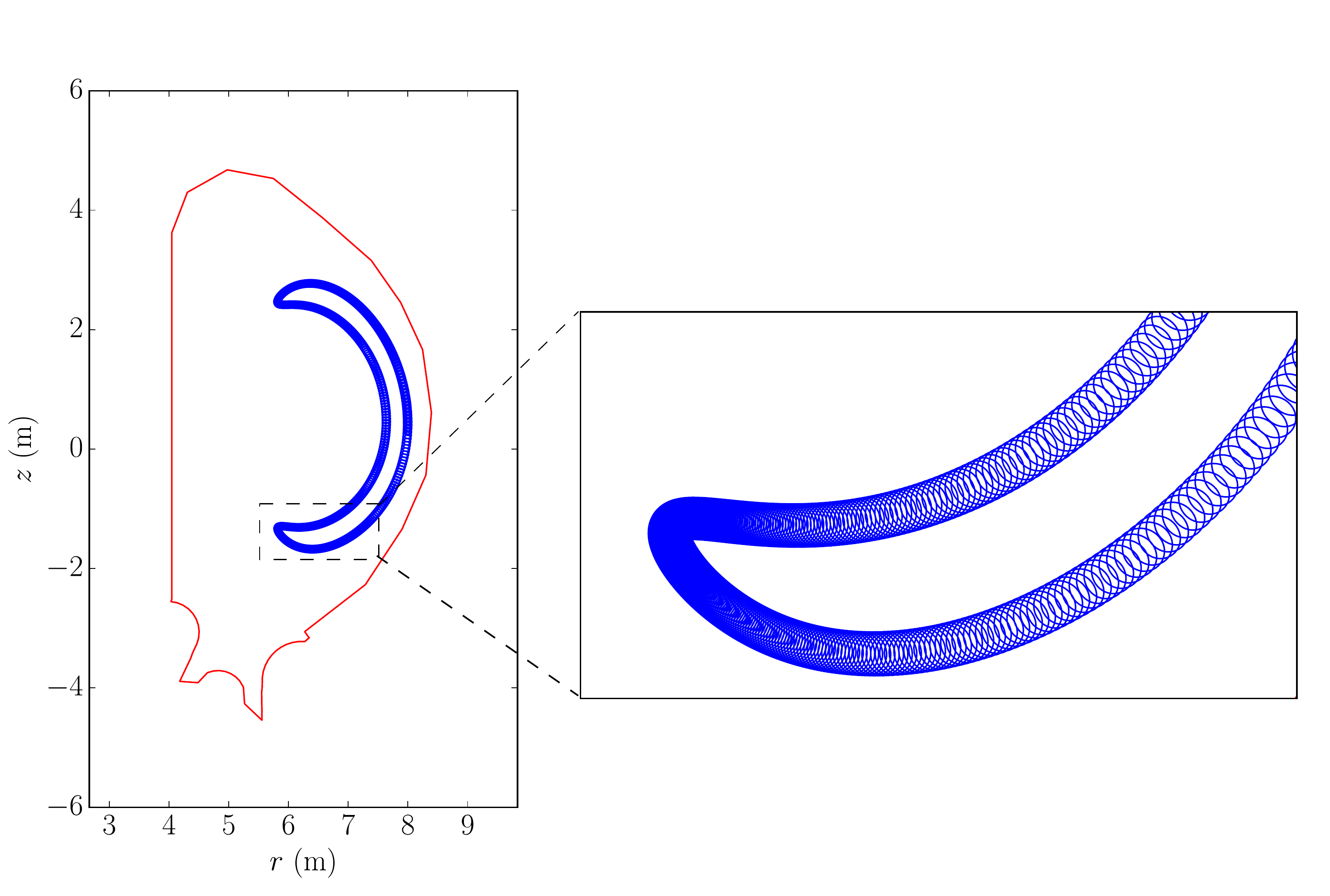}
    \caption{The motion of a charged particle in a ITER-like magnetic field, projected on a poloidal cross-section of the tokamak. The particle gyrates rapidly around its guiding center while slowly drifting along a banana shaped trajectory (Source~\cite{ref:Bsc}).}
    \label{fig:gyroEqMot}    
\end{figure}

Particles in a tokamak plasma have a fast and small scale gyration orbit while slowly drifting across flux surfaces. An example is depicted in Fig.~\ref{fig:gyroEqMot} where the drifts create a banana shaped topology in the poloidal cross-section of the tokamak. The turbulence we are interested have structures in the order of the ion Larmor radius $\rho_i$ but fluctuates on a much longer time-scale compared to the rapid gyration of the particles. On the other hand, the fluctuations are much faster compared to transport of particles and heat over the minor radius. This separation of scales is used in the gyrokinetic formalism to simplify the Fokker-Planck equation.

While drifting slowly across the flux surfaces, particles in a tokamak plasma move very fast along field lines and consequently the response to possible parallel perturbation is much quicker than perpendicular. The turbulent fluctuations will be elongated along the field lines such that $\lambda_{\perp}/\lambda_{\parallel} = k_{\parallel}/k_{\perp} \ll 1$, where $\lambda$ is the wavelength, $k_{\perp}, k_{\parallel}$ is the wavenumber perpendicular and parallel to the magnetic field $\vecB$ respectively. If the tokamak is large enough, we would expect the size of the turbulent fluctuations to be much smaller compared to the corresponding variation in the equilibrium quantities. 

We  assume the full distribution function can be expanded, ${f}_s =  F_s + \delta f_s$, into a slowly varying part $F_s$ in time and space,  and rapidly varying part $\delta f_s$ such that $\delta f_s/f_s \ll 1$. Separating the fields in similar way we then impose the following (gyrokinetic) ordering of the fluctuating quantities
\eqre{
\frac{|\delta \varphi|}{|\varphi|} \sim \frac{|\delta \vecB|}{|\vecB|} \sim \frac{|\delta f_s|}{|F_s|} \sim 
\frac{\kpar}{\kperp} \sim \frac{\omega}{\Omega_s} \sim \frac{\rho_s}{a} = \rho_*,
\label{eq:order}
}
where $\rho_* \ll 1$. 

Expanding Eq.~\eqref{eq:studTurb:FP} in orders of $\rho_*$ we may dismiss the variation in the slow equilibrium parts and focus on the turbulence instead. In particular expanding to $\rho_*^2$ will lead us to the gyrokinetic equation. 

\subsection{The gyrokinetic equation}
The derivation of the gyrokinetic equation has been extensively covered before and here only  the main assumptions and techniques used in the derivation are presented. We follow the work done by Abel et al.~\cite{ref:Abel} with the additional simplification of neglecting equilibrium flow and toroidal rotation of the plasma.
 
Using the ordering postulated in~\eqref{eq:order}, it will be convenient to expand the mean part and the fluctuating part of the distribution function as follows
\eqre{
F_s = F_{0s} + F_{1s} + F_{2s} + \cdots, \\
\delta f_s = \delta f_{1s} + \delta f_{2s} + \cdots,
\label{eq:distExp}
}
such that $F_s \sim f_s, F_{1s} \sim \delta f_{1s} \sim \epsilon f_s, $ etc. Note here that $F_{0s}, F_{1s},  \dots$ are all slowly varying and are seen as constant in time, while $\delta f_{1s}, \delta f_{2s}, \dots$ are rapidly varying on a small spatial and temporal scale. Expanding the kinetic equation~\eqref{eq:studTurb:FP} to first order in $\rho_*$, one finds that the zeroth order distribution function $F_{0s}$ is a Maxwellian
\eqre{
F_{0s} = n_s\left[\frac{m_s}{2\pi T_s}\right]^{3/2} \text{exp}\left[-\frac{\varepsilon_s}{T_s}\right]. \\
\label{eq:MaxW}
}  
In deriving Eq.~\eqref{eq:MaxW} we have introduced a new coordinate system. The components of this coordinate system are the gyrophase $\vartheta$, guiding centre position $\vecR_s$, the particle energy $\varepsilon$, the magnetic moment $\mu_s$ and the sign of the parallel velocity $\sigma$ which are defined as
\eqre{
\label{eq:studTurb:vars}
\vecR_s = \vecr - \frac{\vecb \times \vecv}{\Omega_s}, \qquad \varepsilon = \frac{1}{2}m_sv^2, \\ \mu_s = \frac{m_s\vperp^2}{2B}, \qquad \sigma = \frac{\vpar}{|\vpar|},
}
where $\vecb = \vecB/|\vecB|$ is the unit vector in the direction of the magnetic field. As mentioned in the introduction, the particles in the plasma will gyrate in circular orbits with radius $\rho_s$ and a gyration frequency $\Omega_s$. The guiding centre position $\vecR_s$ is the centre of this orbit.  

We define the average of a quantity $g$ over the gyrophase (gyroaveraging) at fixed $\vecR_s$ 
\eqre{
\label{eq:studturb:gyroav}
\left < g\right>_{\vecR} = \frac{1}{2\pi} \int \text{d}\vartheta g(\vecR_s,\varepsilon,\mu_s,\vartheta,\sigma),
}
which will be denoted as $\left<g\right>_{\vecR}$. Taking the first order expansion of the kinetic Eq.~\eqref{eq:studTurb:FP}, performing an averaging over fluctuations and a gyroaverage according to Eq.~\eqref{eq:studturb:gyroav} we obtain, from Eq.~\eqref{eq:order}, the first order fluctuating distribution function 
\eqre{
\delta f_{1s} = -\frac{Z_se}{T_s}\delta \varphi(\vecr)F_{0s} + h_s(\vecR_s,\mu_s,\varepsilon_s,t).
\label{eq:f1Exp}
}
Here $h_s = h_s(t,\vecR_s,\mu_s,\epsilon_s)$ is a constant of integration, and is the gyrophase independent part of the first order fluctuations. By gyroaveraging we have reduced the six-dimensional Fokker-Planck equation to a problem in five dimensions instead. The term $h_s$ captures the effect of turbulence and is the one we try to determine through the gyrokinetic equation. Setting $h_s = 0$ is referred to as an adiabatic response for the species $s$. This means that the species will react very rapidly compared to the characteristic time scale of the perturbation and is often a good approximation for electrons. 

Keeping only the fluctuating parts of the second order expansion of the Fokker-Planck equation and gyroaveraging yields the gyrokinetic equation
\eqre{
\label{eq:studTurb:GK}
\dpd{h_s}{t} + \left(\vpar\vecb + \vecV_{\text{Ds}} + \left < \vecV_\chi\right >_{\vecR} \right)
\cdot \dpd{h_s}{\vecR_s} - \left<C_L[h_s]\right>_{\vecR} = \\ = \frac{Z_seF_{0s}}{T_s}\dpd{\left<\chi \right>_{\vecR}}{t} - \dpd{F_{0s}}{\psi}\left<\vecV_\chi\right>_{\vecR} \cdot \nabla \psi.
}

In Eq.~\eqref{eq:studTurb:GK} $\left<C_L[h_s]\right>_{\vecR}$ is the collision operator linearised about the zeroth order equilibrium distribution function $F_{0_s}$~\cite{ref:Abel2008}. Then
\eqre{
\label{eq:studTurb:pot}
\chi = \delta \varphi - \frac{1}{c}\vecv \cdot \delta \vecA,
}
is the fluctuating turbulent potential and
\eqre{
\label{eq:studTurb:vel}
\left <\mathbf{V}_\chi\right >_{\vecR} = \frac{c}{\overline B}\vecb\times\dpd{\left<\chi\right>_{\vecR}}{\vecR_s}, \\
\vecV_{\text{Ds}} = \frac{\vecb}{\Omega_s} \times \left [\vpar^2 \vecb \cdot \nabla\vecb + \frac{1}{2}\vperp^2 \nabla \ln \overline B\right],
}
where bar denotes the equilibrium part of the magnetic field, $\vecB = \overline \vecB + \delta \vecB$. In this work we treat the linear gyrokinetic equation and therefore neglect the nonlinear term $ \left< \vecV_\chi \right>_{\vecR}\cdot \partial h_s/\partial \vecR_s$ in Eq.~\eqref{eq:studTurb:GK}. In Eq.~\eqref{eq:studTurb:vel} the second expression represents the magnetic drift velocity. Integrating Eq.~\eqref{eq:f1Exp} and using quasineutrality we may relate the electrostatic potential to $h_s$ using
\eqre{
\label{eq:studTurb:QN}
\sum_s\frac{Z_s^2e}{T_s}\delta \varphi(\vecr)n_s =\sum_s Z_s \int\dv \left<h_s\right>_{\vecr}.
}
The electromagnetic part is obtained by taking the fluctuating component of Ampère's law
\eqre{
 \nabla^2\delta\vecA = \frac{4\pi}{c}\sum_s Z_s e \int\dv \left<\vecv h_s\right>_{\vecr},
\label{eq:fluctAmp}
} 
which includes both parallel $\delta B_{\parallel}$ and perpendicular $\delta B_{\perp}$ fluctuations. Intuitively $\delta B_{\parallel}$ can be understood as a compressional magnetic perturbation and $\delta B_{\perp}$ as a sheared magnetic perturbation. As will be demonstrated in Chapter~\ref{CH:analysis} only $\delta B_{\perp}$ will be important for the cases studied in this work. 

Together with the gyrokinetic Eq.~\eqref{eq:studTurb:GK}, Eqs.~\eqref{eq:studTurb:QN} and~\eqref{eq:fluctAmp} form a closed system which can be solved to investigate the evolution of the turbulent distribution function $h_s$.

\section{A brief introduction to \texttt{GS2}} \label{sec:studTurb:GS2}
To model the turbulence the gyrokinetic equation has to be solved for each species independently. The species' combined effect on the electrostatic and magnetic potentials are then given from Eqs.~\eqref{eq:studTurb:QN} and~\eqref{eq:fluctAmp}. In this work we choose to do this numerically using the gyrokinetic continuum code \texttt{GS2}~\cite{ref:GS22009}. \texttt{GS2} solves the linear set of equations by an implicit algorithm, then has the possibility to add the nonlinear terms and collisions. 

One of the complications in solving the gyrokinetic equation is the complex magnetic field configuration present in a typical fusion device but also the resolution required to resolve the turbulent structures. To simplify the problem of geometry and relax the requirement on the resolution \texttt{GS2} implements a set of field-line following coordinates. This set takes advantage of the anisotropy in the turbulent structures and reduces the required parallel resolution. The gyrokinetic equation is solved in a small region of the plasma local to some flux surface. In addition to the assumption done in deriving the gyrokinetic equation, \texttt{GS2} assumes that the gradients of the equilibrium quantities are constant in this small region. This is called \textit{the local approximation.}

The aim with this section is to outline the input and properties of \texttt{GS2} of relevance for this work. A more detailed documentation can be found in for example~\cite{ref:Edmund2012}.

\subsection{Physical input parameters to \texttt{GS2}}
The physical parameters describe the magnetic geometry, species and plasma composition used in the \texttt{GS2} simulations. All input quantities to \texttt{GS2} are dimensionless and normalised to some reference value. The reference length, density, mass, temperature and magnetic field used in this work are all given in Table~\ref{table:StudTurb:refVals}.

\begin{table}[h]
\centering
\caption{The reference density, temperature, length, magnetic field and mass, used in this work. All \texttt{GS2} parameters are normalised by these values.}
\label{table:StudTurb:refVals}
\begin{tabular}{@{}llllll@{}}
\toprule
$n_{ref}[\unit{10^{19}/m^3}]$ & $T_{ref}[\unit{keV}]$ & $a = L_{ref}[\unit{m}]$ & $B_{ref}[\unit{T}]$ & $m_{ref}[\unit{amu}]$ \\
\addlinespace[0.5em]
$2.95$  & $3.2$        & $0.96$     & 3.4   & 2.014     \\ \bottomrule
\end{tabular}
\end{table}

The physical input parameters in the gyrokinetic simulations are often representative of an experimental discharge. All simulations performed to this point have used the physical parameters of the 73224 JET discharge, presented in Table~\ref{table:StudTurb:73param} and are based on the values reported by Bravenec et al.~\cite{ref:Bravenec}. The choice is motivated by the large reduction in the nonlinear heat fluxes presented by Citrin et al.~\cite{ref:Citrin} as a consequence of a high fast ion density.  
\begin{table}
\def\arraystretch{2}\tabcolsep=6pt
\caption{Parameters for the 73224 discharge studied in detail in this work obtained from Bravenec et al~\cite{ref:Bravenec}. Note that here we have specified $s_{\delta}$ and $s_{\kappa}$ using their \texttt{GS2} definitions. Also this discharge had a non-zero flow shear, but for simplicity this term has been dropped which is indicated by the bold value of $\gamma_{E\times B}$.}
\label{table:StudTurb:73param}
\centering
\begin{tabular}{@{}ll|ll|ll|ll@{}}
\toprule
$r[\unit{m}]$               & 0.36  & $T_i/T_e$            & 1.0    & $a/L_{n,\text{He3}}$             & 0.5028 & $\hat s$                    & 0.523        \\
$a[\unit{m}]$               & 0.96  & $T_{\text{imp}}/T_e$ & 1.0    & $a/L_{T,i}$                      & 3.56   & $\kappa$                    & 1.26         \\
$n_e[\unit{10^{19}m^{-3}}]$ & 2.95  & $T_{\text{FD}}/T_e$  & 9.8    & $a/L_{T,e}$                      & 2.23   & $s_{\kappa}$                & 0.08         \\
$T_e[\unit{keV}]$           & 3.2   & $T_{\text{He3}}/T_e$ & 6.9    & $a/L_{T,imp}$                    & 3.56   & $s_{\delta}$                & 0.085        \\
$n_i/n_e$            & 0.648 & $a/L_{n,i}$          & 0.006  & $a/L_{T,\text{FD}}$              & 1.0326 & $\delta$                    & 0.030        \\
$n_{imp}/n_e$        & 0.025 & $a/L_{n,e}$          & 0.422  & $a/L_{T,\text{He3}}$             & 7.4074 & $\beta_e$                   & 0.0033        \\
$n_{FD}/n_e$         & 0.06  & $a/L_{n,\text{imp}}$ & 0.422  & $q$                              & 1.74   & $Z_{eff}$                   & 1.9          \\
$n_{He3}/n_e$        & 0.07  & $a/L_{n,\text{FD}}$  & 4.7228 & $\Delta = \text{d}R_0/\text{d}r$ & -0.14  & $\gamma_{E\times B}[a/c_s]$ & $\mathbf{0}$ \\
                     &       &                      &        &                                  &        & $R_0/a$                     & 3.12         \\ \bottomrule
\end{tabular}
\end{table} 
%
The 73224 JET discharge contains five species. The thermal species are electrons, deuterium ions, carbon impurities and the energetic species are fast deuterium (FD) and fast helium-3 (He). The fast deuterium has a high density gradient. In line with the discussion in Chapter~\ref{CH:FIin} these fast ions are the result of NBI heating. Fast helium-3 has a high temperature gradient instead and represents ICRH heating. In the analysis later in Chapter~\ref{CH:analysis} many scans will be performed by changing the characteristics of both FD and He particles simultaneously. At one point, however, we will make a distinction between these fast ions and investigate their separate effects on the growth rates. 

The profiles of the discharge were presented earlier in Fig~\ref{fig:Geometry:sp_NT} but there without the contribution from the He ions and a slightly smaller density of FD ($n_f = 0.05n_e$ instead of $n_f = 0.06n_e$). The total fast ion density in this discharge is significantly larger $n_f = 0.13n_e$.

The \texttt{GS2} tool supports a wide range of models for specifying the magnetic geometry. In this work the Miller model presented in the previous chapter, is implemented. The Miller parameters for the 73224 discharge are computed at the radial position $r = \SI{0.36}{m}$ specified in \texttt{GS2} with normalised radius $\rho = r/a$.

Rotational flow shear may be included in the simulations and originally the JET discharge had non-zero rotational flow shear. For simplicity we neglect this parameter throughout this work. 
Simulations with \texttt{GS2} can be both electrostatic, without including fluctuations in the magnetic field, or electromagnetic. In the latter case an additional parameter has to be considered in the \texttt{GS2} input, which is the thermal to magnetic energy ratio 
\eqre{
\label{eq:studturb:beta}
\beta = \frac{8\pi nT}{B_{\text{ref}}^2}.
}
The $\beta$ quantity is related to the efficiency of the confinement, expressing ``how well the magnetic field confines the motion of the particles''.  This parameter is specified for the reference species which is then automatically used to compute the total $\beta$. For a given value of the reference $\beta_{\text{ref}}$, the total $\beta$ is equal to $\beta = \beta_{\text{ref}}/n_{\text{ref}}T_{\text{ref}}\sum_s n_sT_s$. By default the JET discharge uses $\beta_{\text{ref}} = 0.0033$. Electrostatic simulations are equivalent to $\beta = 0$ obtained by assuming infinitely ``stiff'' magnetic field lines and taking the limit $\text{B}_{\text{ref}} \rightarrow \infty$. The pressure gradient of the species in the plasma enters through the radial derivative of $\beta:$ $\dif{\beta}/\dif{r} = \left(\beta /p \right)\partial p/\partial r$, where the magnetic field is seen as constant.

Collisions in \texttt{GS2} are implemented with an approximate form of the Fokker-Planck collision operator, derived by Abel et al.~\cite{ref:Abel2008}. Collisions are specified for each species separately and includes ion-ion, electron-ion and electron-electron collisions. For a species $s$ colliding with $s'$ the collision frequency is calculated with ~\cite{ref:Helander}
\eqre{
\nu_{ss} = \frac{\sqrt{2}\pi n_s Z_s^2e^2\ln \Lambda}{\sqrt{m_s}T_s^{3/2}},
\label{eq:vnewk}
}
where $\ln \Lambda$ denotes the Coulomb logarithm that is a parameter in the range 15-20 for the plasmas of interest. Computation of collisions between electrons and ions are simplified since in this case the collision operator is independent of the ion mass. The influence of all ions on the collision frequency is given by some fictive species with an effective charge 
\eqre{
\label{eq:studturb:zeff}
Z_{\text{eff}} = \frac{\sum _{j\ne e} n_jZ_j^2}{\sum _{j\ne e} n_jZ_j},
}
where the sum runs over all species, excluding the electrons~\cite{ref:Helander}. 

In addition to these input parameters, \texttt{GS2} requires the selection of a number resolution parameters. These are presented in the next section.

\subsection{Resolution parameters}\label{sec:studTurb:ResParam} 
A complete description of the turbulent fluctuations is given by Fourier modes with radial $k_{x}$ and poloidal $k_{y}$ wavenumbers. The linear gyrokinetic equation is then further simplified since it may be solved for each Fourier mode separately. In this work we will,  unless otherwise stated, only look at $k_{x} = 0$ which implies that $k_y = k_{\perp}$ (at $\theta = 0$ where $\theta$ is the poloidal angle). In the poloidal $k_y$ direction mainly the most unstable mode associated to the dominant growth rate $\gamma$ is considered.  We will find $k_{y} \sim 0.4/\rho_i$ corresponding to wavelengths of a microinstability $\lambda \sim 15\rho_i$ .

As mentioned earlier an important subtlety which has to be dealt with is the resolution of the algorithm for the base case presented in Table~\ref{table:StudTurb:73param}. By choosing field-line following coordinates and doing simulations local to some flux surface, the requirement is relaxed. Nevertheless the parameters responsible for the resolution still have to be selected carefully to resolve the physical mechanisms of interest. In linear simulations we deal with six parameters describing the parallel resolution, the velocity-space grid and the resolution in time. In Table~\ref{table:StudTurb:resparam} we present the resolution parameters used in this work. 

Each of the parameters in Table~\ref{table:StudTurb:resparam} specify different kinds of resolution. The length in the parallel direction along the field line is selected with $nperiod$. This length is divided into $ntheta \times nperiod$ number of grid points. The number of trapped and untrapped pitch angles $\xi \equiv v_{\parallel}/v$ sets the velocity space resolution. Trapped pitch angles are automatically calculated based on the value of $ntheta$. The number of untrapped pitch angles in velocity space is given by $2*ngauss$  and \textit{negrid} sets the number of energy grid points $\epsilon$. Finally \textit{delt} is simply the time step in units of $a/v_{t,i}$ and $omegatol$ is the accuracy of the converged growth rates. 

\begin{table}[h]
\centering
\caption{Values of the seven resolution parameters used in the simulations.}
\label{table:StudTurb:resparam}
\begin{tabular}{@{}llllll@{}}
\toprule
$nperiod$ & $omegatol$  & $delt \left[\unit{a/v_{t,i}}\right]$ & $ntheta$ & $ngauss$ & $negrid$   \\ \midrule
\addlinespace[0.5em]
11       & $2e^{-4}$         & 0.03    & 58      & 8      & 36     \\ \bottomrule
\end{tabular}
\end{table}

\begin{figure}[h]
 \setlength\figureheight{0.2\textheight}
 \setlength\figurewidth{0.9\textwidth}
    \centering
    \input{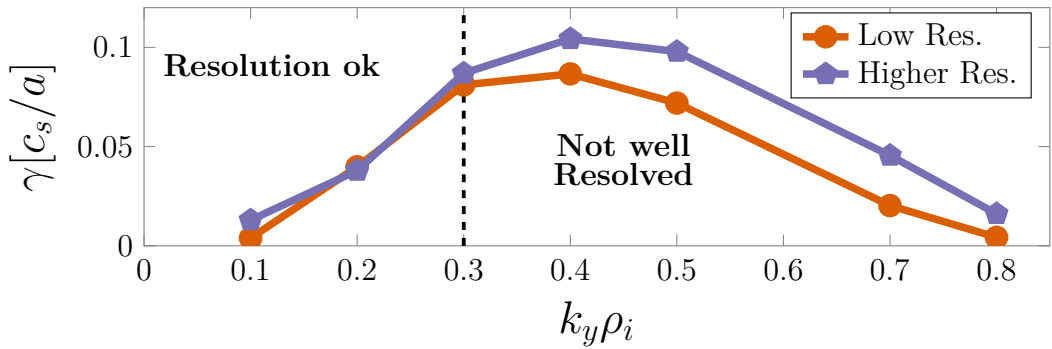}
    \caption{Scan in $k_y\rho_i$ for two cases. One with low resolution using $nperiod=3,  ntheta=16, delt=\SI{1.0}{a/v_{t,i}}, omegatol=2e^{-4}, ngauss=8, negrid = 9$ (orange circles) and a second (purple pentagons) with the resolution parameters presented in Table~\ref{table:StudTurb:resparam}. When increasing the resolution, growth rates at small scales (high $k_y\rho_i$) change the most and are therefore most difficult to resolve. We will find that these small scales are particularly sensitive to $delt, nperiod, ntheta$ and $negrid$.}
\label{fig:studTurb:motivRes:aky}
\end{figure}

In selecting these parameters the following procedure has been carried out. For a given scan in  $k_y\rho_i$ we select a default case with relatively low resolution, and identify the value of $k_y\rho_i$ which is most difficult to resolve. This is demonstrated in Fig.~\ref{fig:studTurb:motivRes:aky} where growth rates for a scan in $k_y\rho_i$ are shown for two cases. One with low resolution, and a second using higher resolution. At high scales ($k_y \rho_i < 0.4$) the parameters do not change the result much. As smaller scales are reached ($k_y \rho_i \ge 0.4$) the values of the growth rates is more sensitive for a change in the resolution parameters. We pick $k_y\rho_i = 0.7$ as a representative case for the region in the figure, which is most difficult to resolve.
Using this value of $k_y\rho_i$ we proceed by increasing one resolution parameter at a time until the simulated growth rate does not change too much. Ideally the growth rate should be constant, but here we relax this requirement by stating that growth rates should not change more than $\pm 15\%$ when the resolution parameters are increased by a sufficient amount, defined as follows. For $ngauss, nperiod$ this increase is two units (since they can only take certain discrete values), $\Delta delt \sim 50\%$ and for $negrid, ntheta$ the increase is at least 25\%. The latter two have been selected based on the variation of the growth rates observed when the resolution parameters are increased. The result of the simulations for each of the resolution parameters can be found in Appendix~\ref{app:D}. The results show that the number of untrapped pitch angles, controlled by $ngauss$ does not change the growth rates much, motivating the low value presented in Table~\ref{table:StudTurb:resparam}. The number of energy grid points, $negrid$ seems to be relatively well converged around $negrid = 36$. Next, $nperiod$ first decreases the growth rates substantially but seems to saturate around $nperiod = 11$ and beyond. Since $ntheta$ is closely related to $nperiod$ we consider variation in $ntheta$ at $nperiod = 11$.  We find $ntheta = 58$ correspond to relatively well converged growth rates. Finally an appropriate value  for the time step $delt$ is found. This parameter is important to resolve fast processes which might affect the growth rate. As the time step is decreased the growth rate increases rapidly at first, but then seem to saturate around $delt = \SI{0.03}{a/v_{t,i}}$. 

 It is important to stress that the resolution parameters obtained in this procedure are by no means ideal. They are merely a ``good'' choice to obtain relatively well resolved growth rates while keeping a tolerable computational time (30-50 CPU hours per simulation). However, we should note that in this work we are mostly concerned with general trends in the growth rates, which we have confirmed, remain the same if the resolution is further increased.
 
\section{Microinstabilities}\label{sec:studTurb:ITG}
 Microinstabilities are instabilities with wavelengths comparable to the ion Larmor radius $\rho_i$ and take the free energy available in a fusion plasma to drive turbulent fluctuations. In the category of small scale instabilities there are many candidates for driving the turbulence, but regarding the turbulence causing perpendicular transport of ion heat, the ion temperature gradient mode has been recognised as being the most important~\cite{ref:Estrada}. 

\subsection{The ITG instability}
\begin{figure}[h]
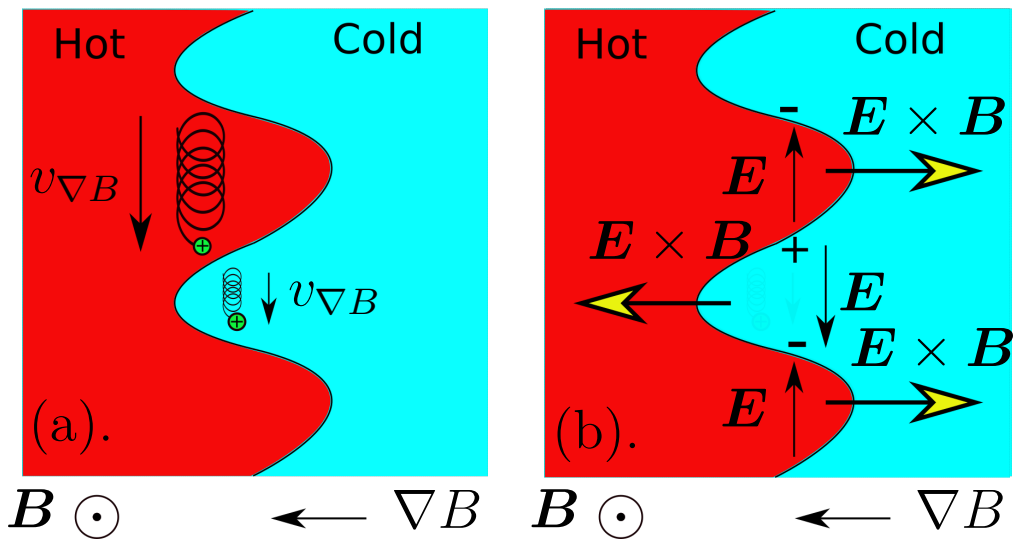

    \centering
\begin{subfigure}[b]{0.45\textwidth}
\resizebox{3.0in}{!}{\input{ITG7a.pdf_tex}}    
\phantomcaption
\label{fig:studTurb:ITGa}
\end{subfigure}
\begin{subfigure}[b]{0.45\textwidth}
\raisebox{0.0cm}{\resizebox{3.0in}{!}{\input{ITG7b.pdf_tex}}}
\phantomcaption
\label{fig:studTurb:ITGb}
\end{subfigure}
    \caption{Illustration of the ITG instability. A temperature perturbation between a hot and a cold region in the plasma generates a charge separation because of different $\nabla B$ drifts between the two regions (a). This creates an electric field and an $\vecE \times \vecB$ drift which for the given direction of $\nabla B$ works to enhance the instability (b).}
\label{fig:studTurb:ITG}
\end{figure}

Particles in a typical fusion plasma are subject to various drifts. In the presence of non-zero electric and magnetic fields the $\vecE \times \vecB$ drift is given by
\eqre{
\label{eq:StudTurb:ExB}
\vecV_E = \frac{1}{c}\frac{\vecE \times \vecB}{B^2}.
}
In the presence of spatially inhomogeneous magnetic fields we also have a $\nabla B$-drift
\eqre{
\label{eq:StudTurb:gradBD}
\vecV_{\nabla B} = \frac{v_{\perp}^2}{2\Omega} \frac{\vecB \times \nabla B}{B^2}.
}
Electrons and ions will move in opposite direction following from the charge dependence in $\Omega$. Consider the boundary between a hotter and a colder region in the plasma. Imagine there is a temperature perturbation in this boundary as shown in Fig.~\ref{fig:studTurb:ITGa}. The gyrating particles in the hotter region have a higher perpendicular velocity $v_{\perp}$ and consequently the $\nabla B$ drift as given by Eq.~\eqref{eq:StudTurb:gradBD} will also be larger compared to the drift of particles in the colder region. The temperature perturbation creates a longitudinal density wave in the direction of the $\nabla B$ drift, alternating between a dense and a sparse region of charged particles. The difference in charge density generates an electric field depicted in Fig.~\ref{fig:studTurb:ITGb}. If $\nabla T$ is aligned and points in the same direction as  $\nabla B$ the $\vecE \times \vecB$ drift in Eq.~\eqref{eq:StudTurb:ExB} will increase the temperature perturbation leading to an exponentially growing instability. If on the other hand $\nabla B$ points in the opposite direction of $\nabla T$ the $\vecE \times \vecB$ will stabilise the temperature perturbation instead. These two regions of $\nabla B$ are known as favourable or unfavourable curvature regions corresponding to the inboard and outboard side of the tokamak respectively~\cite{ref:Freidberg2007}. In Fig.~\ref{fig:studTurb:ITG} the magnetic field and its gradient are oriented such that the $\vecE \times \vecB$ drift enhances the instability. This is an example of unfavourable curvature.

\subsection{Dispersion relation for describing the ITG mode}\label{sec:DispRel1}
From the gyrokinetic ordering in Section~\ref{sec:studTurb:gk} we recall that turbulent fluctuations are strongly elongated along the field lines and their characteristic frequency $\omega$ is lower compared to the ion gyrofrequency $\Omega_i$. The quasineutrality condition holds therefore for the fluctuating density as well, $\sum_s Z_s\delta n_s = 0$. With the gyrokinetic Eq.~\eqref{eq:studTurb:GK} in Fourier space and quasineutrality condition for the fluctuating density we can derive a dispersion relation relating  $\tilde \omega = \omega + i\gamma$ to wavenumbers $k$. Given this relationship useful information on the characteristics of the ITG wave can be retrieved including how possible unstable growth scales with various species and plasma parameters. In this section a calculation for a plasma with only electrons and singly charged ions, is presented. In Section~\ref{sec:DispRel2} we consider then the modification in the dispersion relation as a consequence of adding a single species of fast ions in the plasma. In addition to the above assumptions, we neglect magnetic fluctuations and collisions for simplicity. These effects will be retained later in the numerical treatment presented in Chapter~\ref{CH:analysis}. 

Derivation of the dispersion relation has been demonstrated before. See the work done by Beer~\cite{ref:Beer1995} and Farid et al.~\cite{ref:Farid2001} for some examples. In this section we follow closely the derivation presented by Hammet~\cite{ref:gregNotes}. 

In what follows we will make frequent use of the Fourier transform~\cite{ref:hansen2014} and write the fluctuating quantities as a superposition of plane waves. For a function $\vecf(\vecx,t)$ we make a transformation from real space in $\vecx, t$ coordinates to the coordinates $\veck, \omega$ with
\eqre{
\label{eq:studturb:Fdef}
\mathscr{F}\left[\vecf(\vecx,t)\right](\omega,t) = \frac{1}{\left(2\pi\right)^4}\int \text{d}^3x\int \dif{t}\vecf(\vecx,t)e^{-i \left(\veck \cdot \vecx - \omega t\right)}.
}
For our case all derivatives are with respect to the guiding centre position $\vecR_s$ and we use $\vecR_s$ instead of $\vecx$ in the plane wave decomposition in Eq.~\eqref{eq:studturb:Fdef}. Given Eq.~\eqref{eq:studturb:Fdef} the Fourier transform of time and spatial derivatives of $h_s$ becomes

\eqre{
\label{eq:studturb:derF}
\mathscr{F}\left[\dod{h_s}{\vecR_s}\right] = i\veck\mathscr{F}\left[h_s\right], \qquad
\mathscr{F}\left[\dod{h_s}{t}\right] = -i\omega\mathscr{F}\left[h_s\right].
}

The starting point for finding a dispersion relation is to Fourier transform the gyrokinetic Eq.~\eqref{eq:studTurb:GK}
\eqre{
\label{eq:StudTurb:F1}
i\left(-\omega + \vpar\kpar + \omega_{dv}\right)h_s = \mathscr{F}\left[\text{RHS}\left( \text{Eq.~\eqref{eq:studTurb:GK}}\right)\right],
}
where Eq.~\eqref{eq:studturb:Fdef} has been invoked for the derivatives. We have introduced the drift frequency $\omega_{dv}$ as the Fourier transform (Eq.~\eqref{eq:studturb:Fdef}) of $\vecV_{Ds}\cdot \dif{h_s}/\dif{\vecR_s}$ in Eq.~\eqref{eq:studTurb:vel}. On the outboard side we get $\omega_{dv} = \omega_d\left(v_{\parallel}^2 + v_{\perp}^2/2\right)/v_{t,s}^2$, with $\omega_d = -\rho_s k_y v_{t,s}$. Only linear physics are considered so the nonlinear term $ \left < \vecV_\chi \right>_{\vecR}\cdot \partial h_s/\partial \vecR_s$ is neglected. ITG is an electrostatic mode in the sense that it is appears in the solution of the gyrokinetic equation, even when $\delta \vecB$ fluctuations are ignored. In this case the drift and fluctuating terms reduce to 
\eqre{
\label{eq:StudTurb:drift1}
\chi = \delta \varphi, \quad
\left <V_\chi\right >_{\vecR} = \frac{c}{\overline B}\vecb\times\dpd{\left<\delta\varphi\right>_{\vecR}}{\vecR_s}.
}
Taking the Fourier transform of $\left <V_\chi\right >_{\vecR}$ in Eq.~\eqref{eq:StudTurb:drift1} we obtain
\eq{
\mathscr{F}\left[\left <V_\chi\right >_{\vecR}\right] = -i \omega_*\frac{e J_0(k_y\rho_i) \delta \varphi }{T_s} \cdot L_{n,i},
}
where $\omega_* = \omega_d R/L_{n,i}$. The density gradient scale length $L_{n,i}$ is defined $1/L_{n,i} = -\dif{\ln n_i}/\dif{r}$. The zeroth order Bessel function $J_0(k_y\rho_i)$ represents the gyroaveraging operation in Fourier space.  Often one simply sets $J_0 = 1$ which is equivalent to neglecting Larmor orbit effects. In this simple treatment we make this assumption. 

Evaluating $\nabla F_0$ and using the fact that in the linear equation, $\left<\varphi \right>_{\vecR}$ is fluctuating with the same frequency $\omega$ as $h_s$, the RHS in Eq.~\eqref{eq:StudTurb:F1} reduces to  
\eqre{
\label{eq:studturb:RHS}
\mathscr{F}\left[\text{RHS}\left( \text{Eq.~\eqref{eq:studTurb:GK}}\right)\right] = i\left(\omega_{*v}^T - \omega\right) Z_se\delta \varphi F_{0s}/T_s,
}
where $\omega_{*v}^T = \omega_* \left[1 + \eta_i\left(v^2/v_t^2 - 3/2\right)\right] $ and $\eta_i= L_{n,i}/L_{T,i}$ where  $1/L_{T,i} = -\dif{\ln T_i}/\dif{r}$ is the inverse temperature gradient scale length. Combining Eqs.~\eqref{eq:StudTurb:F1} and~\eqref{eq:studturb:RHS}
yields the total Fourier transform of the gyrokinetic equation
\eqre{
\label{eq:studTurb:resp}
i\left(-\omega + \vpar\kpar + \omega_d\right)h_s = i\left(\omega_{*v}^T - \omega \right) \frac{Z_seF_{0s}}{T_s} \delta \varphi \implies \\
h_s = \frac{\omega-\omega_{*v}^T }{\omega - \vpar\kpar - \omega_d}\frac{Z_seF_{0s}}{T_s} \delta \varphi.
}
To this point the response in Eq.~\eqref{eq:studTurb:resp} is valid for any species. In what follows we will assume a plasma consisting of only electrons and singly charged ions. The ITG mode has a high frequency compared to the parallel ion response, but is much slower than the parallel response of electrons which are assumed to be adiabatic. The frequency of interest is $\vpar k_{\parallel} \sim \kpar v_{t,i} \ll \left(\omega, \omega_{*v}^T\right) \ll k_{\parallel}v_{t,e}$, where $v_{t,i/e}$ is the ion/electron thermal velocity.  The ion response simplifies to
\eqre{
\label{eq:studTurb:resp2}
h_i = \frac{\omega-\omega_{*v}^T}{\omega-\omega_d}\frac{eF_{0i}}{T_i}  \delta \varphi.
}
The ion and electron response are linked with Eq.~\eqref{eq:studTurb:QN} and we obtain
\eqre{
\label{eq:stduTurb:QN}
\frac{e\delta\varphi}{T_e}n_e = -\frac{e\delta\varphi}{T_i}n_i + \int h_i \text{d}^3v,
}
again with adiabatic electrons ($h_e = 0$) and $J_0=1$ for the long wavelength result. 
In the cold plasma approximation $\omega \gg \omega_{d}$ and we can approximate the fraction in Eq.~\eqref{eq:studTurb:resp2} to second order 
\eqre{
\label{eq:studTurb:exp}
\frac{\omega - \omega_{*v}^T }{\omega - \omega_d} \approx 1 - \frac{\omega_{*v}^T}{\omega} +  \frac{\omega_d}{\omega} - \frac{\omega_{*v}^T \omega_d}{\omega^2} + \frac{\omega_d^2}{\omega^2}.
}
Retaining up to second order terms in Eq.~\eqref{eq:studTurb:exp} and using Eq.~\eqref{eq:stduTurb:QN} we obtain 
\eqre{
\label{eq:stduTurb:QN:eval}
\frac{e\delta\varphi}{T_e}n_e = -\frac{e\delta\varphi}{T_i}n_i + \\ \frac{e\delta \varphi}{T_i}\left( \int \text{d}^3v F_{0i} - \int \text{d}^3v F_{0i}\frac{\omega_{*v}^T}{\omega} +  \int \text{d}^3v F_{0i}\frac{\omega_d}{\omega} -  \int \text{d}^3v F_{0i}\frac{\omega_{*v}^T \omega_d}{\omega^2} +  \int \text{d}^3v F_{0i}\frac{\omega_d^2}{\omega^2}\right).
}
Each integral in Eq.~\eqref{eq:stduTurb:QN:eval} involves moments of the Maxwellian distribution function which may be evaluated using the identity $\int\text{d}^3v F_{0,i}/n_{0,i}v_x^{2l} = v_t^{2l}(2l-1)(2l-3)(2l-5)\cdots 5\cdot 3 \cdot 1$~\cite{ref:gregNotes} for the spatial coordinate $x$. Evaluating each integral for both $x$ and $y$ components a dispersion relation is found
\eqre{
\label{eq:disprel0}
\frac{1}{\tau} = -\frac{\omega_*}{\omega} + \frac{2\omega_d}{\omega}-\frac{\omega_d\omega_*2(1+\eta_i)}{\omega^2} + \frac{7\omega_d^2}{\omega^2},
}
where $\tau = T_e/T_i$. The imaginary part of Eq.~\eqref{eq:disprel0} gives the growth rate
\eqre{ 
\label{eq:studTurb:disp1}
\gamma = \frac{\tau}{2}\sqrt{4\omega_d^2 - \frac{4\omega_d}{\tau}\left( 2\eta_i \omega_* - 7\omega_d + 2\omega_*\right) - 4\omega_*\omega_d + \omega_*^2}.
}
The term proportional to the temperature gradient, obtained by taking the high temperature gradient limit, gives
\eqre{
\label{eq:studTurb:Gpropto}
\gamma \propto \sqrt{\tau\eta_i2\omega_*\omega_d} = \sqrt{\frac{\tau}{2}}k_{y}\rho_i\frac{v_t}{\sqrt{L_{T,i}}}.
}
In Eq.~\eqref{eq:studTurb:Gpropto} the increase in the growth rates with the ion temperature gradient is evident. For simplicity we take the flat density limit ($L_{n,i} \rightarrow \infty$) such that $\omega_* \rightarrow 0$.  Solving the equation $\gamma = 0$ for the ion temperature gradient we obtain the threshold for instability
\eqre{
\label{eq:studTurb:crit1}
\frac{R}{L_{T,i}} = \frac{1}{2}\left(7 + \tau\right).
}
Below this threshold this simple derivation predicts $\gamma = 0$ and the ITG mode is stable. For higher temperature gradients the mode becomes unstable and grows with the growth rate $\gamma$ (Eq.~\eqref{eq:studTurb:disp1}). If the critical value of the ion temperature gradient leading to positive growth rates is increased,  the plasma can sustain a higher temperature gradient without driving ITG unstable and generate large turbulent heat fluxes. The term \textit{stiffness} is often used in the literature and refers to the scaling of the heat flux with $1/L_{T,i}$, but may linearly be thought of the corresponding scaling with the growth rate $\gamma$ in Eq.~\eqref{eq:studTurb:disp1}. A stiff profile will lead to a high $\gamma$ for a small increase in $1/L_{T,i}$.

\begin{figure}[h]
 \setlength\figureheight{0.2\textheight}
 \setlength\figurewidth{0.9\textwidth}
      \centering
    \input{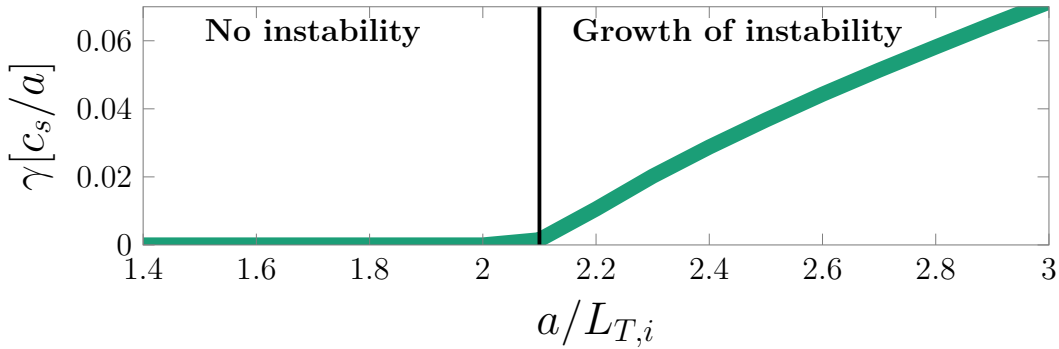}
    \caption{Growth rates for a scan in the main ion temperature gradient. Above some critical value of the ion temperature gradient the ITG mode grows unstable as indicated by positive growth rates $\gamma$. Only electrostatic fluctuations have been included.}
\label{fig:studTurb:Ref_tg}
  \end{figure}

The scaling of the growth rates with the ion temperature gradient in Eq.~\eqref{eq:studTurb:Gpropto} and the critical threshold predicted by Eq.~\eqref{eq:studTurb:crit1} are the two main characteristics of ITG which distinguishes it from other modes in a similar frequency range. These theoretical predictions can be contrasted with numerical results obtained from running \texttt{GS2} for the JET discharge 73224 case presented in Section~\ref{sec:studTurb:GS2}. Growth rates for varying ion temperature gradients are shown in Fig.~\ref{fig:studTurb:Ref_tg}. 

The growth rate is zero until $a/L_{T,i} \approx 2.1$ where a roughly linear increment of $\gamma$ with the ion temperature gradient is seen. For $\tau = 1$, $R = \SI{3.12}{m}$ and $a = \SI{0.96}{m}$ the prediction from the analytical model in Eq.~\eqref{eq:studTurb:crit1} is $a/L_{T,i,\text{crit}} \approx 1.24$, below the numerical result. If increased further (not shown) the growth rate follows a square-root dependence on $a/L_{T,i}$ as predicted by the theory.

\begin{figure}[h]
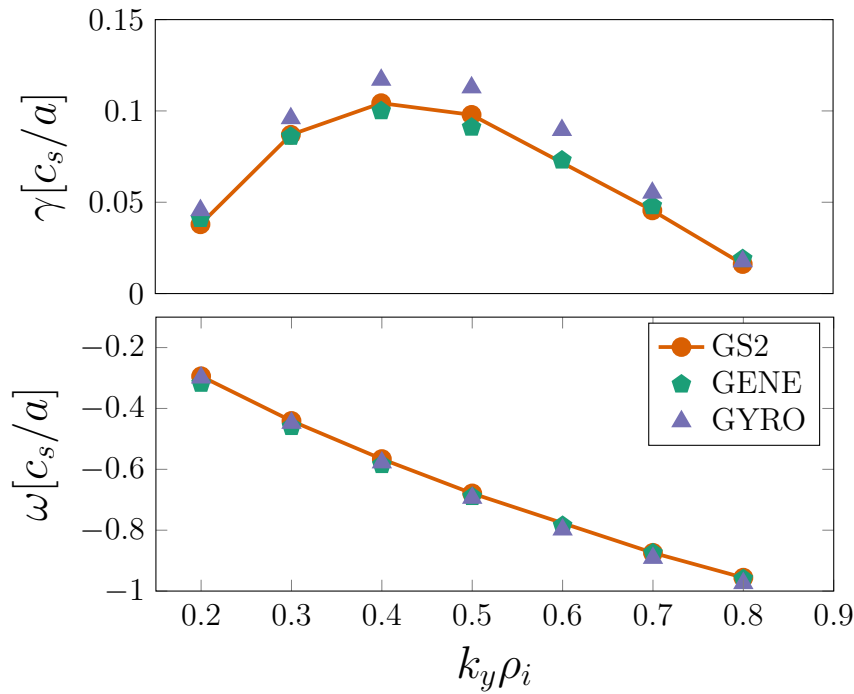

 \setlength\figureheight{0.22\textheight}
 \setlength\figurewidth{0.7\textwidth}
     \centering
    \input{../../figures/Ref/ref_aky_growth.tex}\\
    \hspace{0.29cm}
    \input{../../figures/Ref/ref_aky_omega.tex}
    \caption{Comparison of linear growth rates (top) and frequencies (bottom) for \texttt{GS2} and corresponding results from \texttt{GENE} and \texttt{GYRO}. The latter two include flow shear, while in \texttt{GS2} this term has been neglected. Also the \texttt{GS2} simulations use much higher values of the resolution parameters as presented in Table~\ref{table:StudTurb:resparam}. The simulations have included both electrostatic and magnetic fluctuations.}
\label{fig:StudTurb:Ref}
  \end{figure}

The presence of a peak growth rate is clearly seen in Fig.~\ref{fig:StudTurb:Ref} where growth rates and real frequencies are shown for a scan in $k_y\rho_i$. The results are verified with a comparison of corresponding results obtained from \texttt{GENE} and \texttt{GYRO} as reported by Bravenec et al.~\cite{ref:Bravenec}\footnote{Thanks to R. Bravenec for providing the \texttt{GENE} and \texttt{GYRO} data.}. In this case both fluctuations in the electric and magnetic field have been included. All codes solve the same gyrokinetic equation, but because of different numerical algorithms and implementations, some deviation is possible. For the frequency, the \texttt{GS2} output has been multiplied by a minus sign to match the definition of $\omega^*$ (and the definition used in deriving the dispersion relation in Eq.~\eqref{eq:disprel0}). The result is in good agreement between the codes even though flow shear was included in the \texttt{GENE} and \texttt{GYRO} simulations. These are also less resolved compared to our \texttt{GS2} simulations.   

The quantity $k_y\rho_i$ is the wavelength of the fluctuations, decreasing to the right in the figure. At $k_y\rho_i \sim 1$ the wavelength are in the order of the ion Larmor radius. A peak growth rate at $k_y \rho_i \sim 0.4$ can be seen. We should note that the theoretical model in Eq.~\eqref{eq:studTurb:Gpropto} wrongly predicts a linear scaling with $k_y\rho_i$ without a peak in the growth rate. The reason is most likely that we dropped finite Larmor orbit effects and used $J_0 = 1$. All simulations will unless stated otherwise be performed at  $k_y \rho_i \sim 0.4$ for the peak growth rate. In some cases the location of this peak might be shifted to $k_y \rho_i \sim 0.5$ instead and in that case, all simulations will be modified accordingly. 
 
\subsection{Effects of fast ions on the critical threshold}\label{sec:DispRel2}
The expression for the critical threshold in Eq.~\eqref{eq:studTurb:crit1} was derived for a plasma consisting of electrons and ions only. We want to investigate how the fast ions change this expression. 

The fast ions enter through the quasineutrality condition 
\eqre{
\label{eq:studTurb:eqQN2}
\frac{e\delta \varphi}{T_e}n_e = (1-n_f/n_i)\left(-\frac{e\delta \varphi}{T_i}n_i + \int h_i \text{d}^3v\right) + n_f/n_i \left(-\frac{e\delta \varphi}{T_f}n_f + \int h_f \text{d}^3v\right),
}
and are again singly charged and where $n_f$ is the fast ion density. Given Eq.~\eqref{eq:studTurb:eqQN2} Liljeström~\cite{ref:Liljestrom1990} derived a modified expression for the critical threshold at low $n_f/n_i$. In deriving this expression Liljeström have assumed $\omega/\omega_{Df} \ll 1$  which is valid since the fast ion population has a very high temperature $T_f/T_e \gg 1$. They have also neglected Larmor orbit effects, although the fast ions, because of their high energy, will have a large Larmor radius such that $k_y\rho_i \sim 1$ and the zeroth order Bessel function will be far from one. To examine the consequences of the various simplifications we compare the analytical findings with numerical results.

\begin{figure}[h]
 \setlength\figureheight{0.20\textheight}
 \setlength\figurewidth{0.9\textwidth}
    \input{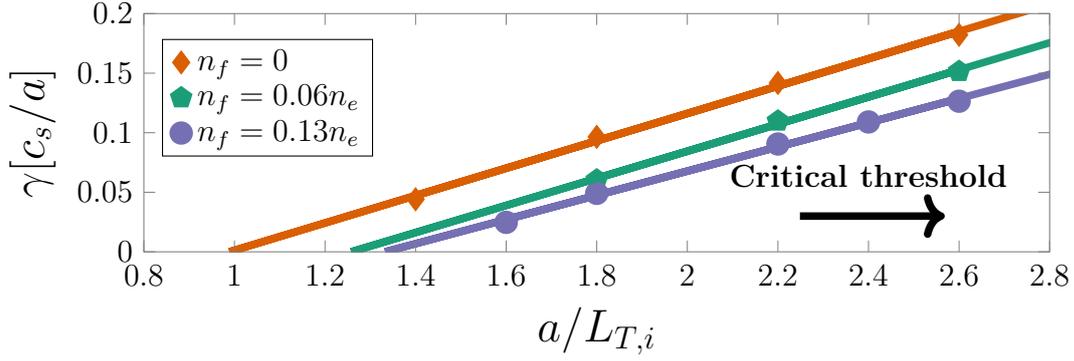}
    \caption{Scan of the growth rate $\gamma$ in $a/L_{T,i}$ for no (orange diamonds), intermediate (cyan pentagons) and high (purple circles) values of the fast ion density. Only electrostatic fluctuations have been included.}
    \label{fig:studTurb:dens}
  \end{figure}

Growth rates as a function of the main ion temperature gradients, for several fast ion densities are shown in Fig.~\ref{fig:studTurb:dens}. A linear fit of the results has been made to illustrate stiffness and the value of the critical ion temperature gradient threshold. Only electrostatic fluctuations have been included. Higher fast ion densities leads to higher value of the critical threshold. 
We note that the slope of the line which we interpret as the stiffness of the temperature profile, does not change much. The main effect is instead the shift in the threshold and therefore the result derived by Liljeström is of particular interest.

\begin{figure}[h]
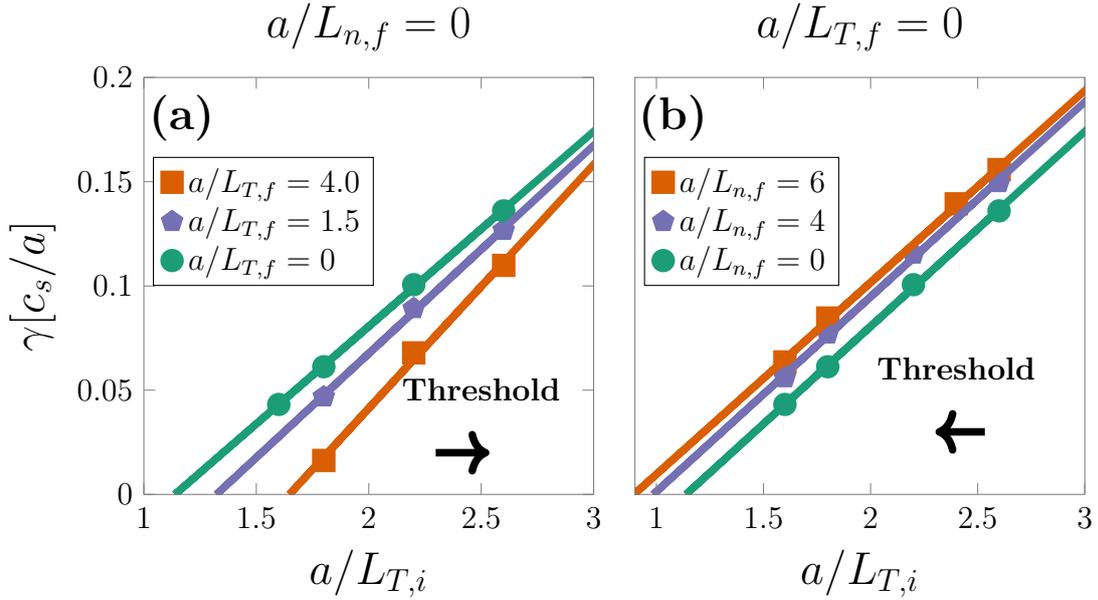

 \setlength\figureheight{0.3\textheight}
 \setlength\figurewidth{0.5\textwidth}       
\begin{subfigure}[b]{0.4\textwidth}          
    \input{../../figures/ESfin/tg/tprim_growth.tex}
    \phantomcaption
    \label{fig:studTurb:tprim}
        \end{subfigure}
        \hspace{2cm}
        \begin{subfigure}[b]{0.4\textwidth}    
   \raisebox{0.0cm}{\input{../../figures/ESfin/tg/fprim_growth.tex}}
       \phantomcaption
    \label{fig:studTurb:fprim}
             \end{subfigure}
    \caption{Scan in  $a/L_{T,i}$ for no (orange squares), intermediate (purple pentagons) and high (cyan circles) values of the fast ion  temperature (left) and density gradient (right) respectively.}
    \label{fig:studTurb:prims}
  \end{figure}

The fast ions are characterised by either high density (NBI) or temperature (ICRH) gradients. Liljeström suggests the fast ion temperature gradient should increase the critical main ion temperature gradient threshold and is therefore stabilising. Increasing the density gradient is instead destabilising since the threshold for instability occurs at smaller values of the ion temperature gradient. This agrees well with simulations presented in Fig.~\ref{fig:studTurb:tprim} and Fig.~\ref{fig:studTurb:fprim} where growth rates for varying main ion temperature gradients are shown. While the theoretical model presented by Liljeström predicts both temperature and density gradients to change the magnitude of the growth rate equally, it is clear from the numerical results in Fig.~\ref{fig:studTurb:prims} that this is not the case. With regard to the absolute change in the critical threshold in Fig.~\ref{fig:studTurb:prims} it seems the fast ion temperature gradient has a stronger influence compared to the fast ion density gradient. The fast ion density gradient has to be increased more compared to the fast ion temperature gradient, to obtain a similar change in critical threshold.

In generating Fig.~\ref{fig:studTurb:prims} we changed the fast ion temperature gradient of fast helium only, while keeping the other parameters, including the fast ion temperature gradient of fast deuterium, fixed. Then, for the density gradient we vary instead only the fast deuterium density gradient, while keeping all the other parameters including the fast ion density gradient of fast helium, fixed. A difference between these two species is their charge. Based on the model developed by Liljeström, we expect that any stabilising or destabilising effect by either increasing the fast ion density, density gradient or temperature gradient should be amplified by a factor of $Z$. This might explain the difference (in magnitude) change of the growth rates for an equal variation in the fast ion temperature and density gradients. Fast helium is doubly charged, while fast deuterium has $Z = 1$. The role of their gradients might still be of equal importance, but the effect is amplified by the fast ion charge leading to a stronger change in the growth rate when the fast helium temperature gradients has been modified.

\begin{figure}[h]
 \setlength\figureheight{0.2\textheight}
 \setlength\figurewidth{0.9\textwidth}
    \centering
    \input{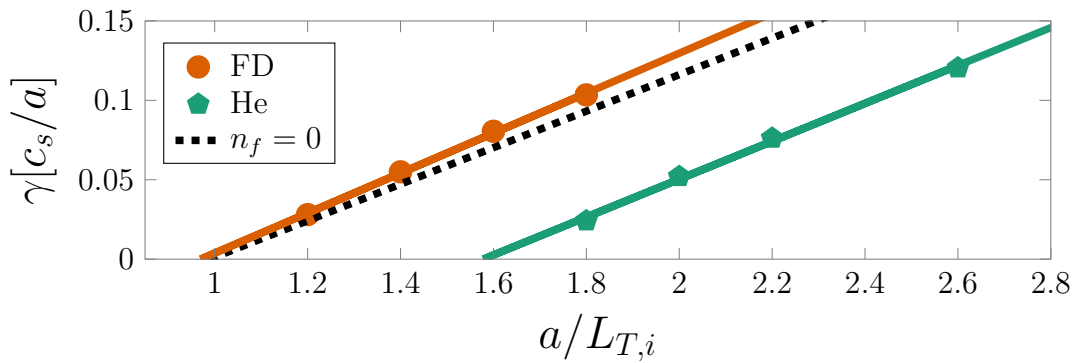}
    \caption{Comparison of growth rate and mode frequency for two cases: (1) only including fast deuterium in the simulations (orange circles) and (2) only fast helium (cyan pentagons). Also shown is the growth rates when fast ions have been removed entirely from the list of species (dashed black line) in Table~\ref{table:StudTurb:73param}. Fast deuterium is slightly destabilising and increases the growth rates because of the large density gradient.}
            \label{fig:studTurb:HeFD} 
  \end{figure}

We may examine the separate role of NBI  (fast deuterium) and ICRH (fast helium) generated fast ions. Given the theoretical and numerical results observed to this point we predict fast deuterium to be less stabilising compared to fast helium, indicated by a smaller value of the critical ion temperature gradient leading to positive growth rates. A simulation comparing critical threshold of only fast deuterium or only fast helium is shown in Fig.~\ref{fig:studTurb:HeFD}. 
It is clear that fast helium is far better at stabilising the ITG mode as compared to fast deuterium. This is indicated by the significantly larger values of the critical ion temperature gradient leading to positive growth rates. Also shown are the growth rates when fast ions have been removed entirely from the list of species in Table~\ref{table:StudTurb:73param}. Including only fast deuterium leads to a slight decrease in the critical ion temperature gradient threshold and increases the slope of the linear fit. Overall this indicates that adding fast deuterium is leading to larger growth rates compared to not having any fast ions at all. In other words, fast deuterium is destabilising the ITG mode instead of stabilising it. We should remind that these simulations and the theoretical models have assumed only electrostatic fluctuations while neglecting possible fluctuations in the magnetic field. In the next chapter we focus on numerical solutions of the gyrokinetic equations and consider the modification in the role of fast ions when magnetic fluctuations are included. 

\clearpage
\setlength{\parskip}{0pt} 

\chapter{Gyrokinetic simulations}\label{CH:analysis}
Using the tools presented in earlier chapters we may now turn to the question we wish to answer with this work: what is the role of fast ions in stabilising ITG turbulence? In this chapter we perform linear gyrokinetic simulations with \texttt{GS2} to explain the role of fast ions in stabilising the ITG mode. Our reference case consists of the physical parameters of the JET discharge and the resolution parameters, both presented in Section~\ref{sec:studTurb:GS2}. The analysis is organised as follows. 
Directly in the beginning of Section~\ref{sec:analysis:s1} the changes in ITG growth rates as a consequence of adding fast ions with varying density, are presented. We show how the fast ions reduce the peak growth rates and how they increase the critical value of the ion temperature gradient. Also, we present how they enhance the already strong finite $\beta$ stabilisation. We proceed with explaining the role of individual fast ion parameters in the observed reduction of the growth rates. In Section~\ref{sec:analysis:beta} the interaction of the fast ion density and temperature gradients with plasma $\beta$, is analysed. The previous chapter ended with the electrostatic effects of the gradients, but here we investigate this further by including also magnetic fluctuations. In Section~\ref{sec:analysis:temp} a separate, electrostatic study of the large temperature of the fast ion population, is done. Then we continue in Section~\ref{sec:analysis:isotop} with examining the role of fast ion mass and charge in reducing the growth rates. We summarise our findings by presenting the differences between using NBI and ICRH generated fast ions. In Section~\ref{sec:analysis:dil} the role of secondary effects, is investigated. These are the change in electron density profile because of quasineutrality, the fast ion contribution to $\beta$ and its radial derivative. Including only secondary effects is more commonly known as \textit{dilution}. Before ending this chapter we link back to the introduction and attempt to use the quasi-linear model for explaining the large reduction in the nonlinearly saturated heat flux.
\section{Fast ion stabilisation of the ITG mode}\label{sec:analysis:s1}
 The gyrokinetic simulations are performed using three different densities of fast ions: $n_{\text{He}} = n_{\text{FD}} = 0.0$ ($n_f = 0$),
 $n_{\text{He}} = n_{\text{FD}} = 0.03n_e$ ($n_f = 0.06n_e$) and $n_{\text{He}} = 0.07n_e, n_{\text{FD}} = 0.06n_e$ ($n_f = 0.13n_e$). The last is the reference case presented in Table~\ref{table:StudTurb:73param}. For each change in fast ion density we modify the electron density profile to account for quasineutrality, plasma $\beta$ (Eq.~\eqref{eq:studturb:beta}) along with its gradient, the effective charge $Z_{\text{eff}}$ (Eq.~\eqref{eq:studturb:zeff}) and the collision frequencies (Eq.~\eqref{eq:vnewk}).
  The geometry is fixed in order to isolate other effects of fast ions. The previous chapter ended with some examples of gyrokinetic simulations of the fast ion effects on the ITG growth rates, including only fluctuations in the electric field. In this section the magnetic field fluctuations are retained.
 
\begin{figure}[h]
 \setlength\figureheight{0.25\textheight}
 \setlength\figurewidth{0.9\textwidth}
    \input{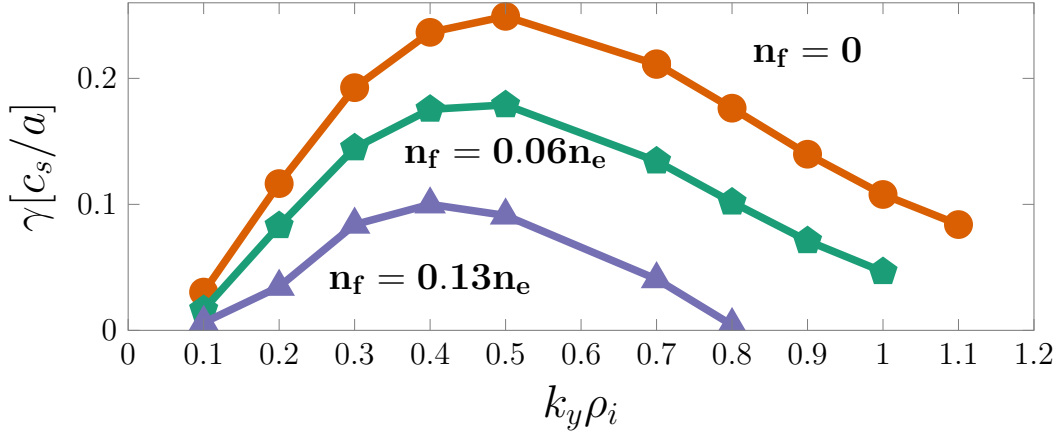}
    \caption{$k_y \rho_i$ spectrum of growth rates. Scans are performed for fast ion densities $n_{\text{He}} = n_{\text{FD}} = 0.0$ ($n_f = 0$, orange circles),
 $n_{\text{He}} = n_{\text{FD}} = 0.03n_e$ ($n_f = 0.06n_e$, cyan pentagons) and $n_{\text{He}} = 0.07n_e, n_{\text{FD}} = 0.06n_e$ ($n_f = 0.13n_e$, purple triangles). The last is the reference case presented in Table~\ref{table:StudTurb:73param}.}
    \label{fig:analysis:aky1}
  \end{figure}

A $k_y\rho_i$ spectrum for each of the three fast ion densities is shown in Fig.~\ref{fig:analysis:aky1}. The spectrum has a parabolic shape with a peak at roughly $k_y\rho_i = 0.5$. Without fast ions the peak growth rate is $\gamma \sim \SI{0.25}{c_s/a}$ which decreases when the fast ion density is increased, reaching $\gamma \sim \SI{0.1}{c_s/a}$ at $n_f = 0.13n_e$ and $k_y\rho_i \sim 0.4$.
For the plasma regime described by the reference parameters in Table~\ref{table:StudTurb:73param} the ITG instability is  the dominant instability, creating the growth rates seen in Fig.~\ref{fig:analysis:aky1}. The instability is driven by the main ion temperature gradient, example of which we have already seen in Section~\ref{sec:studTurb:ITG}. 
\begin{figure}[h]
 \setlength\figureheight{0.25\textheight}
 \setlength\figurewidth{0.9\textwidth}
    \input{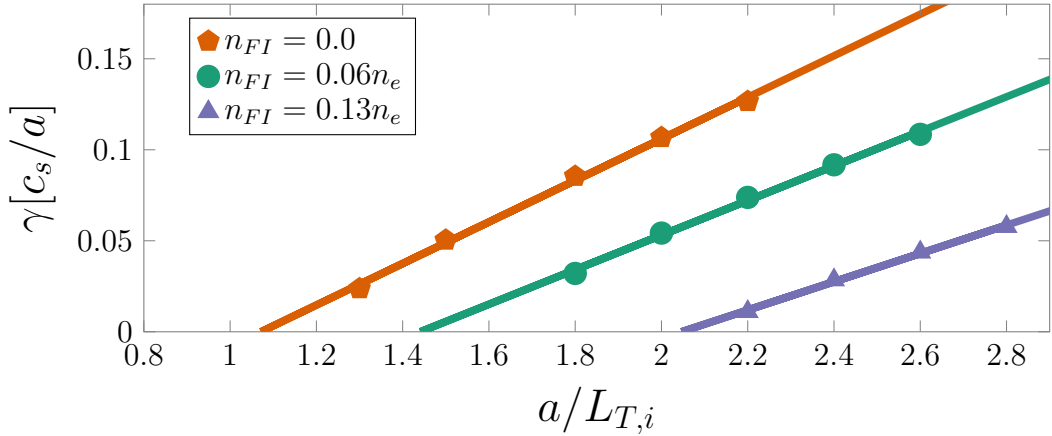}
    \caption{Scans in the ion temperature gradient for varying fast ion density. Simulations are performed for the peak growth rates at $k_y \rho_i \approx 0.4$ as seen in Fig.~\ref{fig:analysis:aky1}. Due to the computational cost of performing these simulations, only a few points are shown. A linear fit is then used to demonstrate the approximate value of the critical threshold. If we compare with the electrostatic results in Fig.~\ref{fig:studTurb:dens} we note a significant increase in the critical threshold.}
    \label{fig:analysis:tg1}
  \end{figure} 
In Fig.~\ref{fig:analysis:tg1} we make a scan in the main ion temperature gradient for the peak growth rate ($k_y\rho_i  = 0.4$) of the three cases with varying fast ion densities shown in Fig.~\ref{fig:analysis:aky1}. In all simulations that follow $k_y\rho_i  = 0.4$ will be used by default.  We see that the effects of fast ions is to increase the threshold of instability. A linear fit is also presented to demonstrate the approximate value of the critical threshold. In contrast to Fig.~\ref{fig:studTurb:dens} where only electrostatic fluctuations were considered, the increase in the critical ion temperature gradient threshold because of fast ions, is
much bigger. Clearly the stabilising effect of fast ions is enhanced when magnetic
fluctuations are included. We will see that the large increase in the critical ion
temperature gradient is due to electromagnetic effects. In Fig.~\ref{fig:analysis:tg1}
also a slight decrease in the slope of the linear fit can be noticed for large fast ion
densities ($n_f = 0.13n_e$).

\subsection{The role of $\beta$}\label{sec:analysis:beta}
Recall that the thermal to magnetic pressure ratio $\beta$ in Eq.~\eqref{eq:studturb:beta} separates electrostatic from electromagnetic simulations. To examine the role of electromagnetic effects a scan in $\beta$ is appropriate. We remind that in \texttt{GS2} variation in the total $\beta$ is achieved by modifying the reference value $\beta_{\text{ref}}$.
 
\begin{figure}[h]
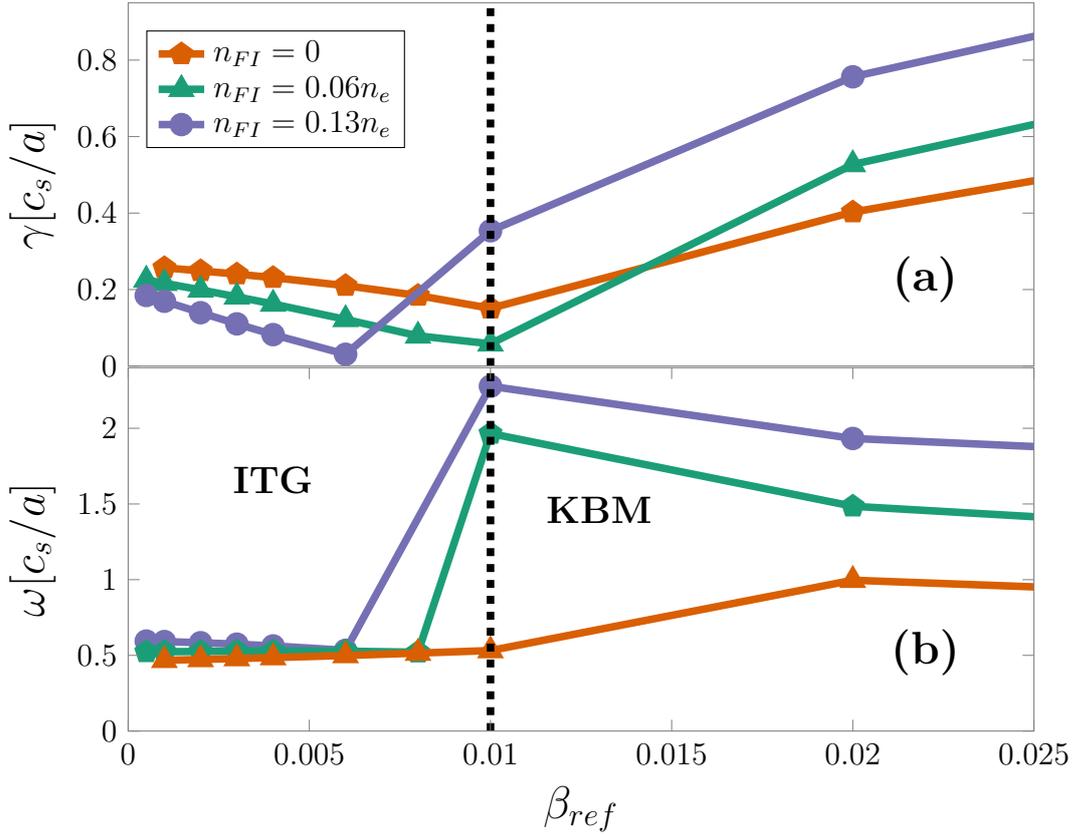

 \setlength\figureheight{0.27\textheight}
 \setlength\figurewidth{0.9\textwidth}
    \centering
         \begin{subfigure}[b]{\textwidth}
            \input{../../figures/EMfin/betaFin/dens_growth.tex}
            \phantomcaption
            \label{fig:analysis:beta1_a}            
        \end{subfigure}\\
    \vspace{-0.8cm}
    \hspace{0.54cm}
             \begin{subfigure}[b]{\textwidth}
            \input{../../figures/EMfin/betaFin/dens_omega.tex}
            \phantomcaption
            \label{fig:analysis:beta1_b}            
            \end{subfigure}
    \caption{Growth rates (top) and mode frequencies (bottom) for a scan in the reference $\beta$ for varying fast ion density. Again with $k_y \rho_i \approx 0.4$ and fast ion densities as shown previously in Fig.~\ref{fig:analysis:aky1}: $n_f = 0$ (cyan pentagons), $n_f = 0.06n_e$ (orange triangles) and $n_f = 0.13n_e$ (purple circles). }
    \label{fig:analysis:beta1}
  \end{figure}

In Fig.~\ref{fig:analysis:beta1} the growth rates (Fig.~\ref{fig:analysis:beta1_a}) and mode frequencies (Fig.~\ref{fig:analysis:beta1_b}) for different values of reference $\beta_{\text{ref}}$ are shown. The general characteristics of this figure are the following. Electrostatic simulations are equivalent to setting $\beta_{\text{ref}} = 0$. For non-zero values there is a strong suppression of the growth rates with a maximum reduction at $\beta_{\text{ref}} \approx 0.01$. This strong reduction is believed to be due to the coupling of the ion acoustic wave with the shear Alfvén wave demonstrated for example in Ref.~\cite{ref:Nishimura2007}. Above this value the growth rates instead increase very rapidly. Looking at the corresponding mode frequencies in Fig.~\ref{fig:analysis:beta1_b} we see a high jump for similar values of reference $\beta$ indicating a change in the dominant mode, most likely to the kinetic ballooning mode (KBM)~\cite{ref:Citrin3}. If ITG is dominating up to $\beta \approx 0.01$ KBM is more important for $\beta_{\text{ref}} > 0.01$. This thesis is limited to study only the dynamics of ITG and we focus therefore on $\beta_{\text{ref}} \le 0.01$ where KBM is less important. 
Introducing fast ions makes several improvements. At $\beta_{ref} = 0$, the electrostatic case, the difference between no fast ions and fast ions with a density $n_f = 0.13n_e$ (the reference case) is a reduction in the growth rate by roughly a factor of 25\%. This difference increases with $\beta$ before the threshold of the next instability. The stabilising effect can be separated in two stages. First is an enhanced effect of $\beta$ measured by the steeper slope when fast ions are added. For a given increase in $\beta$ the reduction in the growth rates is larger with fast ions compared to without fast ions. Second is a left-shift of the threshold such that lower growth rates are reached for the same value of $\beta$ (before the threshold of instability). It is also interesting to see the order of this effect: first a steeper slope, then a shift in the threshold as the fast ion density is increased. Fig.~\ref{fig:analysis:beta1} makes it clear that the effect of fast ions is stronger in electromagnetic simulations, compared to electrostatic, in agreement with the larger increase in the critical ion temperature gradient threshold seen in Fig.~\ref{fig:analysis:tg1}. In this section we will investigate which of the fast ion parameters causes the enhanced $\beta$ stabilisation. 

\begin{figure}[h]
 \setlength\figureheight{0.2\textheight}
 \setlength\figurewidth{0.9\textwidth}
      \centering
    \input{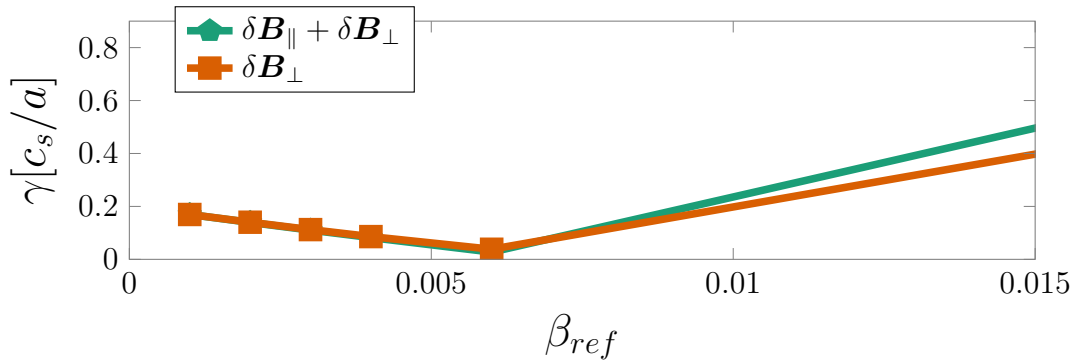}
    \caption{Scan in $\beta_{ref}$ for two cases. The first includes only perpendicular magnetic fluctuations (orange squares) while the second also contains parallel magnetic field fluctuations (cyan pentagons). Below $\beta_{\text{ref}} = 0.007$ the two cases are practically indistinguishable.}
    \label{fig:analysis:fapar}
  \end{figure}

Before beginning with the analysis, recall that the magnetic fluctuations can be separated into compressional (parallel) perturbations $\delta \vecB_{\parallel}$ and sheared (perpendicular) perturbations $\delta \vecB_{\perp}$. In Fig.~\ref{fig:analysis:fapar} we perform a scan in $\beta_{\text{ref}}$ for two cases. One includes only $\delta \vecB_{\perp}$ fluctuations and another including also $\delta \vecB_{\parallel}$ fluctuations.
At low values of $\beta_{\text{ref}}$, $\delta \vecB_{\parallel}$ does not seem to make any difference. Only above the critical threshold where KBM is destabilised does parallel  $\delta \vecB_{\parallel}$ fluctuations start to matter. Since the 73224 discharge is at $\beta_{\text{ref}} = 0.0033$, we note from Fig.~\ref{fig:analysis:fapar} that $\delta \vecB_{\perp}$ fluctuations are, in our case, most important to include. 

Two crucial fast ion parameters are their high density and temperature gradients. The effects of these parameters were already presented in Section~\ref{sec:DispRel2} in the previous chapter but there without including magnetic fluctuations. Our observations were compared with the analytical model derived by Liljeström~\cite{ref:Liljestrom1990} which was in agreement with our findings. From the main ion temperature gradient scan in Fig.~\ref{fig:studTurb:prims} we concluded a stabilising effect on the growth rates because of the fast ion temperature gradient, as it increases the critical value of the main ion temperature gradient which generates the non-zero growth rates of the ITG instability. Increasing the density gradient on the other hand, reduced the threshold for instability, as was shown in Fig.~\ref{fig:studTurb:prims}. In other words, the density gradient of fast ions is destabilising. In this section these electrostatic findings are generalised by including magnetic fluctuations and by investigating the role of the fast ion gradients on the $\beta$-stabilisation. 

\begin{figure}[h]
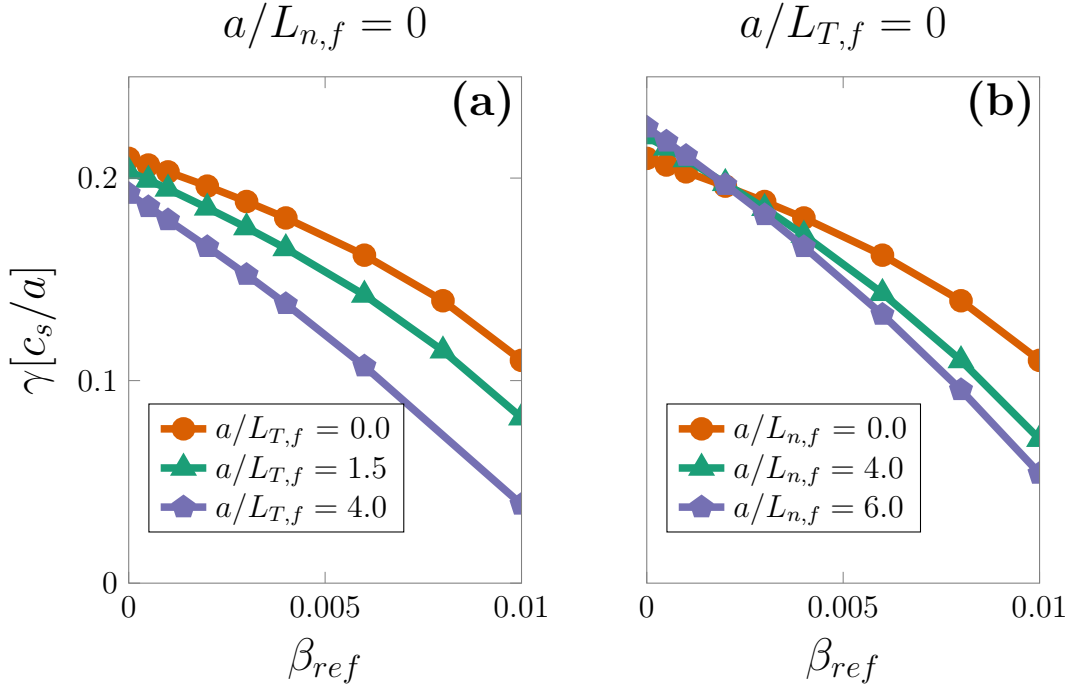

 \setlength\figureheight{0.35\textheight}
 \setlength\figurewidth{0.45\textwidth}
    \centering
             \begin{subfigure}[b]{0.4\textwidth}
\input{../../figures/EMfin/beta/tprim_growth.tex}        
            \phantomcaption
           \label{fig:analysis:EMBeta_tprim}
 \end{subfigure}
 \hspace{2cm}
              \begin{subfigure}[b]{0.4\textwidth}
    \input{../../figures/EMfin/beta/fprim_growth.tex}
            \phantomcaption
            \label{fig:analysis:EMBeta_fprim}
 \end{subfigure}        
    \caption{Left figure: Scan in $\beta_{\text{ref}}$ for several values of the fast ion temperature gradients: $a/L_{T,\text{He}} = 0$ (orange circles), $a/L_{T,\text{He}} = 1.5$ (cyan triangles) and $a/L_{T,\text{He}} = 4$ (purple pentagons). The fast deuterium parameters are kept fixed and the density gradient has been removed for both energetic species. Right figure: same scan but for several values of the fast ion density gradients instead: $a/L_{n,\text{FD}} = 0$, $a/L_{n,\text{FD}} = 4$ and $a/L_{n,\text{FD}} = 6$. The fast helium parameters are kept fixed and the temperature gradient has been removed for both energetic species. }
    \label{fig:analysis:EMBeta_prim}
  \end{figure}

Let us begin with the fast ion temperature gradient. In Fig.~\ref{fig:analysis:EMBeta_tprim} a scan in $\beta$ is shown for three values of fast helium temperature gradients: $a/L_{T,\text{He}} = 0$, $a/L_{T,\text{He}} = 1.5$, $a/L_{T,\text{He}} = 4.0$. The fast deuterium temperature gradient is kept fixed, and the density gradient is set to zero for both species. 
At $\beta_{\text{ref}} = 0$ the temperature gradient is stabilising as seen in previous electrostatic scans. Increasing $\beta$ leads to an even stronger stabilising effect because of the fast helium temperature gradient. A similar scan, with varying fast ion density gradient in Fig.~\ref{fig:analysis:EMBeta_fprim} shows a completely different behaviour. We show results for three values of fast deuterium density gradients: $a/L_{n,\text{FD}} = 0$, $a/L_{n,\text{FD}} = 4.0$, $a/L_{n,\text{FD}} = 6.0$.  Higher values of the fast ion density gradients compared to fast ion temperature gradients are required to observe a clear change in the growth rates. The fast helium density gradient is kept fixed, and the temperature gradient is set to zero for both energetic species. 

At $\beta = 0$, without magnetic fluctuations, the fast ion density gradient is clearly destabilising as previously discussed in Section~\ref{sec:DispRel2}. But this role changes for higher $\beta$. Above $\beta_{\text{ref}} \approx 0.003$ the fast ion density gradients leaves the growth rates unchanged and for higher $\beta_{\text{ref}}$ it even becomes stabilising.

Overall it is possible to conclude that the fast ion temperature and density gradients enhance finite $\beta$ stabilisation, in the sense that increasing the fast ion gradients leads to a greater reduction in the growth rates, for a given change in $\beta$ below the threshold of the KBM instability. Having a larger $\beta$ is of particular interest for NBI generated fast ion with high density gradients. If these fast ions are destabilising at low $\beta$ their role is reversed at higher $\beta$ and together with ICRH generated fast ions they contribute in significantly reducing growth rates of ITG. 
\subsection{Contrasting fast ions with thermal impurities}\label{sec:analysis:temp}
In addition to their high gradients, the fast ion population is characterised by a large temperature. In electromagnetic simulations, when $\beta$ is included, the effect of temperature is to increase the pressure, which propagates to an increase in $\beta$ and reduced growth rates (if $\beta$ is not too high). The temperature of the fast ion population is important also for other reasons. 

The analytical predictions in Section~\ref{sec:DispRel2} and all our numerical results have shown a different role of the fast ion temperature gradient in contrast to the main ion temperature gradient. The first leads to a decrease in the growth rates, while the second is what drives ITG unstable and increases the growth rates. Because of their high temperature it has been suggested that energetic ions do not participate in the dynamics of ITG ~\cite{ref:Tardini2007} and an increase in their temperature gradient should therefore not necessarily complement the main ion temperature gradient in increasing the growth rates of the instability. This claim is assessed in this section. To prevent $\beta$ from overshadowing other possible effects of increasing the temperature, we keep only electrostatic fluctuations using $\beta = 0$.

\begin{figure}[h]
 \setlength\figureheight{0.2\textheight}
 \setlength\figurewidth{0.9\textwidth}
 \centering
        \input{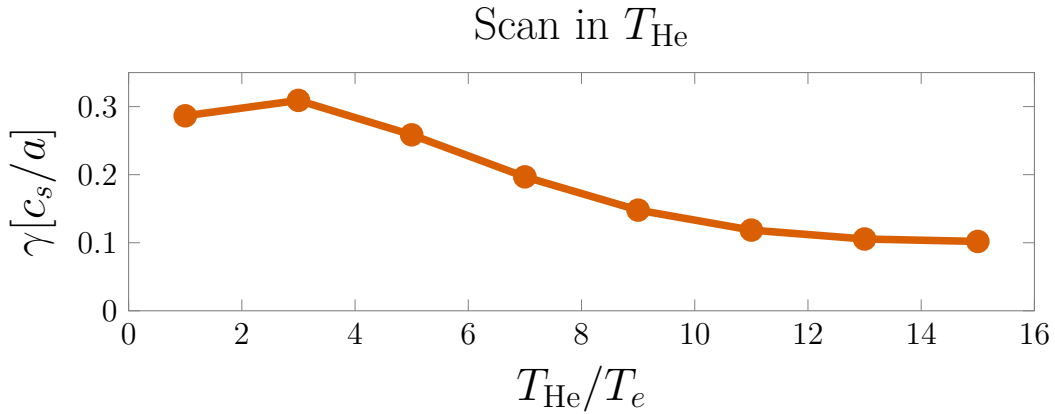}
    \caption{Growth rates for increasing fast helium temperature. Below $T_{\text{He}} \approx 3T_e $ increasing the fast helium temperature is destabilising. For higher values growth rates start to reduce until $T_{\text{He}} \approx 14T_e$ is reached, where they seem to saturate around a value of $\gamma = \SI{0.1}{c_s/a}$. The simulations have not included magnetic fluctuations.}
    \label{fig:analysis:temp}
  \end{figure}

In Fig.~\ref{fig:analysis:temp} a scan in the fast helium temperature is shown for the range $2T_e \le T_{\text{He3}} \le 15T_e$. All the other parameters are of the reference case in Table~\ref{table:StudTurb:73param} and once again with $k_y \rho_i = 0.4$.
In the beginning of the figure, increasing the fast helium temperature is destabilising as is indicated by the increase in the growth rates for $T_{\text{He}}<3T_e$. At $T_{\text{He}}=3T_e$ a peak of $\gamma \approx \SI{0.3}{c_s/a}$ is reached and afterwards the growth rates are effectively reduced until $T_{\text{He}} \approx 14T_e$ where they saturate around $\gamma = \SI{0.1}{c_s/a}$. Apart from $\beta$, the high energy of the fast ions (in the nominal case $T_{\text{He}} = 6.9T_e$) is stabilising by itself. The result in Fig.~\ref{fig:analysis:temp} suggest that the effect of helium is first similar to the main ions (destabilising) but after $T \approx 3T_e$ the helium impurity has to be treated differently from the other, thermal ions. 

\begin{figure}[H]
 \setlength\figureheight{0.25\textheight}
 \setlength\figurewidth{0.9\textwidth}
    \centering
        \input{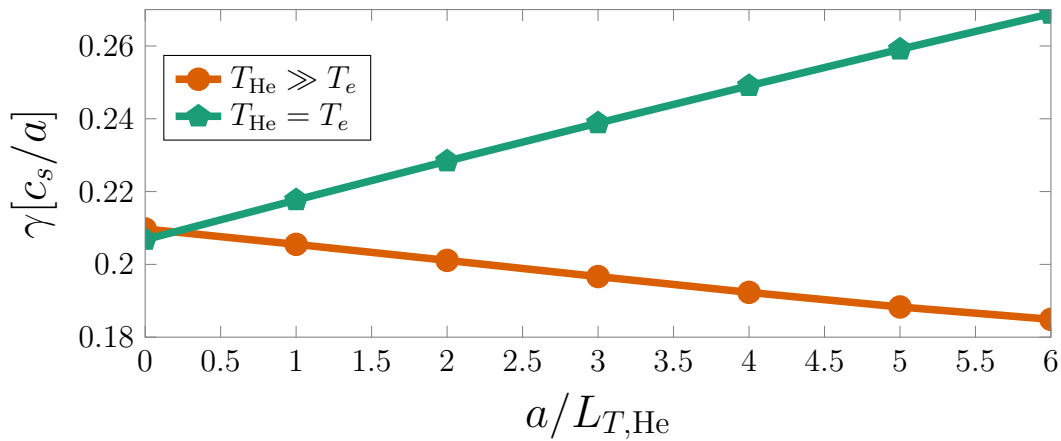}
    \caption{Scan in the helium ion temperature gradient for two cases. One (cyan pentagons) with a helium ion temperature equal to the electron temperature (in which case the fast ion is no longer fast) and a second (orange circles) with the original, high temperature of helium ($T_{\text{He}} = 6.9T_e$). In contrast to previous findings we now find the temperature gradient to be destabilising.}
    \label{fig:analysis:tg2}
  \end{figure}

Let us examine how the role of the helium temperature gradient changes with the temperature. In Fig.~\ref{fig:analysis:tg2} we present a scan in the fast helium temperature gradient for two cases: one where $T_{\text{He}}$ is the nominal value ($T_{\text{He}} \gg T_e$) and a second case with $T_{\text{He}} = T_e$. 
There is a distinct difference between the growth rates for the two cases. Increasing the helium temperature gradient leads to lower growth rates for $T_{\text{He}} \gg T_e$, while for $T_{\text{He}} = T_e$ the growth rates are increased. This shows that fast ions have a completely different role compared to thermal impurities. A question to ask is: is the distinction between thermal and energetic ions possible to make  also when magnetic fluctuations are considered? In Section~\ref{sec:analysis:beta} we found the stabilising effect of the fast helium temperature gradient to be enhanced in the high $\beta$ regime, while the main ion temperature gradient remains destabilising (see for example Fig.~\ref{fig:analysis:tg1}). Therefore the interpretation of Fig.~\ref{fig:analysis:tg2} is expected to remain also when magnetic fluctuations have been included. 

To this point the effects of the fast ion gradients and temperature have been analysed. The latter both in terms of the contribution to total $\beta$ and also in the coupling to the temperature gradient of the energetic species. This leaves the fast ion density and charge, which are discussed in the next section.
\subsection{Isotope effects}\label{sec:analysis:isotop}
\begin{figure}[H]
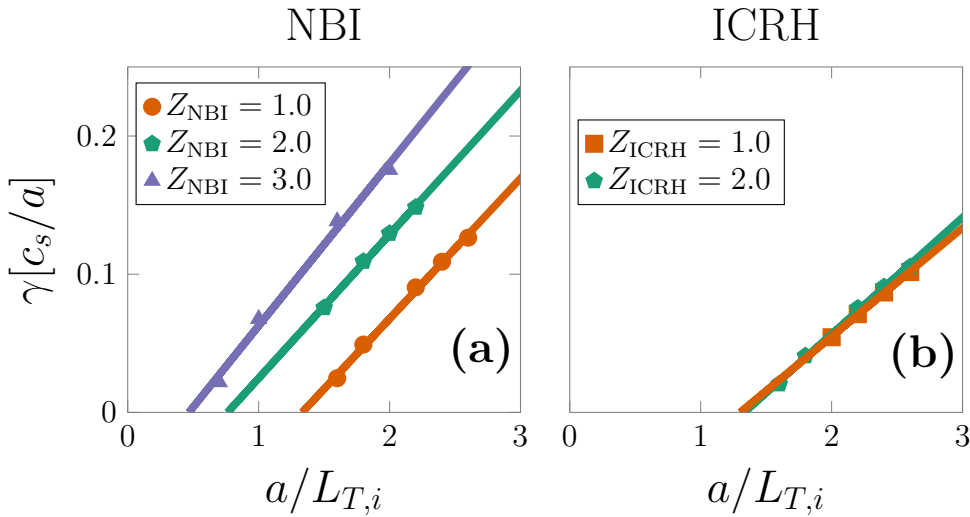

 \setlength\figureheight{0.26\textheight}
 \setlength\figurewidth{0.45\textwidth}
   \centering
               \begin{subfigure}[b]{0.4\textwidth} 
    \input{../../figures/ESfin/tg/charge.tex}
            \phantomcaption{}
           \label{fig:analysis:tg4_a}
\end{subfigure}
 \hspace{1cm}
               \begin{subfigure}[b]{0.4\textwidth} 
    \input{../../figures/ESfin/tg/charge_ICRH.tex}
            \phantomcaption{}
            \label{fig:analysis:tg4_b}
 \end{subfigure}
    \caption{The effect of charge on the stiffness and critical threshold. Each scan is separately performed for NBI (left) and ICRH (right) heated fast ions, using the parameters of fast deuterium and fast helium respectively (except for the charge).}
    \label{fig:analysis:tg4}
  \end{figure}

The model discussed in Section~\ref{sec:DispRel2} predicts the charge to enhance the stabilising (or destabilising) effects of the fast ion temperature or density gradients. Because of their different parameters, we separate between NBI and ICRH generated fast ions. These have all the parameters of fast deuterium and fast helium respectively, except for the charge which we will vary. In Fig.~\ref{fig:analysis:tg4_a} a scan in the ion temperature gradient for $Z_{\text{NBI}} = 1$, $Z_{\text{NBI}} = 2$ and $Z_{\text{NBI}} = 3$, is presented. The fast helium parameters (ICRH) are kept fixed. The same but for varying ICRH charge and constant fast deuterium parameters is shown in~\ref{fig:analysis:tg4_b}. We obtain the reference case (Table~\ref{table:StudTurb:73param}) when $Z_{\text{NBI}} = 1$ and $Z_{\text{ICRH}} = 2$ in the two figures respectively. For simplicity, all scans in this section are electrostatic ($\beta = 0$).
 Increasing the charge of NBI heated fast ion enhances the destabilising effect of the density gradient as is seen by the stronger reduction in the critical threshold in Fig.~\ref{fig:analysis:tg4_a}. This confirms the idea of the charge multiplying the destabilising effect of the density gradient. The same conclusion cannot be drawn for the high fast ion temperature gradient (ICRH). In the step from $Z_{ICRH} = 1$ to $Z_{ICRH} = 2$ the growth rates are more or less unchanged. We might recall the discussion in Section~\ref{sec:studTurb:ITG} where it was suggested that the stronger effect of the helium temperature gradient compared to the fast deuterium density gradient could be explained by the double charge of helium. Given the conclusions in this section this statement does not seem to hold, rather it is indeed the temperature gradient itself which is more important in changing the ITG growth rates compared to the fast ion  density gradient. 
 
\begin{figure}[h]
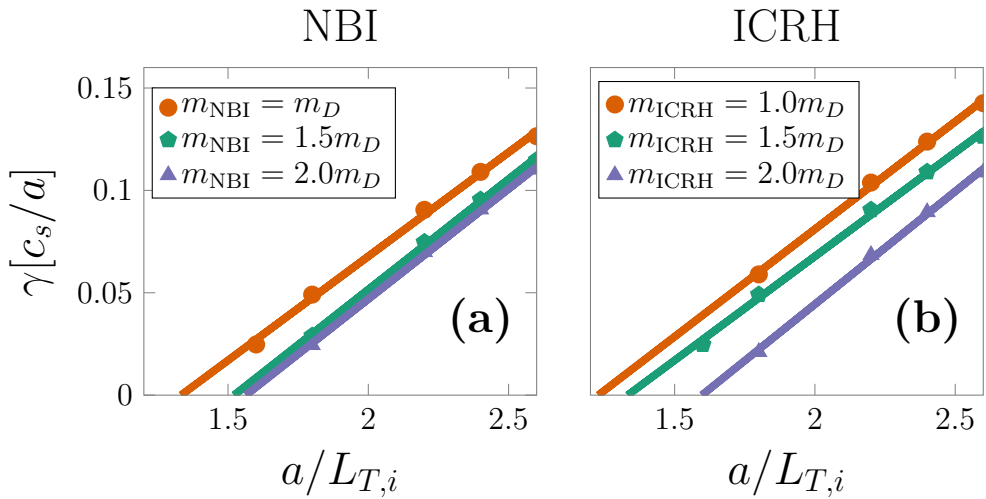

 \setlength\figureheight{0.25\textheight}
 \setlength\figurewidth{0.45\textwidth}
                 \begin{subfigure}[b]{0.3\textwidth}  
    \input{../../figures/ESfin/tg/mass.tex}
            \phantomcaption
                \label{fig:analysis:tg5_a}
   \end{subfigure}
    \hspace{3.0cm}
                   \begin{subfigure}[b]{0.3\textwidth}   
    \input{../../figures/ESfin/tg/mass_ICRH.tex}
            \phantomcaption
           \label{fig:analysis:tg5_b}
   \end{subfigure}
    \caption{The effect of mass on the stiffness and critical threshold. Each scan is separately performed for NBI (left) and ICRH (right) heated fast ions, using the parameters of fast deuterium and fast helium respectively (except for the mass).}
    \label{fig:analysis:tg5}
  \end{figure}

 Let us proceed with performing a similar scan but for varying isotope mass of NBI and ICRH generated fast ions. In Fig.~\ref{fig:analysis:tg5_a} we vary the mass of NBI generated fast ions, again using all the parameters of fast deuterium in Table~\ref{table:StudTurb:73param}, but the mass. Similarly in Fig.~\ref{fig:analysis:tg5_b} we vary the mass of ICRH generated fast ions. In both cases, the other hot ion species parameters are kept constant just as in Fig.~\ref{fig:analysis:tg4}.

The result is the same for both NBI and ICRH generated fast ions: increasing the fast ion mass leads to lower growth rates indicated by increasing the threshold in the ion temperature gradient. The stabilising effect is larger in the case of ICRH, Fig.~\ref{fig:analysis:tg5_b} as compared to NBI in Fig.~\ref{fig:analysis:tg5_a}.
   
\begin{figure}[h]
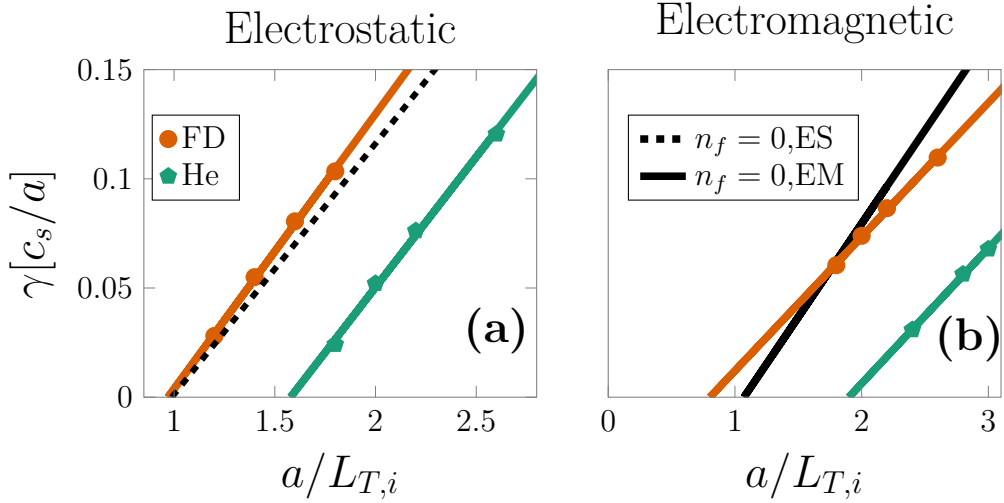

 \setlength\figureheight{0.25\textheight}
 \setlength\figurewidth{0.45\textwidth}
                 \begin{subfigure}[b]{0.3\textwidth}      
    \input{../../figures/ESfin/tg/He_vs_FD_growth1.tex}
            \phantomcaption
            \label{fig:analysis:comp_a}
            \end{subfigure}
    \hspace{3cm}            
                 \begin{subfigure}[b]{0.3\textwidth}      
              \input{../../figures/EMfin/tgFin/comp.tex}
            \phantomcaption
            \label{fig:analysis:comp_b}
            \end{subfigure}
    \caption{Comparison of growth rate and mode frequency for two cases: (1) only including fast deuterium in the simulations (orange circles) and (2) only fast helium (cyan pentagons). Results are shown without $\delta \vecB$ fluctuations (left) and with (right).}
    \label{fig:analysis:comp}
  \end{figure}

To conclude our findings in this and previous sections we compare the overall difference in the growth rates for a scan in the ion temperature gradient, when using either fast deuterium or only fast helium.  In Fig.~\ref{fig:analysis:comp_a} the growth rates obtained when using the parameters from Table~\ref{table:StudTurb:73param} and removing fast helium from the list (orange circles) or fast deuterium (cyan pentagons), is shown. Again, fluctuations in the magnetic field are not considered.
Fig.~\ref{fig:analysis:comp_a} was discussed already in Section~\ref{sec:DispRel2} where the slightly destabilising effect of fast deuterium (even compared to the case without fast ions ($n_f = 0$) was pointed out. The suggested reason is the high density gradient of the species. Fast helium, because of the large temperature gradient increases the critical ion temperature gradient threshold and is stabilising the ITG mode. We proceed with comparing this electrostatic result with the equivalent cases when magnetic fluctuations are also included, presented in Fig.~\ref{fig:analysis:comp_b}. Again, the high temperature gradient species (He) is most effective in reducing the growth rates. The role of magnetic fluctuations and $\beta$
is to increase the critical threshold even further, from $a/L_{T,i} \approx 1.5$ (electrostatic) to $a/L_{T,i} \approx 2.0$ (electromagnetic). In agreement with the scan in $\beta_{\text{ref}}$ in Fig.~\ref{fig:analysis:EMBeta_tprim} including magnetic fluctuations enhances the stabilising influence of the fast ion temperature gradient. The critical ion temperature gradient threshold for the high density gradient species (FD) is more or less unchanged when magnetic fluctuations are also considered. In fact it is slightly reduced suggesting that the NBI generated FD ions are even more destabilising compared to the case with only electrostatic fluctuations. This statement seem to hold for small values of the ion temperature gradient but as soon as larger values are reached, a stabilising influence of FD can be observed. For comparison we show the case with electromagnetic fluctuations but without fast ions $n_f = 0$ (black dashed line). Including FD has reduced the slope of the linear fit and leads therefore to a stabilising influence for sufficiently high ion temperature gradient. For the nominal parameters of the 73224 discharge, $a/L_{T,i} = 3.56$ and FD is therefore contributing in reducing the ITG growth rates. 

Apart from the gradients, it should be noted that our ICRH generated fast ions have a slightly larger density, mass, temperature and charge as compared to our NBI generated fast ions. All these effects, as we have seen, contribute to enhancing the stabilisation even further. Their difference in density and temperature leads to a different contribution to the total pressure and therefore a different contribution to total $\beta$. In the end, this implies a further stabilising effect when transitioning from the electrostatic cases in Fig.~\ref{fig:analysis:comp_a} and electromagnetic in Fig.~\ref{fig:analysis:comp_b}, as is indicated by the increase in the slope. 
\subsection{Role of secondary effects}\label{sec:analysis:dil}
  \begin{figure}[h]
    \hspace{-1cm}
\input{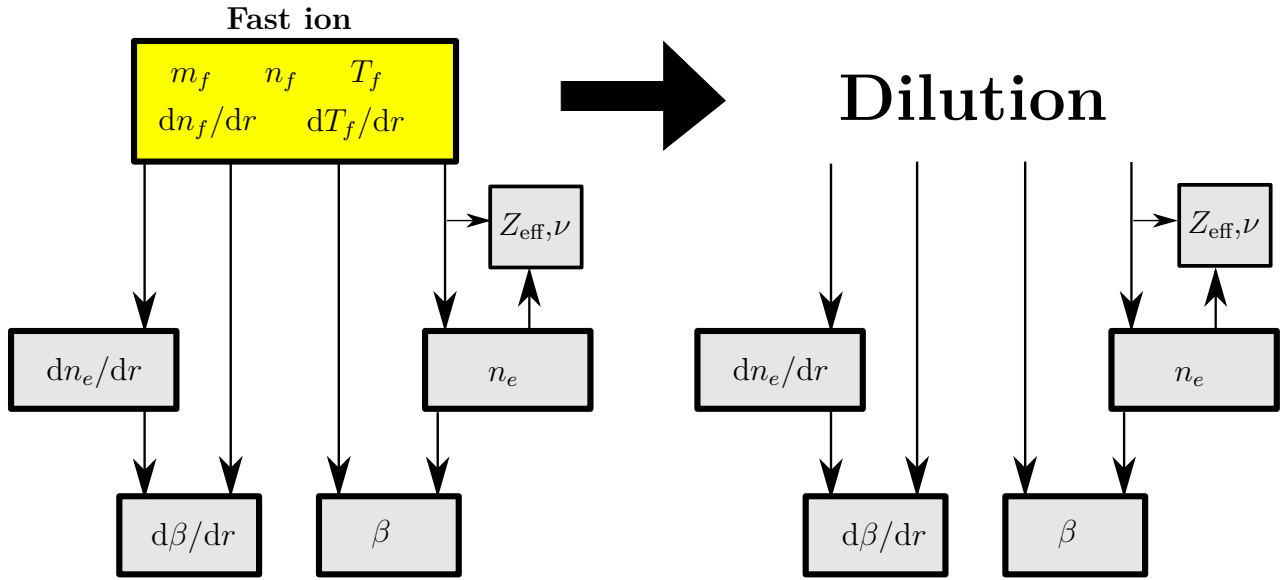}
    \caption{Illustration of the various secondary changes which has to be included for a change in the fast ion parameters. These are the electron density, density gradient, $Z_\text{eff}$, collision frequencies $\nu$, $\beta$ and its radial derivative. If we keep all the secondary changes because of the presence of fast ions, but remove them from the list of species in the \texttt{GS2} simulation, we get \textit{dilution}.}
    \label{fig:analysis:second}
  \end{figure}
In the previous section when the parameters of the fast ions were changed, we also needed to recalculate the electron density and its gradient (to satisfy quasineutrality), the collision frequency, the effective charge $Z_{\text{eff}}$, the radial derivative of $\beta$, and $\beta$ itself. We call the changes in all these parameters as the \textit{secondary effects} of the fast ions. The secondary effects of fast ions can equally well be accounted for by varying the electron and/or ion parameters without actually having to include the fast ion species in the simulation. Only including the secondary effects of the fast ions is known as \textit{dilution} and is illustrated in Fig.~\ref{fig:analysis:second}. In this section we will determine the role of dilution in the reduction of the ITG growth rates observed in the previous sections. 

\begin{figure}[h]
 \setlength\figureheight{0.25\textheight}
 \setlength\figurewidth{0.9\textwidth}
          \centering
              \input{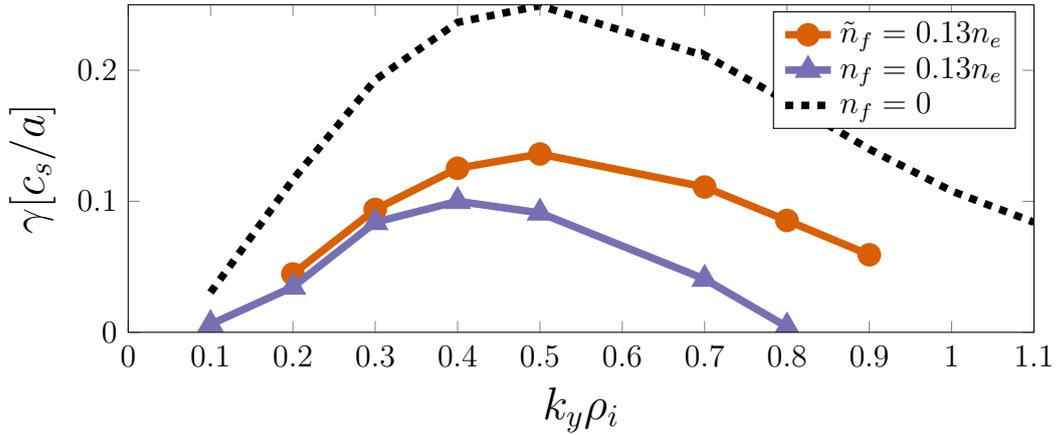}
    \caption{Scan in $k_y\rho_i$ for three sets of input parameters. The first (purple triangles) is the reference case including both the direct and secondary effects of the fast ions. The second ($\tilde n_f = 0.13n_e$, orange circles) only keeps the secondary effects of the fast ions and finally in $n_f = 0$ (dashed black line) fast ions have been entirely removed.}
    \label{fig:analysis:dil1}
  \end{figure}

The role of dilution in stabilising ITG can be evaluated by comparing the growth rates obtained for the reference ($n_f = 0.13n_e$) and the case when all the secondary effects (presented in Fig.~\ref{fig:analysis:second}) are included, but the fast ion themselves are removed as an active species. Electromagnetic results are shown in Fig.~\ref{fig:analysis:dil1} for a scan in $k_y\rho_i$. We show  dilution (denoted by the $\sim$ symbol) $\tilde n_f = 0.13n_e$, when the fast ions have been removed, but their secondary effects remains. The other case with $n_f = 0.13n_e$ is the same, but now with fast ions as an active kinetic species. For comparison we also present growth rates obtained without including fast ions at all, $n_f = 0$ (see initial Fig.~\ref{fig:analysis:aky1}) which includes removing their secondary effects. 

\begin{figure}[h]
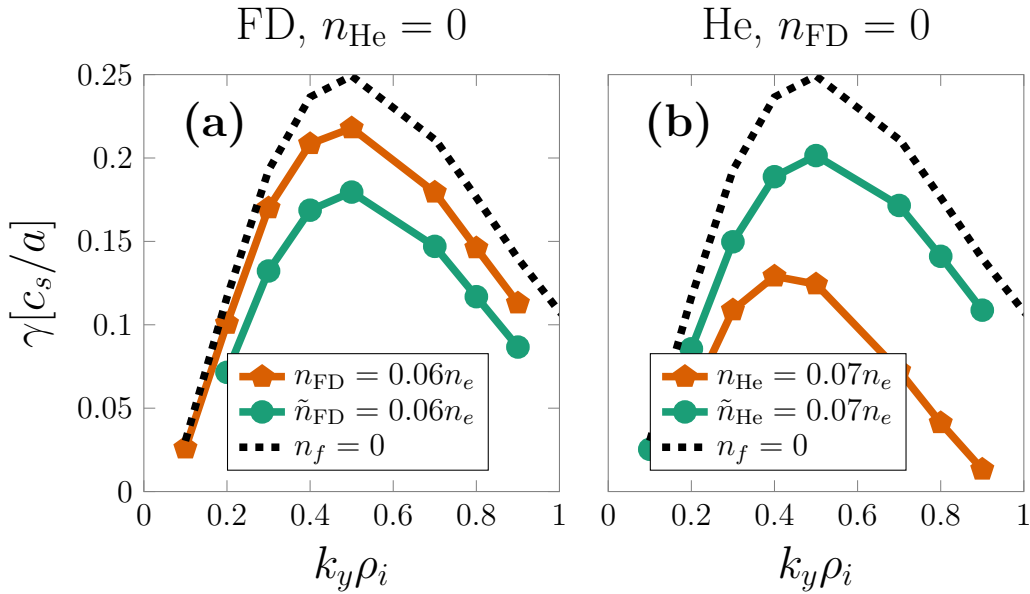

 \setlength\figureheight{0.3\textheight}
 \setlength\figurewidth{0.47\textwidth}
                           \begin{subfigure}[b]{0.3\textwidth}      
              \input{../../figures/EMfin/dilution/nofi_FD_aky.tex}
            \phantomcaption
            \label{fig:analysis:compFD}
            \end{subfigure}
    \hspace{3cm}               
                               \begin{subfigure}[b]{0.3\textwidth}      
              \input{../../figures/EMfin/dilution/nofi_He_aky.tex}
            \phantomcaption
            \label{fig:analysis:compHe}
            \end{subfigure}
    \caption{Scan in $k_y\rho_i$ for three sets of input parameters. Left: no fast ions (dashed black line), with FD (orange pentagons) and dilution using the parameters of FD only (cyan circles). He has been completely removed. Right: same  but using He and completely removing FD. Simulations have included magnetic fluctuations.}
    \label{fig:analysis:compSp}
  \end{figure}

   Without fast ions the peak growth rates are $\gamma \approx \SI{0.25}{c_s/a}$. With fast ions this drops down to $\gamma \approx \SI{0.1}{c_s/a}$. Dilution reduces the peak growth rate to $\gamma \approx \SI{0.13}{c_s/a}$ which corresponds to 85 \% of the total stabilising effect of the fast ions. Dilution alone does not quite account for all of the observed stabilisation, but for the given case it seems to be dominating. It should be mentioned that Fig.~\ref{fig:analysis:dil1} is slightly misleading since two fast ions of different characteristics are considered. Both contribute to stabilise the growth rate in terms of dilution but their kinetic effects modify the growth rates in different ways. A better evaluation  is to consider each fast ion species independently. For this purpose the same three cases as in Fig.~\ref{fig:analysis:dil1} are repeated but only including either FD (Fig.~\ref{fig:analysis:compFD}) or He (Fig.~\ref{fig:analysis:compHe}).
Secondary and direct effects of FD reduces the growth rate only slightly (from $\gamma \approx \SI{0.25}{c_s/a}$ down to $\gamma \approx \SI{0.2}{c_s/a}$). Dilution on the other hand brings the growth rates further down to $\gamma \approx \SI{0.18}{c_s/a}$. While the secondary effects are stabilising the ITG growth rates, the direct effects of FD seems to be destabilising. The simulations have included magnetic fluctuations using $\beta_{\text{ref}} = 0.0033$ such that from Fig.~\ref{fig:analysis:EMBeta_fprim} we would expect a very small, but slightly stabilising influence of the fast deuterium density gradient. On the other hand, it should be pointed out that while in Fig.~\ref{fig:analysis:EMBeta_fprim} He (but without the temperature gradient) was also present, in Fig.~\ref{fig:analysis:compFD} this species has been removed entirely. Removing He from the simulations performed previously in Fig.~\ref{fig:analysis:EMBeta_fprim} might make the the density gradient of FD slightly destabilising. Apart from the density gradient there is also the direct effects of fast ion density and charge which have not been studied here, and might lead to a slight increase in the growth rates as is seen in Fig.~\ref{fig:analysis:compFD} when comparing the dilution and full FD case. 

Including only He leads to the different result shown in Fig.~\ref{fig:analysis:compHe}. Again dilution reduces the growth rates. For He this reduction is slightly less compared to FD in Fig.~\ref{fig:analysis:compFD}. A possible explanation is the higher temperature of FD which leads to a larger contribution in total $\beta$. Including also the direct, kinetic effects of fast He leads to a large drop in the growth rate. Given all the results to this point and associated discussions, we explain the large reduction when direct effects of He are included with the large He temperature gradient. 

Combining these two separate studies a more correct evaluation of the role of dilution in Fig.~\ref{fig:analysis:dil1} can be done. Both FD and He provide with a roughly equal contribution to dilution which stabilises ITG and reduces the growth rates. For FD this is the main stabilising effect, while its direct effects lead to a slight increase in the growth rates. With both dilution and kinetic effects FD contributes to only $\sim 25\% $ of the total stabilisation by fast ions. The rest comes from the ICRH generated He fast ions. A significant part is again because of dilution, but of equal (if not greater) importance is the large temperature gradient of He. Adding both the small reduction because of FD $(\Delta \gamma \approx \SI{0.05}{c_s/a})$ and the much larger reduction because of He $(\Delta \gamma \approx \SI{0.14}{c_s/a})$ we obtain the total reduction of the fast ions as seen in Fig.~\ref{fig:analysis:dil1}. Once again the fact that ICRH generated fast ions are most important in stabilising the ITG growth rates, has been confirmed.

\begin{figure}[h]
 \setlength\figureheight{0.2\textheight}
 \setlength\figurewidth{0.7\textwidth}
    \centering
    \input{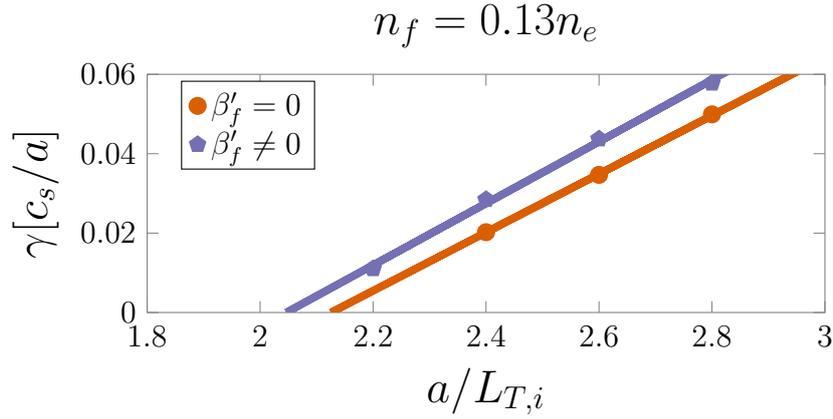}
    \caption{Scan in the ion temperature gradient. Results are shown including all secondary effects (purple pentagons) and without fast ion contribution to $\beta'$ (orange circles). A slight shift in the critical threshold is seen when $\beta_f'$ is included.}
    \label{fig:analysis:dil3}
  \end{figure}

Dilution in effect encapsulates a range of parameters shown in Fig.~\ref{fig:analysis:second}. We have confirmed a small effect because of the change in $Z_{\text{eff}}$ and collision frequencies. Also the fast ion contribution to the total radial derivative of $\beta$ only leads to a slight decrease in the critical threshold and does not affect the growth rates of ITG much. The latter is demonstrated in Fig.~\ref{fig:analysis:dil3} where a scan in the ion temperature gradient with and without the fast ion contribution to $\beta'$, is shown. Here $\beta_f = \beta_{\text{ref}}n_fT_f/n_{\text{ref}}T_{\text{ref}}$ . The effect from the fast ion contribution to $\beta$ itself, is more important.

\begin{figure}[H]
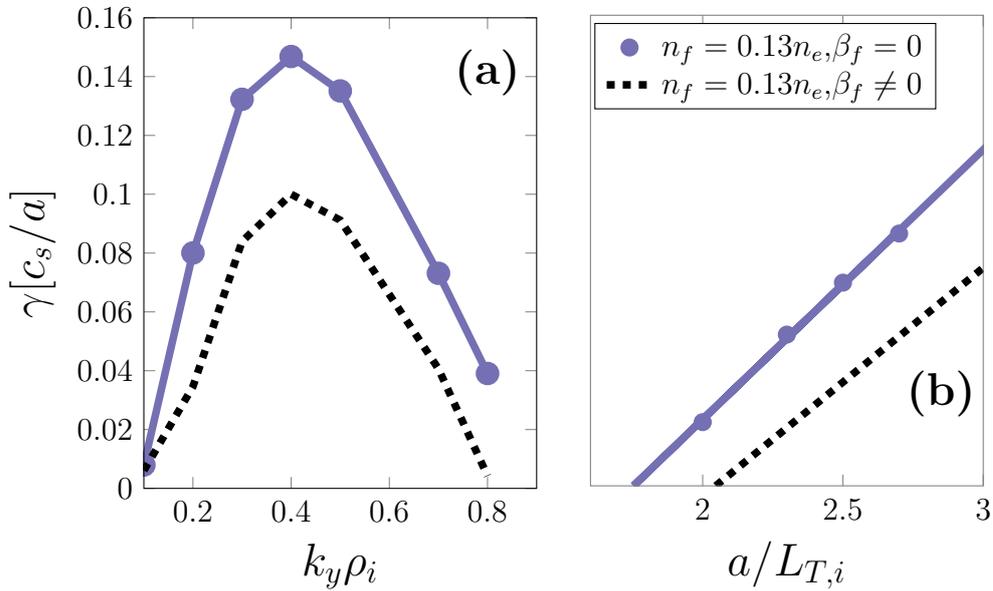

 \setlength\figureheight{0.33\textheight}
 \setlength\figurewidth{0.45\textwidth}
         \begin{subfigure}[b]{0.3\textwidth}
            \input{../../figures/EMfin/akyFin/beta_dens_growth.tex}
            \phantomcaption
            \label{fig:analysis:dil4_a}            
            \end{subfigure} 
        \hspace{3cm}                       
           \begin{subfigure}[b]{0.3\textwidth}
    \input{../../figures/EMfin/tgFin/beta_dens_growth.tex}
           \phantomcaption
            \label{fig:analysis:dil4_b}            
            \end{subfigure}   
    \caption{Scan in $k_y\rho_i$ (left) and ion temperature gradient (right) for varying fast ion density, but without their contribution to total $\beta$, ($\beta_f = 0$). For comparison we repeat the reference case with $n_f = 0.13n_e$ and $\beta_f = 0$ (purple circles) including fast ion contribution to total $\beta$ (dashed black line).}
    \label{fig:analysis:dil4}
  \end{figure}

We compare the reference case with and without fast ion contribution to $\beta$ in Fig.~\ref{fig:analysis:dil4}. The scan in $k_y\rho_i$ is shown in Fig.~\ref{fig:analysis:dil4_a} and for the ion temperature gradient in Fig.~\ref{fig:analysis:dil4_b}. Comparing with the reference  $n_f = 0.13n_e , \beta_f \ne 0$ reveals roughly a 35 \% reduction in the growth rates due to the fast ion contribution to total $\beta$ alone. In Fig.~\ref{fig:analysis:tg1} it is seen that $\beta_f$ increases the critical threshold of the ITG instability and reduces the slope of the linear fit. A large fraction of the stabilising effect because of dilution in Fig.~\ref{fig:analysis:dil1} is because of $\beta_f$ and thus electromagnetic effects. 

The analysis of dilution is complete by examining the role of quasineutrality.

\subsubsection*{Quasineutrality}\label{sec:analysis:QN}
\begin{figure}[h]
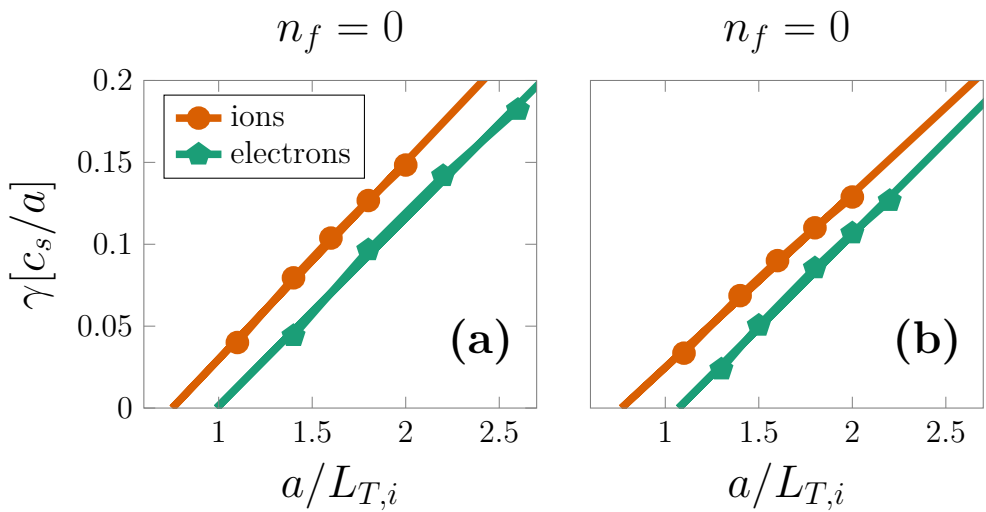

 \setlength\figureheight{0.25\textheight}
 \setlength\figurewidth{0.45\textwidth}
         \begin{subfigure}[b]{0.3\textwidth}
    \input{../../figures/ESfin/dilution/QN1.tex}
            \phantomcaption
            \label{fig:analysis:qn1}            
            \end{subfigure}  
            \hspace{3cm}                               
             \begin{subfigure}[b]{0.3\textwidth}        
     \input{../../figures/ESfin/dilution/QN1_EM.tex}
            \phantomcaption
            \label{fig:analysis:qn2}            
            \end{subfigure}      
    \caption{Scan in the ion temperature gradient, using the parameters of the reference discharge in Table~\ref{table:StudTurb:73param} but without fast ions. To remain quasineutral we have recalculated the main ion (orange circles) and electron (cyan pentagons) parameters respectively.}
    \label{fig:analysis:qn}
  \end{figure}

To account for a change in fast ion density and its gradient the density profiles of electrons and/or ions has to be modified for remaining quasineutral. To this point we have modified the electron parameters with
\eqre{
\label{eq:analysis:QN}
n_{e} = -\frac{\sum_{j\ne e,f} n_jZ_j + Z_{f}n_{f}}{Z_{e}}, \qquad
 \frac{a}{L_{n,e}} = -\frac{\sum_{j\ne e,f} n_jZ_j\frac{a}{L_{n,j}} + Z_{f}\frac{a}{L_{n,f}} n_{f}}{Z_{e}n_{e}},
}
where the sum is over all species excluding electrons and fast ions. As discussed in Chapter~\ref{CH:FIin} another possibility could have been to change the main ion parameters. Since electrons and main ions have different roles in the dynamics of ITG, using ions for satisfying quasineutrality instead might change our results.  We evaluate the difference between using either case in Fig.~\ref{fig:analysis:qn} where a scan in the ion temperature gradient is shown. The fast ion density is set to zero ($n_f = 0$) and density profiles of either main ions (orange circles) or electrons (purple pentagons) are recalculated for remaining quasineutral (along with the other secondary effects in Fig.~\ref{fig:analysis:second}). In Fig.~\ref{fig:analysis:qn1} only electrostatic fluctuations are included while Fig.~\ref{fig:analysis:qn2} also includes fluctuations in the magnetic field. 
\noindent
Using ions leads to a lower value of the critical ion temperature gradient threshold as expected, since there are now more ions that contribute in driving ITG unstable. Electrons are surprisingly not that different.  Given the change of the slope in the linear fit when magnetic fluctuations have been included in Fig.~\ref{fig:analysis:dil4} an even smaller difference is expected for higher main ion temperature gradients. Since  the reference discharge has $a/L_{t,i} = 3.56$ using ions for satisfying quasineutrality instead of electrons, should not change our results.   

\section{Quasi-linear prediction of the nonlinear diffusion coefficients}
We have investigated the role of fast ions in reducing ITG growth rates but the ultimate goal is to explain the large reduction in the nonlinear heat flux demonstrated by Citrin et al.~\cite{ref:Citrin}. For this purpose the exponential growth $\gamma$ of ITG has to be linked to the saturated value of the heat flux. This can be done with quasi-linear theory. 

A quasi-linear model predicts the nonlinear saturated value of the electrostatic potential which is used to estimate nonlinear diffusion coefficients and heat fluxes. It is of particular interest in transport models used in the design of fusion reactors which requires an accurate description of particle and heat transport. Computationally complex and time consuming nonlinear gyrokinetic simulations are not feasible and instead these tools rely on the quasi-linear model for predicting the turbulent transport of heat and particles. The QuaLiKiz predictive tool~\cite{ref:Bourdelle2007} is a well known example which implements a quasi-linear model. In this section we will use a quasi-linear model in the attempt of generalising the deduced role of fast ions in stabilising ITG growth rates to their nonlinear reduction of the heat flux. The quasi-linear model based on the work done by Casati et al.~\cite{ref:Casati} is reviewed.

In the quasi-linear model considered in this thesis, the maximal perpendicular diffusion coefficient, measuring the transport of particles across flux surfaces, is estimated with
\eqre{
\label{eq:analysis:diff}
D_{\perp,\text{max}} \approx \frac{\gamma_{k_{\perp,\text{max}}}}{\left <k_{\perp,\text{max}}^2\right>}.
}
In a simple random walk model the diffusion coefficient is the ratio between the step size squared, and the step time. For turbulence this can be thought of as the time and length a turbulent structure can propagate in the plasma before it dissipates and mixes with the environment. The quasi-linear model uses the \textit{mixing-length} assumption to assume a time step equal to the growth $\gamma_{k_{\perp,\text{max}}}$ of the unstable mode and the step length is the associated averaged perpendicular wavenumber 
\eqre{
\label{eq:analysis:kavg}
\left <k_{\perp}^2\right> = k_{y}^2\left(1+(\hat s-\alpha)^2\left<\theta^2\right>\right),
} 
and the poloidal extent of the mode structure is characterised by the quantity
\eqre{
\label{eq:analysis:thetaAvg}
\left<\theta^2\right> = \frac{\int \theta^2\left |\varphi\right|^2\dif{\theta}}{\int \left |\varphi\right|^2\dif{\theta}}.
}
In Eq.~\eqref{eq:analysis:diff} we use the value of the averaged perpendicular wavenumber $k_{\perp,\text{max}}$ and the associated growth $\gamma_{k_{\perp,\text{max}}}$ that maximises the diffusion coefficient. In Eq.~\eqref{eq:analysis:kavg} the parameter $\alpha$ is defined through~\cite{ref:Tegnered2017}
\eqre{
\label{eq:analysis:MHDa}
\alpha = q^2\sum_j\beta_j\left(\frac{R}{L_{n,j}} + \frac{R}{L_{T,j}}\right),
}
where the sum runs over all (five) species in the plasma. The diffusion coefficient in Eq.~\eqref{eq:analysis:diff} is directly related to the nonlinear saturated electrostatic potential $\varphi_{\text{sat}}$ which in turn is used to estimate nonlinear heat fluxes. With the growth rates, the associated $k_y$ wavenumbers and the electrostatic potentials from our linear simulations we can evaluate the role of fast ions in reducing diffusion coefficients, the saturated value of the electrostatic potential and consequently the heat fluxes. 

\begin{figure}
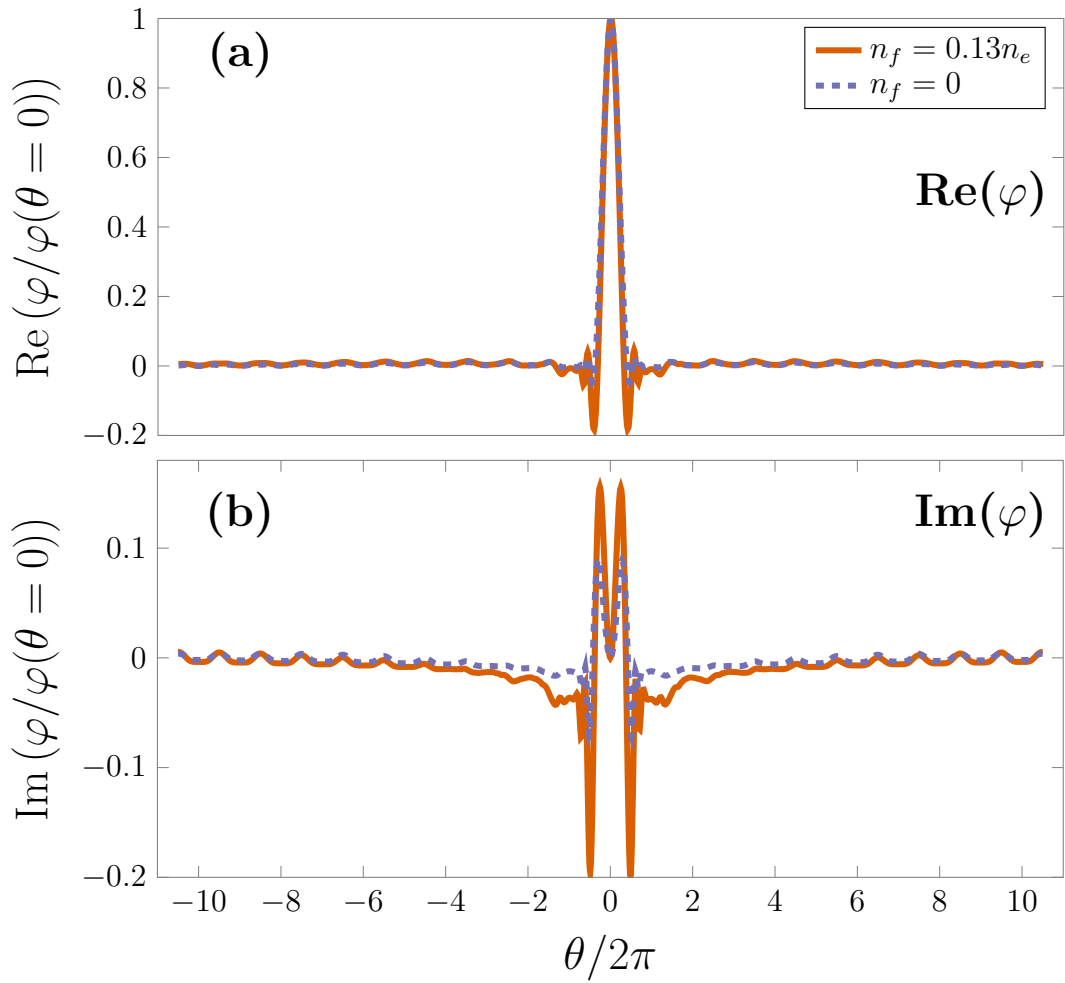

 \setlength\figureheight{0.3\textheight}
 \setlength\figurewidth{0.9\textwidth}
     \centering
     \begin{subfigure}[b]{\textwidth}
\input{../../figures/stuff/realEigenfun.tex}        
\phantomcaption
            \label{fig:analysis:rePhi}
    \end{subfigure}\\
           \hspace{-0.9cm}
                \centering
     \begin{subfigure}[b]{\textwidth}
    \centering
    \input{../../figures/stuff/ImEigenfun.tex}
\phantomcaption
            \label{fig:analysis:imPhi}
    \end{subfigure}   
    \caption{Real and imaginary parts of the normalised electrostatic potential as a function of the angle $\theta$ including fast ions (solid orange) and without fast ions (dashed purple). A dominant peak at $\theta = 0$ can be seen. In a tokamak plasma this corresponds to the outboard midplane where the ITG instability is the worst, as discussed in Section~\ref{sec:studTurb:ITG}.}
    \label{fig:analysis:Phi}
  \end{figure}

Consider Fig.~\ref{fig:analysis:aky1} and take the growth rates for the case without fast ions ($n_f = 0$) and with fast ions ($n_f = 0.13n_e$). The real and imaginary part of the electrostatic potential as a function of $\theta$ for the peak growth rates of the two cases are depicted in Fig.~\ref{fig:analysis:Phi}. We integrate the absolute value of the electrostatic potential $\varphi = \text{Re}(\varphi) + i\text{Im}(\varphi)$ according to Eq.~\eqref{eq:analysis:thetaAvg}, calculate $\alpha$ with Eq.~\eqref{eq:analysis:MHDa} and compute the averaged wavenumber in Eq.~\ref{eq:analysis:kavg}. The geometrical parameters, $\hat s$, $q$ and the gradient scale lengths are as before taken from Table~\ref{table:StudTurb:73param}. We use Eq.~\eqref{eq:analysis:diff} and evaluate the maximum value of the diffusion coefficient for the two cases, with and without fast ions to obtain  $D_{\perp,\text{max},\text{with FI}}/D_{\perp,\text{max},\text{no FI}} \approx 0.41$. The quasi-linear model predicts 59\% reduction in the maximum diffusion coefficient, while for the saturated nonlinear heat flux reported by Citrin et al.~\cite{ref:Citrin} it is approximately 85 \%. The quasi-linear model predicts a large stabilising effect on the heat fluxes because of the fast ions but the estimate is significantly smaller compared to the actual reduction obtained from the nonlinear gyrokinetic simulations.  A large part of the stabilisation by fast ions for the 73224 JET discharge is due to nonlinear interactions which cannot be fully explained with linear physics. This is a discovery shown previously by Citrin et al.~\cite{ref:Citrin}. It appears that to fully capture the role of fast ions in reducing the heat fluxes for this discharge, nonlinear simulations are required. Nevertheless, our findings for the reduction of the ITG growth rates can be used for designing desired nonlinear simulations and fully explain the role of fast ions in reducing the nonlinear heat flux.

We want to stress that the quasi-linear model is in essence an electrostatic theory since only the electrostatic potential is taken into account in calculating averaged perpendicular wavenumbers in Eq.~\eqref{eq:analysis:kavg}. Citrin on the other hand, included electromagnetic fluctuations in the nonlinear simulations of the heat flux. The comparison between the electrostatic quasi-linear model in predicting nonlinear electromagnetic stabilisation of the heat flux is therefore not entirely equitable. A proper comparison would require a more general quasi-linear model to be developed where also the magnetic potential is taken into account in Eq.~\eqref{eq:analysis:thetaAvg}. 

\clearpage

\setlength{\parskip}{0pt} 

\chapter{Summary and Discussion}\label{CH:discussion}

In this work the role of fast ions in stabilising the ion temperature gradient mode (ITG) is investigated. We discuss the properties of this energetic species and how, as a consequence of their high contribution to the total pressure, they change the magnetic geometry. We find a negligible impact on most of the shaping parameters but a change in the derivative of the Shafranov shift and safety factor profile, which both affect the linear growth rates of the ITG instability. The growth rates are computed by solving the linear gyrokinetic equation numerically, using the gyrokinetic continuum code \texttt{GS2}. The reference input parameters to \texttt{GS2} is a model of the 73224 discharge from the JET tokamak, motivated by the strong fast ion stabilisation effect in reducing the nonlinear heat fluxes. The discharge contains two different types of fast ions: fast deuterium with a high density gradient, and fast helium with a high temperature gradient. These are used to model NBI and ICRH heated fast particles respectively.

Using this reference case an in depth study of the ITG stabilisation is performed. We isolate each parameter of the energetic species and study their effects on the growth rates separately. Results reveal the fast ion temperature gradient to be stabilising for high temperatures of the energetic species. The role of the fast ion density gradient depends on plasma $\beta$. For small $\beta$ the fast ion density gradient is destabilising while for larger $\beta$ it becomes stabilising instead. Therefore, at equal pressure, we find ICRH generated fast ions to be more efficient in reducing the growth rates as compared to NBI injected fast ions. The findings are compared with an electrostatic dispersion relation which includes the response of electrons, ions and fast ions. This approximate model successfully predicts the role of the fast ion gradients, at low $\beta$.

The stabilising role of the fast ions is most pronounced at the critical $\beta$ where the kinetic ballooning mode (KBM) is destabilised. In this linear treatment we also find parallel magnetic fluctuations to be negligible compared with including perpendicular magnetic fluctuation for value of $\beta$ below the threshold of KBM. 

The influence of fast ion dilution of the thermal species, is investigated. Dilution explains most of the reduction in growth rates for NBI generated fast ions, but for ICRH the temperature gradient is of equal, if not greater importance. A large part of the stabilising influence from dilution comes from the fast ion contribution to the total $\beta$.
 
 Finally the quasi-linear model is applied in the attempt of using linear results to explain the fast ion stabilisation of heat fluxes.  We find the quasi-linear prediction to not be sufficient in explaining the order-of magnitude decrease in the nonlinear heat fluxes.  
 
\section{Outlook}
Naturally, this work has involved many simplifications which have to be addressed in the future. First of all, the findings presented here are only for one particular discharge and at one radial position. We have treated the problem locally, but equivalent investigations with global geometry should be done to extrapolate our findings to real tokamak scenarios. Next, although it is concluded that the change in the magnetic geometry because of fast ions should not be neglected entirely, we perform all of our gyrokinetic simulations keeping the geometry fixed. Corresponding cases using real geometry which matches the given pressure profiles, could yield very different results. On a similar note, we have mostly isolated the effect of individual parameters, such as fast ions density, temperature, their gradients and contribution to total $\beta$ in our scans, but combinations thereof are equally important and more experimentally relevant. 

Another simplification is that the fast ion species is described with a Maxwellian distribution function, which might be incorrect. We know that the gradient of the distribution function appears in the gyrokinetic equation. Upon calculating this term we find the energy dependence in the Maxwellian distribution function to be coupled to the temperature gradient of the species. Our finding that the fast ion temperature gradient is important for stabilisation motivates the need to use a proper distribution function when describing the fast particle population. A perhaps better choice could be to use a slowing down distribution function~\cite{ref:Wilkie2015}. In any case, \texttt{GS2} is a natural choice for such numerical studies, as its ``alphas'' branch can be used to calculate the turbulent transport of fast ions with arbitrary energy distributions~\cite{ref:Wilkie2017,ref:Pusztai16}. We note that NBI and ICRH fast ions can develop significant velocity space anisotropies, which affect their transport, although it is not accounted for here, for simplicity. These anisotropies can also generate poloidal asymmetries with implications for the transport of highly charged impurities~\cite{ref:Pusztai14,ref:Pusztai2013,ref:Kazakov12}.

The quasi-linear model was applied for predicting nonlinear diffusion coefficients and heat-fluxes. Even if the stabilisation is less strong, the linear dependencies of ITG growth rates on various fast ion parameters may provide insight and ideas for running nonlinear simulations, where the large computational cost prevents equivalent parameter scans to be performed. From our findings the fast ion temperature gradients should correspondingly decrease the diffusion, while the fast ion density gradient should lead to stronger transport (at low $\beta$). But again, the accuracy in the quasi-linear estimate may be limited given the much stronger stabilising effect observed nonlinearly. There are beta sensitive phenomena, such as zonal flows~\cite{ref:Catto17}, which may play a role in the nonlinear stabilisation of ITG turbulence. Additional nonlinear simulations are required and welcome to verify our linear results.
 
Finally, this study is relevant because of the present, but especially future importance of fast ions and their role in successfully operating fusion reactors. Injected fast ions are used to heat the plasma, but are also important for the formation of internal transport barriers (ITBs) which significantly improve the confinement time~\cite{ref:Tardini2007}. Other fast ions which we do not explicitly study here are the products of deuterium-tritium fusion reactions, the energetic alpha particles. These particles are essential for reaching ignition and are the main source of heat in a self-sustaining plasma~\cite{ref:Freidberg2007}. The high energetic alpha particles will therefore have an important role already in ITER and future fusion devices. The characteristics of the alpha particles is similar to NBI, having a high-density gradient. But since their density is significantly smaller the role of dilution is not equally important. Given the dependence of fast ion density gradients with $\beta$ we may however, to some extent, expect the alpha particles to stabilise ITG turbulence at the higher $\beta$ regime, and therefore improve the confinement time. 

Many questions remain to be answered, but with this work we have, once again, confirmed the importance of fast ions in stabilising the turbulence. We increased the knowledge of how and why these fast particles affect ITG turbulence and provided insight into possible nonlinear and perhaps even experimental role of fast ions on ITG turbulence. 

\cleardoublepage
\addcontentsline{toc}{chapter}{Bibliography}
\printbibliography 

\cleardoublepage
\appendix
\setcounter{page}{1}
\pagenumbering{Roman}			

\chapter{Global effects}
\setlength{\parskip}{0pt} 
\section{Input Parameters to \texttt{CHEASE}}\label{app:geom:1}
The non-default input parameters used for running \texttt{CHEASE} are presented in Table~\ref{table:PreSim:Geometry:CHnam}. For definitions and values of default parameters we refer to the complete manuscript~\cite{ref:CHEASE}.
\begin{table}[h]
\def\arraystretch{2}\tabcolsep=6pt
\caption{Values of \texttt{chease\_namelist} variables for computing the toroidal MHD equilibrium.}
\centering
\begin{tabular}{@{}llllllll@{}}
\toprule
\texttt{nsurf}  & 6  & \texttt{ns}    & 30  & \texttt{nmeshd} & 1         & \texttt{nideal} & 9         \\
\texttt{nppfun} & 4  & \texttt{nt}    & 30  & \texttt{npoidd} & 1         & \texttt{epslon} & $1e^{-8}$ \\
\texttt{nfunc}  & 4  & \texttt{nrbox} & 400 & \texttt{dplace} & $-1.5707$ & \texttt{ninsca} & 20        \\
\texttt{neqdsk} & 1  & \texttt{nzbox} & 400 & \texttt{dwidth} & 0.157     & \texttt{ninmap} & 20        \\
\texttt{npsi}   & 40 & \texttt{niso}  & 200 & \texttt{solpdd} & 0.6       & \texttt{nsym}   & 1         \\ \bottomrule
\end{tabular}
\label{table:PreSim:Geometry:CHnam}
\end{table}

\section{Effect of Miller parameters on the growth rates}\label{app:geom:2}
The gyrokinetic simulations for evaluating the role of the Miller parameters on ITG growth rates are presented in this section. To obtain a reasonable starting point the reference Miller parameters for the 73224 JET discharge presented in Table~\ref{table:StudTurb:73param}, are used. For convenience we summarise these reference values in Table~\ref{table:geom:refParam}. The gyrokinetic simualtions are performed in the neighbourhood of these values. All scans include both electric and magnetic fluctuations.

\begin{table}[H]
\centering
\caption{The reference values of the Miller parameters used in the simulations presented in this section. All values are taken from the model of the 73224 JET discharge presented in Table~\ref{table:geom:refParam}.}
\label{table:geom:refParam}
\begin{tabular}{@{}lllllll@{}}
\toprule
$\kappa$ & $\delta$ & $s_{\delta}$ & $s_{\kappa}$ & $q$ & $\hat s$ & $\Delta$ \\ \midrule
1.26     & 0.030   & 0.088    & 0.08    & 1.74 & 0.523    & -0.14     \\ \bottomrule
\end{tabular}
\end{table}

\begin{figure}
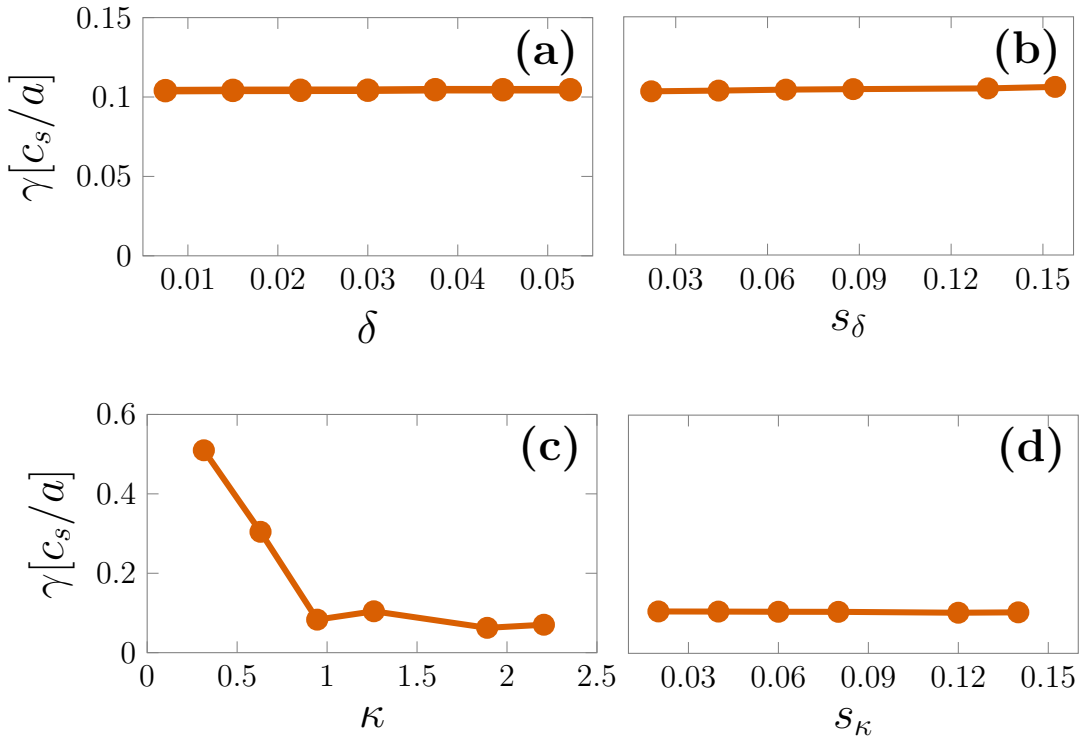

 \setlength\figureheight{0.2\textheight}
 \setlength\figurewidth{0.5\textwidth}
    \hspace{-0.4cm}
    \begin{subfigure}[b]{0.33\textwidth}
    \input{../../figures/Simulations/linearSim/geometry_CH2/delta_growth.tex}
            \phantomcaption
                \label{fig:analysis:delta}
            \end{subfigure}
                \hspace{3cm}
            \begin{subfigure}[b]{0.33\textwidth}
   \raisebox{0.08cm}{\input{../../figures/Simulations/linearSim/geometry_CH2/sdelta.tex}}
            \phantomcaption
                \label{fig:analysis:sdelta}
    \end{subfigure} \\ 
        \begin{subfigure}[b]{0.33\textwidth}
     \input{../../figures/Simulations/linearSim/geometry_CH2/kappa_growth.tex}
            \phantomcaption
            \label{fig:analysis:kappa}
    \end{subfigure}
    \hspace{2.8cm}
        \begin{subfigure}[b]{0.33\textwidth}
     \raisebox{-0.1cm}{\input{../../figures/Simulations/linearSim/geometry_CH2/skappa.tex}}
            \phantomcaption
            \label{fig:analysis:skappa}
    \end{subfigure}
        \caption{Growth rates for varying $\delta$ (a), $s_{\delta}$ (b), $\kappa$ (c) and $s_{\kappa}$ (d) respectively. The scans are performed in the neighbourhood of the reference parameters presented in Table~\ref{table:geom:refParam}.}
\label{fig:analysis:geometry1}
  \end{figure}

\begin{figure}
 \setlength\figureheight{0.2\textheight}
 \setlength\figurewidth{\textwidth}

    \input{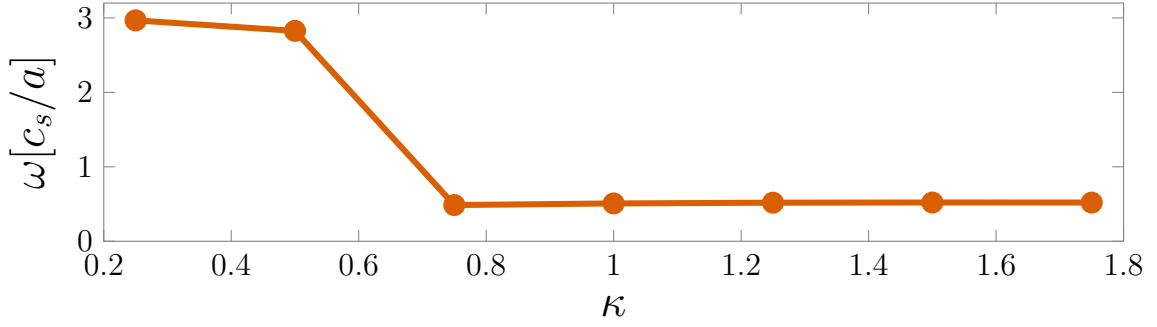}
    \caption{Real frequencies for varying $\kappa$. Above $\kappa = 1$ ITG is the dominant mode leading to the nonzero growth rates seen in Fig.~\ref{fig:analysis:kappa}.}
                \label{fig:analysis:kappaReal}
  \end{figure}

Growth rates for varying $\delta$, $\kappa$ and their derivatives are shown in Fig.~\ref{fig:analysis:geometry1}. Neither $\delta$ (Fig.~\ref{fig:analysis:delta}), $s_{\delta}$ (Fig.~\ref{fig:analysis:sdelta}) nor $s_{\kappa}$ (Fig.~\ref{fig:analysis:skappa}) change the growth rates for the selected range of parameter values. Below $\kappa = 1$ in Fig.~\ref{fig:analysis:kappa} there is a significant increase in the growth rates. Further inspection of the corresponding real frequencies in Fig.~\ref{fig:analysis:kappaReal} indicates a change in the dominant mode for small values of $\kappa$. In equivalent simulations (not shown) without magnetic fluctuations the large increase in both real and imaginary frequencies could not be discovered. The mode leading to the almost discontinuous step in both growth rates (Fig.~\ref{fig:analysis:kappa}) and mode frequencies, is electromagnetic, similarly to the kinetic ballooning mode (mentioned in Section~\ref{sec:analysis:beta}).

\begin{figure}
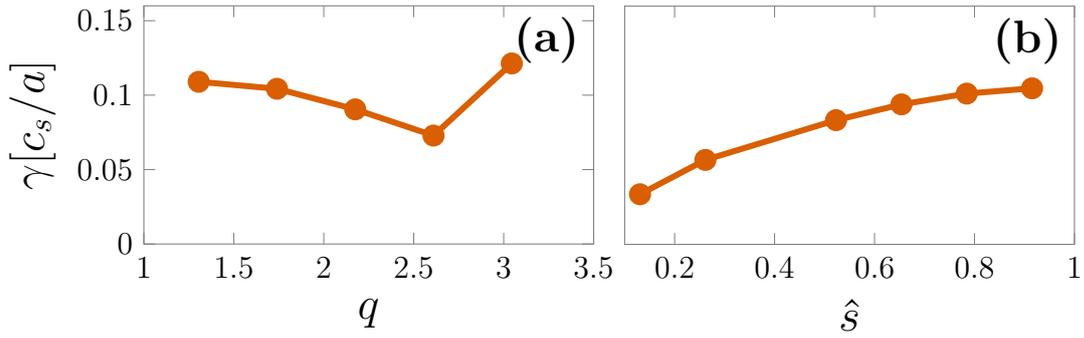
 
 \setlength\figureheight{0.2\textheight}
 \setlength\figurewidth{0.5\textwidth}
            \begin{subfigure}[b]{0.33\textwidth}
    \centering
    \input{../../figures/Simulations/linearSim/geometry_CH2/q_growth.tex}
                            \phantomcaption
                \label{fig:analysis:q}
    \end{subfigure}   
        \hspace{3cm}
        \begin{subfigure}[b]{0.33\textwidth}
      \raisebox{-0.04cm}{\input{../../figures/Simulations/linearSim/geometry_CH2/shat_growth.tex}}
            \phantomcaption
            \label{fig:analysis:shat}
    \end{subfigure}
    \caption{Growth rates for varying safety factor (a) and magnetic shear (b).}
\label{fig:analysis:geometry2}
  \end{figure}

\begin{figure}
 \setlength\figureheight{0.2\textheight}
 \setlength\figurewidth{0.9\textwidth}
     \input{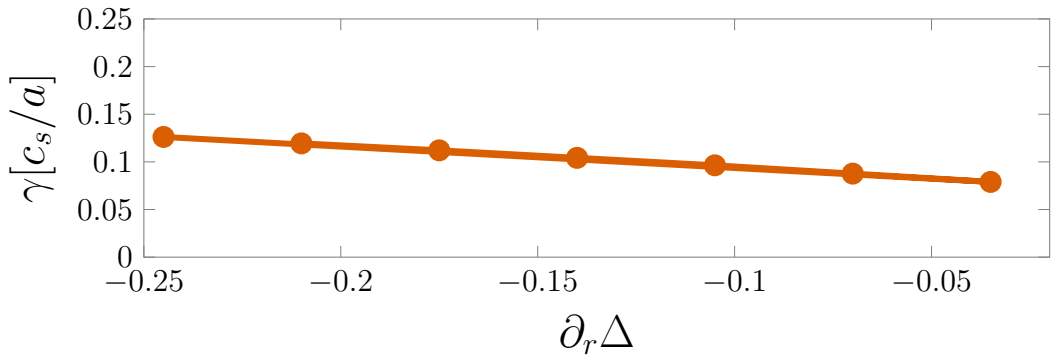}
         \caption{Growth rates for varying radial derivative of the Shafranov shift.}
            \label{fig:analysis:shift}
  \end{figure}

A change in the growth rate can be discovered for varying safety factor and magnetic shear in Figs.~\ref{fig:analysis:q}, \ref{fig:analysis:shat} respectively. Increasing the safety factor from the nominal value $q=1.74$ stabilises first the growth rates, but for $q>2.6$ they start to increase instead. Increasing the magnetic shear is destabilising the ITG mode. Finally, we consider a change in the radial derivative of the Shafranov shift, $\partial_r\Delta$ shown in Fig.~\ref{fig:analysis:shift}. Increasing the shift leads to a decrease in the growth rates. 
 
Given the influence of fast ions with a density $n_f = 0.05n_e$ on the shaping parameters presented in Section~\ref{ref:geom:Eff} and their effects on the growth rates shown in Fig.~\ref{fig:analysis:geometry1} we obtain a negligible effect of $\delta, \kappa, s_{\delta}$ or $s_{\kappa}$. Fast ions slightly reduce the Shafranov shift and according to Fig.~\ref{fig:analysis:shift} this leads to an increase in the growth rates. Finally, the energetic ions slightly decrease the safety factor and increase the magnetic shear which from Fig.~\ref{fig:analysis:geometry2} leads to an increase in the growth rates. Summing up each individual contribution given the change in the various Miller parameters because of fast ions we obtain $\sim 25 \%$ increase in $\gamma$.
\clearpage
\chapter{Studying turbulence and the ITG mode}
\setlength{\parskip}{0pt} 
\section{Scan in the resolution parameters}\label{app:D}
In this section gyrokinetic simulations of ITG growth rates for varying values of the resolution parameters are presented. The result is used in the discussion in Section~\ref{sec:studTurb:ResParam} where the values of these parameters, used in all simulations in this work, are presented. For each scan we vary only one of the five resolution parameters while keeping the others fixed with the values presented in Table~\ref{table:appD:resparamLow}. Since the role of $ntheta$ is closely linked to $nperiod$ the scan in this parameter is performed at a high value of $nperiod=11$ where the growth rates are relatively well resolved. 

\begin{table}[H]
\centering
\caption{Values of the seven resolution parameters used as default in the simulations presented in this section.}
\label{table:appD:resparamLow}
\begin{tabular}{@{}llllll@{}}
\toprule
$nperiod$ & $omegatol$  & $delt \left[\unit{a/v_{t,i}}\right]$ & $ntheta$ & $ngauss$ & $negrid$   \\ \midrule
\addlinespace[0.5em]
3       & $2e^{-3}$         & 1.0    & 16      & 8      & 9     \\ \bottomrule
\end{tabular}
\end{table}

Ideally, we seek to find values of the resolution parameters such that, if they are further increased, no change in the growth rates should be seen. Given the computational cost of running even linear simulations with higher resolution we relax the requirement and investigate when the growth rates seem to saturate. For all scans we present the result in terms of growth rates divided by $\gamma_{low}$. This is the growth rate obtained when using all the parameter values presented in Table~\ref{table:appD:resparamLow}.

\begin{figure}[H]
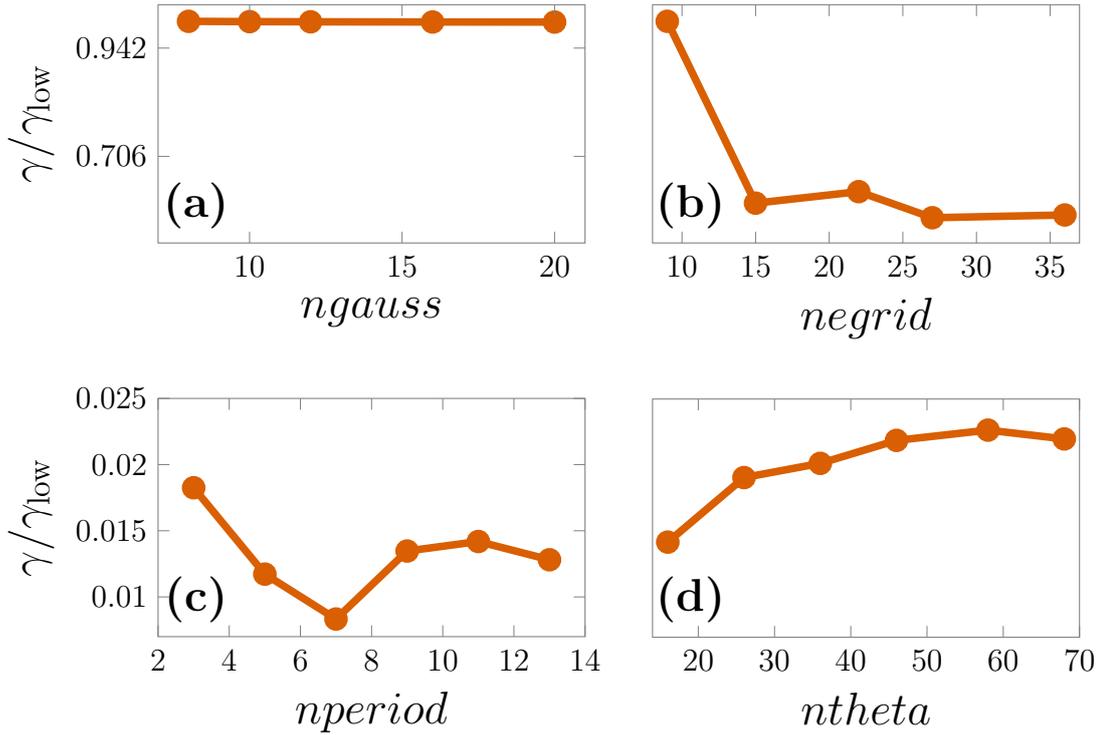

 \setlength\figureheight{0.2\textheight}
 \setlength\figurewidth{0.48\textwidth}
    \centering
   \begin{subfigure}[b]{0.35\textwidth}
       \centering
    \input{../../figures/stuff/ngauss.tex}
        \phantomcaption
        \label{fig:ngauss}
        \end{subfigure}
        \hspace{3cm}
      \begin{subfigure}[b]{0.35\textwidth}
          \centering
    \raisebox{-0.16cm}{\input{../../figures/stuff/negrid.tex}}
        \phantomcaption
        \label{fig:negrid}    
        \end{subfigure}
   \begin{subfigure}[b]{0.35\textwidth}
       \centering
    \input{../../figures/stuff/nperiod.tex}
        \phantomcaption
        \label{fig:nperiod}    
        \end{subfigure}
                \hspace{3cm}
      \begin{subfigure}[b]{0.35\textwidth}
          \centering
    \raisebox{0.11cm}{\input{../../figures/stuff/ntheta.tex}}
        \phantomcaption
        \label{fig:ntheta}    
        \end{subfigure}        
    \caption{Rescaled growth rates for a scan in $ngauss$ (top left) $negrid$ (top right) $nperiod$ (bottom left) and $ntheta$ (bottom right). Growth rates have been rescaled by the initial value, at low resolution as presented in Table~\ref{table:appD:resparamLow}. These constant values are used for all cases, except for the scan in $ntheta$, where a higher value $nperiod = 11$ has been used.}
\label{fig:StudTurb:Res1}
  \end{figure}

Increasing the parameter $ngauss$  as shown in Fig.~\ref{fig:ngauss} does not change the growth rates at all and a low value of $ngauss = 8$ is therefore sufficient to use. Increasing the number of energy grid points $negrid$ in Fig.~\ref{fig:negrid} has first a strong influence on the growth rates, but then seem to saturate at $negrid \approx 36$. The length along the magnetic field, selected with $nperiod$ has a strange influence on the growth rates as seen in Fig.~\ref{fig:nperiod}. First there is a strong reduction, then increase and then a slightly smaller reduction in the growth rates again. A proper investigation would require even higher values of $nperiod$ to make sure the growth rates do not change, but given the large increase in the computational time for a unit increase in $nperiod$ we suffice with a value of $nperiod = 11$. At this value of $nperiod$ we increase $ntheta$ and obtain Fig.~\ref{fig:ntheta}. This parameter leads to an increase in the growth rates but seems to saturate around $ntheta = 58$.

Finally a scan in $delt$, the parameter governing the time resolution important to resolve physics on small time scales, is shown in Fig.~\ref{eq:delt}. Even a small reduction in this parameter also leads to a large increase in the computation time, explaining the few values of $delt$ shown in Fig.~\ref{eq:delt}. It is clear that $delt < 0.1$ would provide completely wrong values of the growth rates. Above this value the growth rates seems to oscillate. We choose to use $delt = \SI{0.03}{a/v_{t,i}}$ to obtain relatively well resolved values. 

  \begin{figure}[h]
 \setlength\figureheight{0.2\textheight}
 \setlength\figurewidth{0.7\textwidth}
    \centering
    \input{../../figures/stuff/delt.tex}
    \caption{Same as in Fig.~\ref{fig:StudTurb:Res1} but for a scan in $delt$.}
\label{eq:delt}
\end{figure}   
\clearpage

\end{document}